\documentclass[preprint]{aastex}
\usepackage{graphicx}
\usepackage{amsmath}
\usepackage[usenames,dvips]{color}

\usepackage[maxfloats=44]{morefloats}

\newcommand{\be}{\begin{equation}}
\newcommand{\bel}[1]{\begin{equation}\label{eq:#1}}
\newcommand{\ee}{\end{equation}}
\newcommand{\bd}{\begin{displaymath}} 
\newcommand{\ed}{\end{displaymath}}   
\newcommand{\bea}{\begin{eqnarray}}
\newcommand{\beal}[1]{\begin{eqnarray}\label{eq:#1}}
\newcommand{\eea}{\end{eqnarray}}

\newcommand{\lsim }{{\lower0.8ex\hbox{$\buildrel <\over\sim$}}}
\newcommand{\gsim }{{\lower0.8ex\hbox{$\buildrel >\over\sim$}}}

\def\simge{\mathrel{%
   \rlap{\raise 0.511ex \hbox{$>$}}{\lower 0.511ex \hbox{$\sim$}}}}
\def\simle{\mathrel{
   \rlap{\raise 0.511ex \hbox{$<$}}{\lower 0.511ex \hbox{$\sim$}}}}

\newcommand{\Msun}{\ifmmode {M_{\odot}}\else${M_{\odot}}$\fi}
\newcommand{\Lsun}{\ifmmode {L_{\odot}}\else${L_{\odot}}$\fi}
\newcommand{\Rsun}{\ifmmode {R_{\odot}}\else${R_{\odot}}$\fi}

\shorttitle{Radiation Feedback Forces in GMCs}
\shortauthors{Raskutti, Ostriker, Skinner}

\begin{document}
\title{Numerical Simulations of Turbulent, Molecular Clouds Regulated by Radiation Feedback Forces I:
Star Formation Rate and Efficiency}  

\author{Sudhir Raskutti, Eve C. Ostriker, and M. Aaron Skinner}
\affil{Department of Astrophysical Sciences, Princeton University, Princeton, NJ 08544, USA}
\email{raskutti@astro.princeton.edu, eco@astro.princeton.edu, askinner@astro.princeton.edu}

\slugcomment{\today}

\begin{abstract}
Radiation feedback from stellar clusters is expected to play a key
role in setting the rate and efficiency of star formation in giant
molecular clouds (GMCs). To investigate how radiation forces influence  
realistic turbulent systems, we have conducted a series of numerical
simulations employing the {\it Hyperion} radiation hydrodynamics
solver, considering the regime that is optically thick to ultraviolet (UV) and
optically thin to infrared (IR) radiation.  Our model clouds cover initial
surface densities between $\Sigma_{\rm cl,0} \sim 10-300~M_{\odot}~{\rm pc^{-2}}$, with varying
initial turbulence. We follow them through
turbulent, self-gravitating collapse, formation of star clusters, and cloud dispersal by
stellar radiation. All our models display a lognormal distribution of
gas surface density $\Sigma$; for an initial virial parameter
$\alpha_{\rm vir,0} = 2$, the lognormal standard deviation is $\sigma_{\rm ln \Sigma} = 1-1.5$ and the
star formation rate coefficient $\varepsilon_{\rm ff,\bar\rho} = 0.3-0.5$, 
both of which are sensitive to turbulence but not radiation feedback. The
net star formation efficiency $\varepsilon_\mathrm{final}$ increases
with $\Sigma_{\rm cl,0}$ and decreases with $\alpha_{\rm vir,0}$.  We
interpret these results via a simple conceptual framework, whereby 
steady star formation increases the radiation force,
such that local gas patches at successively higher $\Sigma$
become unbound. Based on this formalism (with fixed $\sigma_{\rm ln \Sigma}$),
we provide an analytic upper bound on $\varepsilon_\mathrm{final}$,
which is in good agreement with our numerical results. The final star formation
efficiency depends on the distribution of Eddington ratios in the
cloud and is strongly increased by turbulent compression of gas.
\end{abstract}


\maketitle

\section{Introduction}
\label{Sec:Introduction}

Gravitational collapse within Giant Molecular Clouds (GMCs) leads to
star formation, and the radiation force produced by young, hot stars
can be dynamically important in regulating this process. Radiation
forces may contribute to driving turbulence within clouds, and if
strong enough, can halt global collapse and lead to the overall
dispersal of a star forming cloud \citep{Odell1967, Elmegreen1983,
  Scoville2001, KrumholzMatzner2009, Fall2010, Murray2010,
  KrumholzDekel2010, Sales2014,Kim2016}. If radiation forces dominate over
other forms of feedback, they may be responsible for setting both the
mean star formation rate (SFR) and the net star formation efficiency
(SFE) over a cloud's lifetime.

Indications of some form of feedback regulating star formation can be
found in direct observations of star-forming Milky Way Clouds, which
have low observed SFEs 
$M_*/M_{\rm gas}\sim 0.002-0.20$ \citep{CohenKuhi1979,
  Myers1986, MooneySolomon1988, WilliamsMckee1997, Carpenter2000,
  Evans2009, Lada2010, Murray2011, KennicuttEvans2012, Garcia2014},
and clear signatures of disruption by massive stars for clouds at the
higher end of the SFE distribution.  Assessing the net SFE 
over the
lifetime of a cloud ($\varepsilon_{\rm final} \equiv M_{\rm *, final}/M_{\rm gas, init}$)
is observationally challenging, since up to the
time of its dissolution, the cloud's stellar population is secularly
increasing, while at late stages the observed total mass has dropped
below the initial value through gas mass loss.
However, observing the instantaneous SFE across a population of clouds still
provides constraints, assuming a fair sample across different stages of the
cloud lifecycle. If, for example, the SFR is
steady and dispersal of gas takes place rapidly compared to the
lifetime of a cloud, then the observed instantaneous
SFE would uniformly cover the
range between zero and $\varepsilon_{\rm final}$, so that the average observed SFE 
would be roughly half of $\varepsilon_{\rm final}$.

Additional indirect evidence that is cited to explain low SFEs by
feedback in GMCs comes from the long molecular depletion times 
($t_{\rm dep}\equiv M_{\rm mol}/\dot M_* \sim
{\rm Gyr}$ for gas traced by CO) observed in extragalactic studies
\citep{Bigiel2008, Schruba2011, Saintonge2011, Rahman2012, Leroy2013, Genzel2015}. 
If molecular gas is concentrated in GMC-like structures with lifetimes
of a few tens of Myrs \citep{Leisawitz1989, Mizuno2001b, Kawamura2009},
then a low lifetime SFE $\sim 0.01$ would be implied by $t_{\rm
  dep}\sim {\rm Gyr}$, potentially demanding strong feedback.
Alternatively, if molecular clouds have long lifetimes, radiation
forces and other feedback effects could in principle keep $t_{\rm
  dep}$ long compared to the gas freefall time
$t_{\rm  ff}$
\citep[see][]{ZuckermanPalmer1974, KrumholzTan2007, Krumholz2012} if
they were able to sustain high turbulence levels without destroying the cloud 
\citep{KrumholzMckee2005, PadoanNordlund2011, Padoan2012,
  FederrathKlessen2012}. Thus, it has been argued that the
inefficiency in both GMC SFEs and SFRs can be accounted for by massive
stars supporting and then destroying their host GMCs \citep[see
  reviews of][]{MckeeOstriker2007, Padoan2014, Krumholz2014}.

Although  galactic-scale molecular depletion times are
often invoked to constrain GMC-scale star formation processes, a
complication is that extragalactic observations beyond the Local Group
generally do not resolve individual clouds.  When using a beam
corresponding to kpc or larger scales and integrating CO emission,
molecular gas in bound GMCs is not easily distinguished from more
diffuse molecular gas.  Since the feedback mechanisms that control the
dynamics and evolution of diffuse gas (molecular and/or atomic) may be quite
different from those that control dynamics and SFEs of bound GMCs, this
caveat must be kept in mind.  Indeed, well-studied star-forming clouds
in the Gould Belt appear to have molecular $t_{\rm dep}$ much lower --
perhaps by an order of magnitude -- than values indicated by kpc-scale
extragalactic studies  \citep{KennicuttEvans2012}.  This may mean
that much of the molecular gas in extragalactic studies is diffuse (and
non-star-forming), rather than in bound, star-forming GMCs.
Nevertheless, even in local clouds with shorter molecular $t_{\rm dep}$,
the corresponding observed SFE is still 
$< 0.1$ \citep{Lada2010}. 

A number of forms of massive-star feedback have been proposed as
dynamical drivers in the interstellar medium (ISM).  
The dominant feedback mechanism for
driving kinetic energy in the ISM as a whole is likely Type II
supernovae (SNe) \citep{MacLowKlessen2004, ElmegreenScalo2004}. One
useful way to characterize star formation feedback is in terms of the
momentum injection per total mass in stars formed (averaged over the
IMF), and this is likely to be an order of magnitude larger for SNe
than for other forms of feedback such as radiation
(\citealt{OstrikerShetty2011}; see also \citealt{KimOstriker2015,
  IffrigHennebelle2015, WalchNaab2015, Martizzi2015} for assessments of
momentum injection by SNe). However, the delay of SNe by 
$3-30~{\rm Myr}$ means that depending on how long GMCs survive, this
momentum may primarily be deposited in the diffuse (atomic or 
molecular) ISM rather than in
bound, star forming clouds \citep{Matzner2002, Fall2010}. 
Furthermore, blast waves from (early) supernovae that explode within GMCs 
preferentially eject lower-density gas, and may leave higher-density
structures behind.  Thus, even if
other forms of feedback are less intrinsically powerful than SNe, 
they may be more important to GMC evolution.

Forms of feedback associated with earlier stages of stellar evolution
include winds from massive stars, ionizing and non-ionizing radiation,
and outflows and jets from low mass protostars. Some of these have
been proposed as candidates for driving the dynamics of whole GMCs,
while others are expected to have only weak or localized effects
\citep[see, e.g., review of][]{Padoan2014}.
Protostellar jets and outflows are most important in clouds or
cluster-forming clumps that do not contain massive stars
\citep{Quillen2005, Cunningham2006, LiNakamura2006, NakamuraLi2008,
  Wang2010, Hansen2012}. Although shocked winds from massive stars
were initially expected to be important to cloud evolution
\citep{Castor1975, Weaver1977}, the inhomogeneous structure of
turbulent clouds implies that much of the wind energy can escape
\citep{RogersPittard2013}. This may explain the low X-ray luminosity
of observed systems such as the Carina nebula
\citep{Harper-ClarkMurray2009}.

Much of the current work on early feedback instead concentrates on the
dynamics of expanding HII regions, as driven by both the warm ionized
gas \citep{Whitworth1979, Mckee1984, Matzner2002,
KrumholzMatznerMcKee2006, 
Dale2005, Dale2012,Dale2013a, 
Vazquez-Semadeni2010, 
Colin2013, Walch2012}, and by radiation -- either direct
\citep{KrumholzMatzner2009, Fall2010, Murray2010, 
  Sales2014}
or reprocessed by dust \citep{Murray2010,SkinnerOstriker2015}. Reprocessed radiation is expected to
increase in importance relative to direct radiation in high surface
density clouds that are optically thick to infrared (IR). Overall, 
radiation
forces are expected to exceed those from ionized gas pressure only in more
massive, higher surface density clouds
\citep{Fall2010,Kim2016}, and there is
some observational support for this \citep{Lopez2011, Lopez2014,
  Pellegrini2007, Pellegrini2010}.  
Numerical simulations by
\cite{Dale2012, Dale2013a} show that pressure from photoionized gas is
capable of disrupting clouds and expelling substantial gas only when the
escape speed is low compared to the ionized gas sound
speed. \cite{Walch2012} found consistent results for the effects of 
photoionized gas in their simulations
of clouds with varying fractal dimension in the initial density
structure.

While analytic and spherically symmetric numerical models predict that
radiation feedback effects will become dominant in clouds with high
mass and surface density, it is clearly necessary to understand the
effects of strongly inhomogeneous density structure, which is the
hallmark of turbulent GMCs.
\cite{KrumholzThompson2012} and \cite{Davis2014} found for turbulent
disks that radiation and gas tend to be anticorrelated, which reduces
the net force of radiation. A recent analysis by
\cite{ThompsonKrumholz2014} suggests that incorporating the full
lognormal density distribution imposed by turbulence is crucial to
understanding how radiative feedback can drive outflows and 
limit star formation.

Realistic investigation of radiation effects in suppressing star formation
in turbulent, inhomogeneous clouds requires full time-dependent
radiation hydrodynamics (RHD) modeling with self-gravity.
Recently, \citet{SkinnerOstriker2015}
applied the \textit{Hyperion}
RHD code \citep{SkinnerOstriker2013} to study the evolution of massive
clouds that are optically thick to reprocessed radiation. This work
showed that reprocessed radiation only expels significant mass from a
turbulent, initially gravitationally-bound
cloud when the Eddingtion ratio at IR $f_{\rm Edd,*} \equiv
\kappa_{\rm IR}\Psi / (4 \pi G c)$ exceeds unity, where $\kappa_{\rm
  IR}$ is the mean IR opacity and $\Psi$ the mean light-to-mass ratio
of stars. Even at $f_{\rm Edd,*} \sim 1 - 3$, however, the SFE is high,
with $\sim 50 \%$ or more of the original GMC collapsing and accreting
on to star particles. Furthermore, the turbulent structure of the gas
significantly reduces radiation forces, due to a matter-radiation
anti-correlation.

Here, we apply the \textit{Hyperion} code to consider the opposite
limit of clouds with lower surface density, 
in which direct UV dominates over reprocessed radiation. We
consider model GMCs with a wide range of sizes, masses, and initial
virial parameters. Our chief aim is to understand how
self-gravitating, turbulent clouds react dynamically to radiation that emerges
from their densest (collapsed) regions. In particular, we wish to
quantify any radiation effects on reducing the SFR and/or SFE by limiting
local collapse and disrupting clouds.

  In the present work, we consider solely the effects of UV radiation forces  
  on cloud dynamics, a question that has not previously been addressed in
fully three-dimensional models.  For clouds at the lower range of
surface density that we model, pressure forces from photoionized gas
(which we do not treat here) may in real systems be comparable to direct
radiation forces \citep{KrumholzMatzner2009, Fall2010, Lopez2011, Kim2016}.
A more realistic treatment, including the effects of
ionization and heating from radiation in addition to radiation forces,
is necessary for quantifying the relative importance of these processes
but will be deferred to future work.  The present
study is intended to provide a baseline for future more
comprehensive simulations by quantifying how radiation forces
(in the single-scattering approximation)
by themselves affect the evolution of turbulent, self-gravitating clouds.

We begin in Section ~\ref{Sec:Setup} by describing the \textit{Hyperion}
code and the numerical setup of our turbulent clouds. 
In Section \ref{Sec:Fiducial} 
we present an overview of evolution for a fiducial model, as well as 
convergence tests. In Section \ref{Sec:SFE}
we show our results for SFE and SFR across the full suite of 
model clouds, and further analyze the effects of 
the lognormal density distribution on the radiation/gas interaction.
We summarize and discuss our conclusions in context of other theoretical 
work and observations in Section \ref{Sec:Conclusion}.  
Appendix \ref{Sec:Tests} presents additional code tests for idealized 
problems over a range of parameters in the single-scattering regime.

\section{Numerical Setup}
\label{Sec:Setup}

\subsection{Equations and Algorithms}

We run three-dimensional radiation hydrodynamic (RHD) simulations on a
Cartesian grid using the \textit{Hyperion} \citep{SkinnerOstriker2013} 
extension of the \textit{Athena} code \citep{Stone2008}. For this 
application, we solve the following simplified mixed-frame
equations of RHD:
\begin{eqnarray}
  \partial_t \rho + \nabla \cdot (\rho \mathbf{v}) &=& 0, 
\label{Eq:density} \\
  \partial_t (\rho \mathbf{v}) + \nabla \cdot (\rho \mathbf{v} \mathbf{v} + P\mathbb{I}) &=& -\rho\nabla\Phi + \rho\kappa\frac{\mathbf{F}}{c}, 
\label{Eq:RHDMomentum} \\
  \frac{1}{\hat{c}} \,\partial_t \mathcal{E} + \nabla \cdot \left(\frac{\mathbf{F}}{c}\right) &=& -\rho\kappa\mathcal{E} +\mathbb{S}, 
\label{Eq:RHDEnergy} \\
  \frac{1}{\hat{c}} \,\partial_t \left(\frac{\mathbf{F}}{c}\right) + \nabla \cdot \mathbb{P} &=& -\rho\kappa\frac{\mathbf{F}}{c}, 
\label{Eq:RHDRadMomentum}
\end{eqnarray}
where $\rho$, $\mathbf{v}$, and $P$ are the gas density, velocity, and
pressure, and $\Phi$ is the gravitational potential, all evaluated in
the lab frame. We adopt the simplifying assumption of an isothermal
equation of state for the gas with $P = c_s^2 \rho$ (see discussion below).
The variables
$\mathcal{E}$, $\mathbf{F}$, and $\mathbb{P}$ are the radiation energy
density, flux vector, and pressure tensor, respectively, again
evaluated in the lab frame, while $\kappa$ is the frequency-weighted
specific material opacity calculated in the gas rest frame.

\textit{Hyperion} closes the two radiative moment equations above by
adopting the $M_1$ relation \citep{Levermore1984}. This expresses the pressure tensor in
terms of $\mathcal{E}$ and $\mathbf{F}$, with $\mathbb{P} \rightarrow
(1/3)\mathcal{E}\mathbb{I}$ in the diffusion limit ($|\mathbf{F} | / \mathcal{E}c \ll 1$) and 
$\mathbb{P} \rightarrow \mathcal{E}\mathbf{\hat{n}\hat{n}}$ in the streaming limit
($| \mathbf{F} | / \mathcal{E}c \rightarrow 1$), where $\mathbf{\hat{n}}
= \mathbf{F} / |\mathbf{F}|$. We omit radiative emission terms
from gas and dust in Equation~(\ref{Eq:RHDEnergy}), as we are
interested in the limit in which the effects from direct stellar
radiation dominate over IR emission from the dust. This stellar
emission is captured in the term $\mathbb{S}$ in
Equation~(\ref{Eq:RHDEnergy}), which describes radiative emission from
star particles (see below).  Finally, $\hat{c} \neq c$ is a reduced
radiation propagation speed, adopted within the Reduced Speed of Light
Approximation (RSLA) \citep{GnedinAbel2001} to ensure that timesteps updating the radiation
field are not unfeasibly short.

\textit{Hyperion} divides
Equations~(\ref{Eq:density})-(\ref{Eq:RHDRadMomentum}) into gas and
radiation subsystems because these variables are transported on very
different time scales. We therefore separate the two systems and solve
for the gas subsystem using \textit{Athena}'s unsplit Van Leer (VL)
integrator \citep{StoneGardiner2009}, a Godunov finite-volume method
adapted from the MUSCL-Hancock scheme of \cite{Falle1991}. The
hydrodynamic timestep is determined using a radiation-modified CFL
condition with Courant number of $0.4$ (the typical value adopted in 
VL integration schemes) and a radiation-modified
effective sound speed that accounts for the effect of interactions
between the gas and radiation fields, $c_{\rm eff} \equiv
\sqrt{(\gamma P + 4/9\mathcal{E}(1 - {\rm e}^{-\rho \kappa_0 \Delta
    x})) / \rho}$ \citep{Krumholz2007}.

\textit{Hyperion} solves the radiation subsystem using an
operator-split method that separates the radiation source terms into
explicit and implicit terms, with the explicit terms updated together
with the update from the divergence of $\mathbf{F}$ and
$\mathbb{P}$. Here, both the radiation energy and the flux absorption
updates are solved using a standard $\theta$-scheme update with
$\theta=0.51$, very close to second order implicit in time.

The radiation subsytem is similarly solved using a VL scheme, with a
Harten-Lax-van Leer (HLL) Riemann solver \citep{Gonzalez2007} used to
calculate the flux between cells. In this case, the timestep is set by
a CFL condition with the radiation signal speed $\hat{c}$. This
reduced speed of light is chosen so that
radiation timesteps are as long as possible, while still ensuring that
the RSLA does not improperly affect the gas dynamics. We may achieve
this so long as $\hat{c}$ is sufficiently large that the radiation
field approaches equilibrium much faster than characteristic gas time
scales.  For streaming radiation, a practical condition is $\hat{c}
\sim 10v_{\rm max} \gg v_{\rm max}$
\citep[see][]{SkinnerOstriker2013}.  There are then roughly $10$
radiation substeps for each update to the gas subsystem.

In order to avoid unrealistically short timesteps for low density
regions accelerated by a strong radiative flux, there is an
artificially imposed density floor. Cells whose density falls below a
prescribed value at any given timestep are reset to the density floor
with zero momentum. This may add mass to the grid over the course of a
simulation, but in practice the density floor is chosen such that
less than $\sim 0.1 \%$ of the initial cloud mass is artificially
added over the course of a simulation.

Stars are represented within the code by point-mass sink particles
\citep{GongOstriker2013}. Sink particles are formed dynamically when
cells exceed a density threshold $\rho_{\rm th} = 8.86~c_s^2 / (\pi G \Delta
x^2)$ motivated by the \cite{Larson1969} and \cite{Penston1969}
solutions for self-gravitating isothermal collapse. Locations where
sink particles form must also be potential minima \citep{Banerjee2009,
  Federrath2010, Vazquez-Semadeni2011}.
If a cell satisfies these criteria, a sink particle is created at the
center of a control volume of width $3\Delta x$. The sink particle has
initial mass and momentum set by the sum over all control volume
cells. Subsequently, gas is accreted onto the sink particles based on
the HLL flux at the interface between sink control volumes and the
rest of the grid. Gas variables within the control volume are set by
extrapolating values from the surrounding active zones in the grid.

Sink particles are evolved in time using a leapfrog kick-drift-kick
method \citep{Springel2005}, where the particles' positions and
momenta are updated alternately. The position is updated using the
current velocities, while the momentum is updated based on 
gravitational potential differences.  The potential itself is computed
using particle-mesh methods with a Triangular Shaped Cloud (TSC) kernel applied to map each
particle's mass onto the grid \citep{HockneyEastwood1981}. The
combined particle $+$ gas potential is found using a Fourier transform
method on a domain equal to eight times the
computational volume in order to implement vacuum boundary conditions
for $\Phi$ \citep{HockneyEastwood1981}.  Finally, when the control
volumes of two sink particles overlap, they are merged and placed at
the center of mass of the two old particles.

Monochromatic radiation from the sink particles is emitted
isotropically, representing idealized luminous
stellar clusters. The source function $\mathbb{S}=j_{\rm *}/c$ 
of each particle of mass $M_{\rm *}$ takes a
Gaussian shape with 
\begin{equation}
	j_{\rm *}(r) = \frac{L_{\rm *}}{(2\pi \sigma_*^2)^{3/2}} \exp \left( -\frac{r^2}{2\sigma_{\rm *}^2} \right),  
	\label{Eq:jprofile}
\end{equation}
with (fixed) radius $r_{\rm *} = \sqrt{2 {\rm log}2} \sigma_{\rm *} =
1$~pc and fixed luminosity per unit mass $\Psi \equiv L_*/M_*$ typical
of young, luminous clusters. We adopt a fiducial value of $\Psi =
2000~{\rm erg~s^{-1}~g^{-1}}$, characteristic of a fully sampled Kroupa
IMF \citep{Dopita2006}.

We note that undersampling of the IMF can lead to 
both a large stochastic variation \citep{DaSilva2012} and a systematic 
overestimate in the luminosity per unit mass \citep{WeidnerKroupa2006}. 
However, as we discuss in Section~\ref{SubSec:Convergence}, all of our clouds 
are close to fully sampled by the time star formation is complete.
We further discuss the effects of varying $\Psi$
in Section~\ref{SubSec:VaryPsi}.

\subsection{Initial Conditions}
\label{SubSec:Setup}

\begin{deluxetable}{cc}
\tablecaption{Fiducial Parameters}
\tabletypesize{\scriptsize}
\def\arraystretch{0.5}
\tablewidth{0pt}
\tablehead{
\\
\colhead{Parameter} & \colhead{Value}
}
\startdata
\\
$\alpha_{\rm vir,0}$ & 2.0 \\
$r_{\rm 0}$ & $15~{\rm pc}$ \\
$M_{\rm cl,0}$ & $5 \times 10^4~{M}_\odot$ \\
$\Sigma_{\rm cl,0}$ & $70.74~{M}_\odot~{\rm pc}^{-2}$ \\
$t_{\rm ff,0}$ & $4.29~{\rm Myr}$ \\
$v_{\rm RMS}$ & $4.16~{\rm km~s^{-1}}$ \\
$v_{\rm esc}$ & $5.36~{\rm km~s^{-1}}$ \\
$c_s$ & $0.2~{\rm km}~{ \rm s}^{-1}$ \\
$\hat{c}$ & $250~{\rm km}~{ \rm s}^{-1}$ \\
$\Psi$ & $2000~{\rm erg}~{\rm s}^{-1}~{\rm g}^{-1}$ \\
$\kappa$ & $1000~{\rm cm}^{2}~{\rm g}^{-1}$
\enddata
\label{Tab:FiducialParams}
\end{deluxetable}

\begin{figure}
  \centering
  \epsscale{1.}
\includegraphics{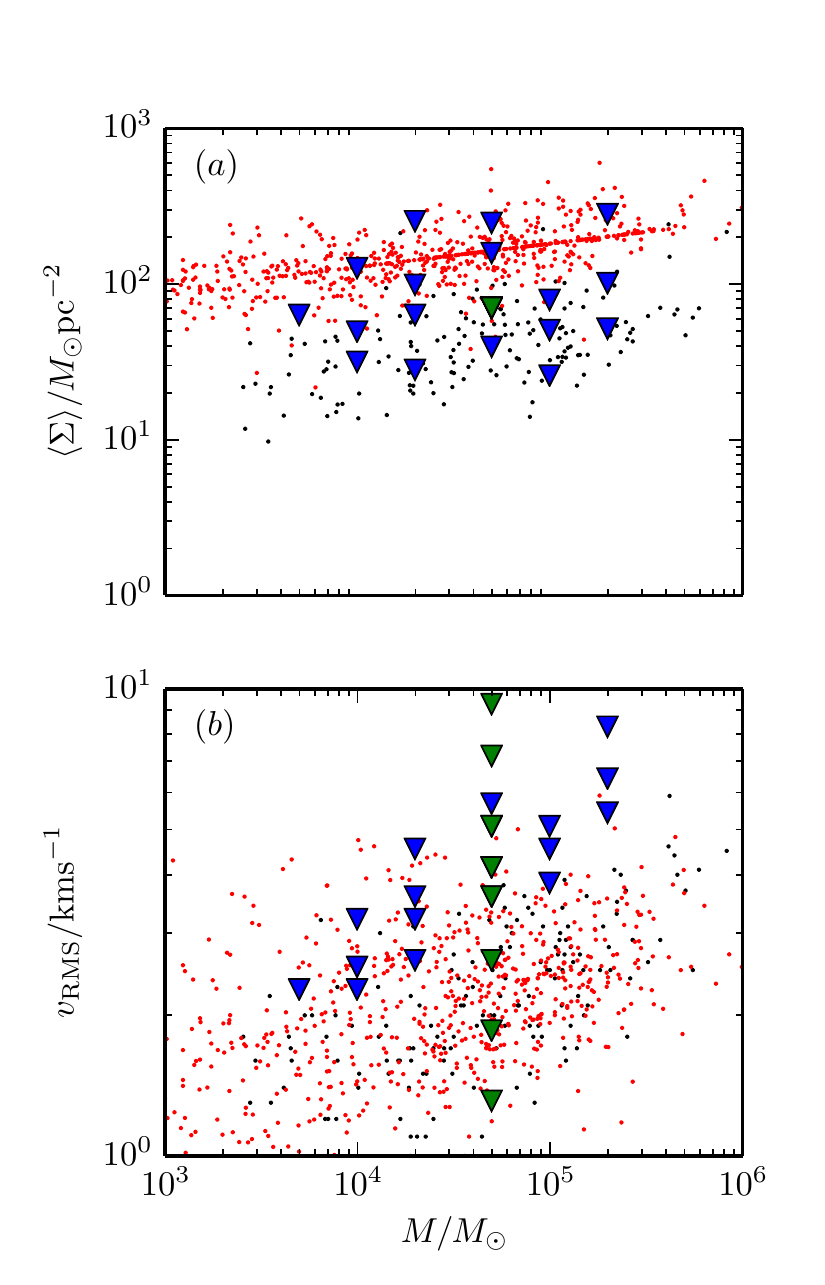}
  \caption{Parameter space of (a) cloud mass vs surface density, and
    (b) cloud mass vs velocity dispersion for our $\Sigma$-series (blue
    triangles) and $\alpha$-series (green triangles) models. For
    comparison, we also show corresponding values for Galactic GMCs
    observed by \cite{Heyer2009} (black) and \cite{Duval2010} (red).}
  \label{Fig:MassRange}
\end{figure}

We consider the evolution of self-gravitating star-forming clouds over
a period of $\sim 4$ initial freefall times. Each cloud is initialized
as a uniform density sphere, with ${\rho}_0 = 3M_{\rm cl,0} / (4 \pi
r_0^3)$, where $r_0$ is the initial cloud radius. The clouds are
centered inside cubic simulation volumes of length $L = 4r_0$ with
outflow boundary conditions, so that we may track the mass expelled
from the cloud by radiation forces. 
The gas surrounding the cloud is initialized at a factor of $10^{3}$
lower than the cloud density, so that the total mass surrounding the
cloud is $\sim 0.015~M_{\rm cl,0}$. The density floor is lower than
this again by a factor of $10$, so that only a small amount of mass is
added to the cloud over the course of a simulation run.

In order to cover a realistic range in cloud surface density, we adopt
a range above and below fiducial values $M_{\rm cl,0} = 5 \times 10^4$
and $r_0 = 15$~pc, which corresponds to a cloud surface density of
$\Sigma_{\rm cl, 0}\equiv M_{\rm cl,0}/(\pi r_{\rm 0}^2)  = 70.7~M_{\odot}~{\rm pc^{-2}}$. Our
$\Sigma$-series consists of a subset of models with initial cloud
masses $M_{\rm cl,0} = 5 \times 10^3, 10^4, 2 \times 10^4, 5 \times
10^4, 10^5, 2 \times 10^5~M_{\odot}$ and radii $r_0 = 5, 8,
10, 15, 20, 25, 35$~pc.  In Figure~\ref{Fig:MassRange} we show the
masses, surface densities and velocity dispersions of the full set of
$\Sigma$-series models compared to the same quantities derived for a
set of 158 Milky Way GMCs measured by \cite{Heyer2009} and a more
extended sample of 580 molecular clouds measured in
\cite{Duval2010}. Similar to \cite{Dale2012}, we cover the high-mass
end of the observed distribution, with our fiducial model being
roughly characteristic of the median observed cloud.

The gas motions in our clouds are initially seeded by a turbulent
velocity field, with power spectrum $v^2(k) \propto k^{-4}$ as is
observed within GMCs \citep[e.g., ][]{Dobbs2013}.
The turbulence 
is initialized as described in \cite{Stone1998, SkinnerOstriker2015}. 
Briefly, we generate a Gaussian random field in Fourier space, such 
that over the range $k \in [2, 64] \times dk$ where $dk = 2\pi / L$, $\delta v_k$ 
is chosen from a Gaussian distribution with variance $P(k) \propto k^{-4}$. 
This field is then transformed back to real space and renormalized in terms 
of the virial parameter $\alpha_{\rm vir,0} \equiv 2 E_K / |E_G|$ so that 
the variance of the velocity distribution obeys 
$\sigma^2 = 2 E_K / M_{\rm cl, 0} = \alpha_{\rm vir,0} E_G / M_{\rm cl, 0}$, 
where $E_K = M_{\rm cl,0} v_{\rm RMS}^2 / 2$ is the total
initial turbulent gas kinetic energy, and $E_G = -3 G M_{\rm cl,0}^2 /
(5 r_0)$ is the cloud's initial gravitational binding energy. 
Finally, the momentum field is forced to have zero mean by subtracting off the 
initial net momentum of the cloud.  The initial turbulent
power spectrum is a mixture of solenoidal and compressive modes.
We further discuss effects of the specific initialization of turbulence
in Sections~\ref{SubSubSec:PDF Seed}
and \ref{SubSec:SFEEvol}.
For the fiducial model, and other models in the $\Sigma-$series, we set
$\alpha_{\rm vir,0} = 2$. We also consider another series of models, the
$\alpha$-series, in which the initial $\alpha_{\rm vir,0}$ is in the
range $0.1$ to $10.0$.

Clouds with $\alpha_{\rm vir} = 2$ are still marginally bound so long as the thermal energy does 
not contribute significantly to the total kinetic energy i.e., $c_s^2 \ll 3GM_{\rm cl,0} / (5r_0)$. 
In practice, for the lowest surface density clouds that we consider, this is satisfied provided 
$c_s \ll 1~{\rm km~s^{-1}}$.
We adopt a constant isothermal sound speed $c_s = 0.2~{\rm km~s^{-1}}$
for all simulations, consistent with a temperature of $T \sim 10$~K,
as is characteristic of most of the mass in observed GMCs
\citep{Scoville1987}. Of course, ionizing UV radiation will heat a
very small fraction of the gas to a much higher temperature, and
non-ionizing radiation can raise the temperature of gas within regions
near stellar sources (within $A_V \sim 1$ where it is absorbed).
However, given the high (FUV) optical depths of clouds, most of the gas is
shielded from both internal and external radiation sources.  For
the regime of the present study, the optical depth to IR is
small, so radiation that is absorbed and reprocessed to IR near sources
subsequently escapes from the cloud without significant re-absorption by dust.
To the extent that regions near sources can be heated above $\sim
10$K, the increase in pressure would limit fragmentation, with
important consequences for the IMF \citep{Krumholz2007, Myers2014}. 
In addition, gas that is heated above the escape speed of the cloud could directly evaporate.

In Table~\ref{Tab:FiducialParams}, we list simulation inputs for our
fiducial model, including the initial cloud mass, radius, and virial
parameter, and the parameters $c_s, \hat{c}, \Psi$ and
$\kappa$. Our standard resolution is 
$256^3$, although we have employed higher and lower resolution grids to test convergence.
We also adopt a fixed opacity $\kappa = 1000~{\rm cm^{2}~g^{-1}}$, consistent with the radiation 
pressure cross sections per H derived from the \cite{WeingartnerDraine2001} dust model \citep{Draine2011}. 

We extend this in Table~\ref{Tab:ModelParams}, to show 
$M_{\rm cl,0}$, $r_{\rm 0}$, and $\alpha_{\rm vir,0}$ for 
all models in our $\Sigma$ and $\alpha$-series. For each
simulation, we show also the initial surface density 
$\Sigma_{\rm cl,  0} \equiv M_{\rm cl, 0} / (\pi r_0^2)$, the initial RMS velocity
dispersion $v_{\rm RMS} = [3 \alpha_{\rm vir,0} G M_{\rm cl, 0} /
  (5r_0)]^{1/2}$, the escape speed at the edge of the cloud $v_{\rm
  esc} = (2 G M_{\rm cl, 0} / r_0)^{1/2}$, the initial gravitational free-fall
time $t_{\rm ff,0} = [3 \pi / (32 G \rho_0)]^{1/2}$ and the initial
hydrogen number density $n_0 = \rho_0 / (1.4 m_p)$, where 
$\rho_0 = 3M_{\rm cl, 0} / (4 \pi r_0^3)$ is the initial gas density and 
we allow for 40\% helium by mass. Model
$\Sigma$-M5E4-R15 (the same as $\alpha$-A2.0), 
shown in bold, is the fiducial model.

\begin{deluxetable}{l|cccccccc}
\tabletypesize{\scriptsize} \tablecolumns{9} \tablewidth{0pt}
\def\arraystretch{1.0}
\tablecaption{Model Parameters}
\tablehead{
 \vspace{-0.2cm} &
\colhead{$\Sigma_{\rm cl,0}$} &
\colhead{$M_{\rm cl,0}$} &
\colhead{$r_{\rm 0}$} &
\colhead{$n_{\rm H,0}$} &
\colhead{$t_{\rm ff,0}$} &
\colhead{$v_{\rm RMS}$} & 
\colhead{$v_{\rm esc}$} & 
 \\ 
\colhead{Model} \vspace{-0.2cm} &
 &
 &
 &
 &
 &
 & 
 & 
\colhead{$\alpha_{\rm vir,0}$} \\
 &
\colhead{[${M}_\odot~{\rm pc}^{-2}$]} &
\colhead{[${M}_\odot$]}&
\colhead{[${\rm pc}$]} &
\colhead{[${\rm cm}^{-3}$]} &
\colhead{[${\rm Myr}$]} &
\colhead{[${\rm km~s^{-1}}$]} &
\colhead{[${\rm km~s^{-1}}$]} &
\colhead{}
}

\startdata
$\Sigma$-M2E4-R25 & 10.19 & $2 \times 10^4$ & 25 & 8.860 & 14.6 & 2.04 & 2.63 & 2.0 \\
$\Sigma$-M5E4-R35 & 12.99 & $5 \times 10^4$ & 35 & 8.072 & 15.3 & 2.72 & 3.51 & 2.0 \\
$\Sigma$-M2E4-R20 & 15.92 & $2 \times 10^4$ & 20 & 17.30 & 10.5 & 2.28 & 2.94 & 2.0 \\
$\Sigma$-M5E4-R25 & 25.46 & $5 \times 10^4$ & 25 & 22.15 & 9.24 & 3.22 & 4.16 & 2.0 \\
$\Sigma$-M1E5-R35 & 25.98 & $1 \times 10^5$ & 35 & 16.14 & 10.8 & 3.85 & 4.97 & 2.0 \\
$\Sigma$-M2E4-R15 & 28.29 & $2 \times 10^4$ & 15 & 41.02 & 6.79 & 2.63 & 3.39 & 2.0 \\
$\Sigma$-M1E4-R10 & 31.83 & $1 \times 10^4$ & 10 & 69.22 & 5.23 & 2.28 & 2.94 & 2.0 \\
$\Sigma$-M5E4-R20 & 39.79 & $5 \times 10^4$ & 20 & 43.26 & 6.61 & 3.60 & 4.65 & 2.0 \\
$\Sigma$-M1E4-R08 & 49.74 & $1 \times 10^4$ & 8 & 135.2 & 3.74 & 2.54 & 3.29 & 2.0 \\
$\Sigma$-M1E5-R25 & 50.93 & $1 \times 10^5$ & 25 & 44.30 & 6.53 & 4.55 & 5.88 & 2.0 \\
$\Sigma$-M2E5-R35 & 51.97 & $2 \times 10^5$ & 35 & 32.29 & 7.65 & 5.44 & 7.02 & 2.0 \\
$\Sigma$-M5E3-R05 & 63.66 & $5 \times 10^3$ & 5 & 276.9 & 2.61 & 2.28 & 2.94 & 2.0 \\
$\Sigma$-M2E4-R10 & 63.66 & $2 \times 10^4$ & 10 & 138.4 & 3.70 & 3.22 & 4.16 & 2.0 \\
${\bf \Sigma}${\bf -M5E4-R15} & {\bf 70.74} & ${\bf 5 \times 10^4}$ & {\bf 15} & {\bf 102.5} 
& {\bf 4.29} & {\bf 4.16} & {\bf 5.36} & {\bf 2.0} \\
$\Sigma$-M1E5-R20 & 79.58 & $1 \times 10^5$ & 20 & 86.52 & 4.67 & 5.09 & 6.57 & 2.0 \\
$\Sigma$-M2E4-R08 & 99.47 & $2 \times 10^4$ & 8 & 270.4 & 2.64 & 3.60 & 4.65 & 2.0 \\
$\Sigma$-M2E5-R25 & 101.9 & $2 \times 10^5$ & 25 & 88.60 & 4.62 & 6.44 & 8.31 & 2.0 \\
$\Sigma$-M1E4-R05 & 127.3 & $1 \times 10^5$ & 5 & 553.7 & 1.85 & 3.22 & 4.16 & 2.0 \\
$\Sigma$-M5E5-R35 & 129.9 & $5 \times 10^5$ & 35 & 80.72 & 4.84 & 8.60 & 11.11 & 2.0 \\
$\Sigma$-M1E5-R15 & 141.5 & $1 \times 10^5$ & 15 & 205.1 & 3.04 & 5.88 & 7.59 & 2.0 \\
$\Sigma$-M5E4-R10 & 159.2 & $5 \times 10^4$ & 10 & 346.1 & 2.34 & 5.09 & 6.57 & 2.0 \\
$\Sigma$-M2E5-R20 & 159.2 & $2 \times 10^5$ & 20 & 173.0 & 3.31 & 7.20 & 9.29 & 2.0 \\
$\Sigma$-M5E4-R08 & 248.7 & $5 \times 10^4$ & 8 & 676.0 & 1.67 & 5.69 & 7.35 & 2.0 \\
$\Sigma$-M2E4-R05 & 254.6 & $2 \times 10^4$ & 5 & 1107 & 1.31 & 4.55 & 5.88 & 2.0 \\
$\Sigma$-M2E5-R15 & 282.9 & $2 \times 10^5$ & 15 & 410.2 & 2.15 & 8.31 & 10.7 & 2.0 \\
\hline
$\alpha$-A0.1 & 70.74 & $5 \times 10^4$ & 15 & 102.5 & 4.29 & 0.93 & 5.36 & 0.1 \\
$\alpha$-A0.2 & 70.74 & $5 \times 10^4$ & 15 & 102.5 & 4.29 & 1.32 & 5.36 & 0.2 \\
$\alpha$-A0.4 & 70.74 & $5 \times 10^4$ & 15 & 102.5 & 4.29 & 1.86 & 5.36 & 0.4 \\
$\alpha$-A0.8 & 70.74 & $5 \times 10^4$ & 15 & 102.5 & 4.29 & 2.63 & 5.36 & 0.8 \\
$\alpha$-A1.5 & 70.74 & $5 \times 10^4$ & 15 & 102.5 & 4.29 & 3.60 & 5.36 & 1.5 \\
${\bf \alpha}${\bf -A2.0} & {\bf 70.74} & ${\bf 5 \times 10^4}$ & {\bf 15} & {\bf 102.5} & {\bf 4.29} & {\bf 4.16} & {\bf 5.36} & {\bf 2.0} \\
$\alpha$-A3.0 & 70.74 & $5 \times 10^4$ & 15 & 102.5 & 4.29 & 5.09 & 5.36 & 3.0 \\
$\alpha$-A6.0 & 70.74 & $5 \times 10^4$ & 15 & 102.5 & 4.29 & 7.21 & 5.36 & 6.0 \\
\vspace{-0.1cm} $\alpha$-A10.0 & 70.74 & $5 \times 10^4$ & 15 & 102.5 & 4.29 & 9.30 & 5.36 & 10.0
\enddata
\tablecomments{Columns display the following information (i) model
  name, (ii) initial cloud surface density, (iii) initial cloud mass,
  (iv) initial cloud radius, (v) initial cloud hydrogen number
  density, assuming a mean atomic weight of $\mu = 1.4$, (vi) initial
  cloud free-fall time, (vi) initial turbulent velocity dispersion,
  (viii) cloud escape velocity from the initial cloud radius, (ix)
  initial virial parameter. The fiducial model is
  shown in bold ($\Sigma$-M5E4-R15 and $\alpha$-A2.0).}
\label{Tab:ModelParams}
\end{deluxetable}

\section{Tests of the Fiducial Model}
\label{Sec:Fiducial}

\subsection{Overview of Time Evolution}
\label{SubSec:FiducialTime}

\begin{figure*}
  \centering
  \epsscale{1}
\includegraphics{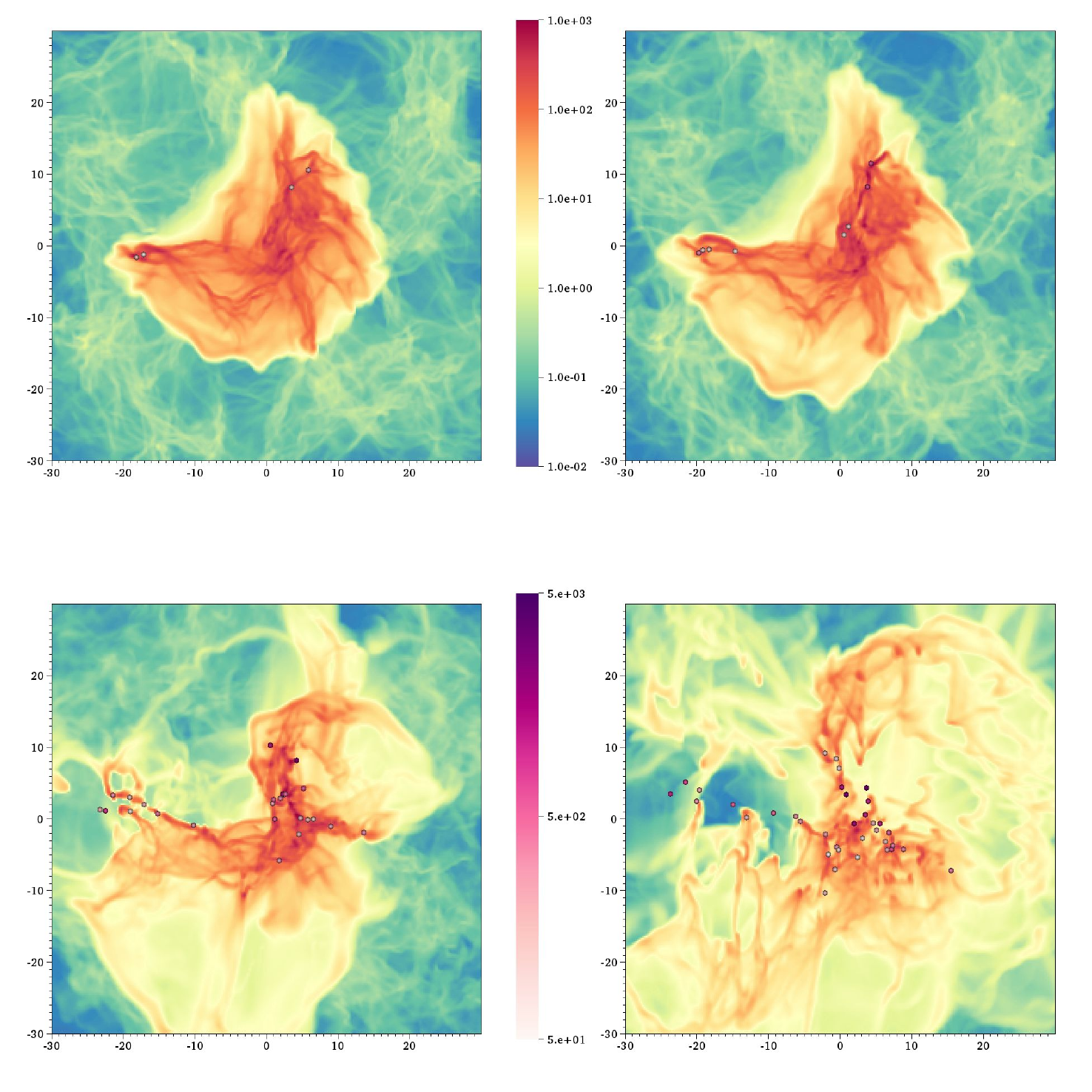}
  \caption{Snapshots from the fiducial model evolution. We show
    column densities in the y-z plane and all star particles projected
    onto the y-z plane, calculated at times $t_2$, $t_{10}$, $t_{50}$ and $t_{90}$, when the total stellar
    mass is $2 \%, 10 \%, 50 \%$ and $90 \%$ of the final value. This
    corresponds to $t / t_{\rm ff,0} = 0.43, 0.59, 1.06$ and $1.57$ as
    shown, where $t_{\rm ff,0} = 4.29$~Myr is the initial free-fall time
    in the cloud. The color scale for the gas column density (top) is
    in units of $M_{\odot}~{\rm pc^{-2}}$ and the color scale for the
    particle mass (bottom) is in units of $M_{\odot}$. The box size is
    $60$~pc, $4$ times the initial cloud radius.}
  \label{Fig:FiducialEvolutionColX}
\end{figure*}

\begin{figure*}
  \centering
  \epsscale{1}
  \includegraphics{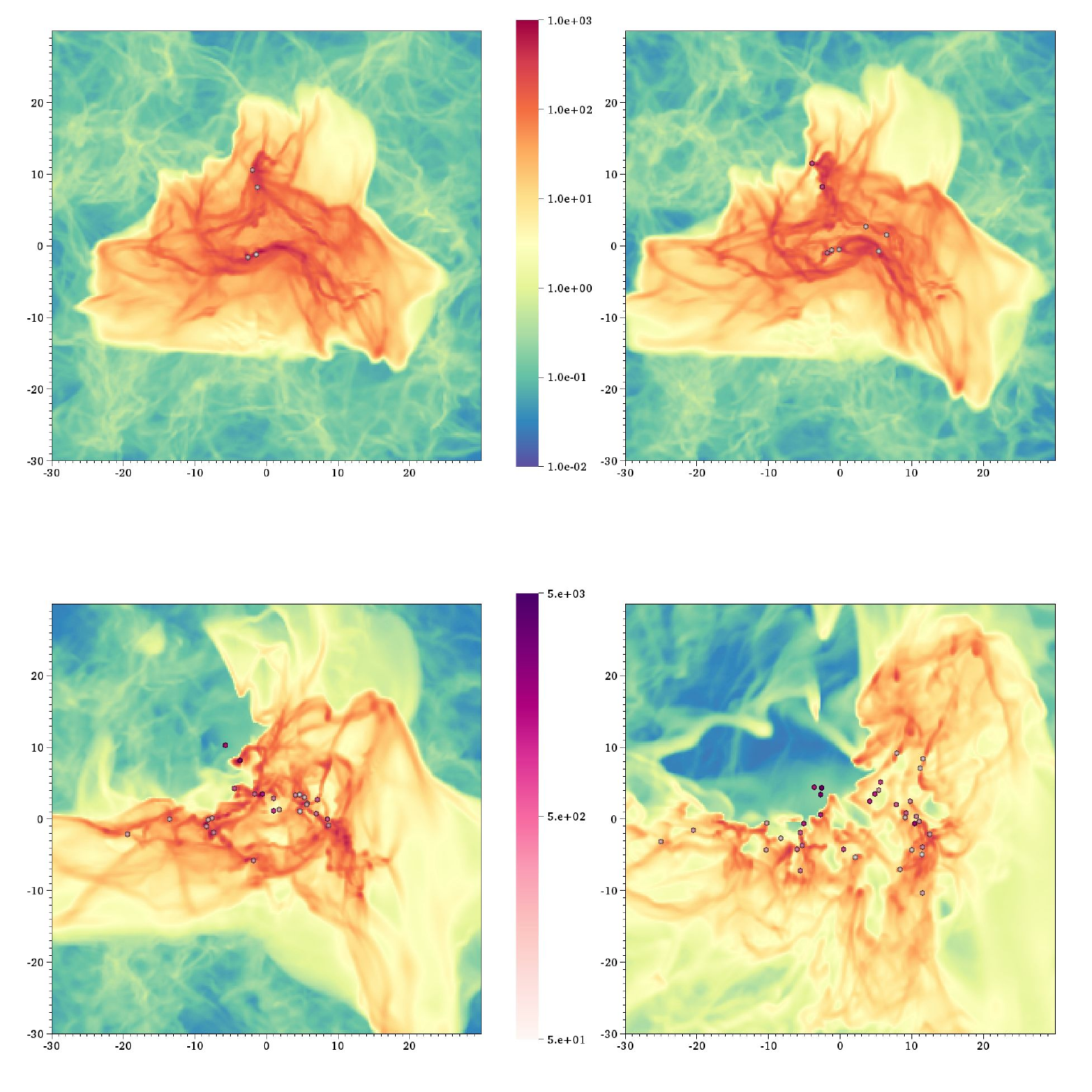}
  \caption{Same as Figure~\ref{Fig:FiducialEvolutionColX} except
    column densities are projected on the x-y plane. The cloud very
    rapidly develops a filamentary structure with mass preferentially
    gathered along the x-axis. Star formation then
    proceeds with the cloud remaining filamentary until a critical
    mass of stars drives away the remaining gas via radiation forces.}
  \label{Fig:FiducialEvolutionColZ}
\end{figure*}

\begin{figure*}
  \centering
  \epsscale{1}
  \includegraphics{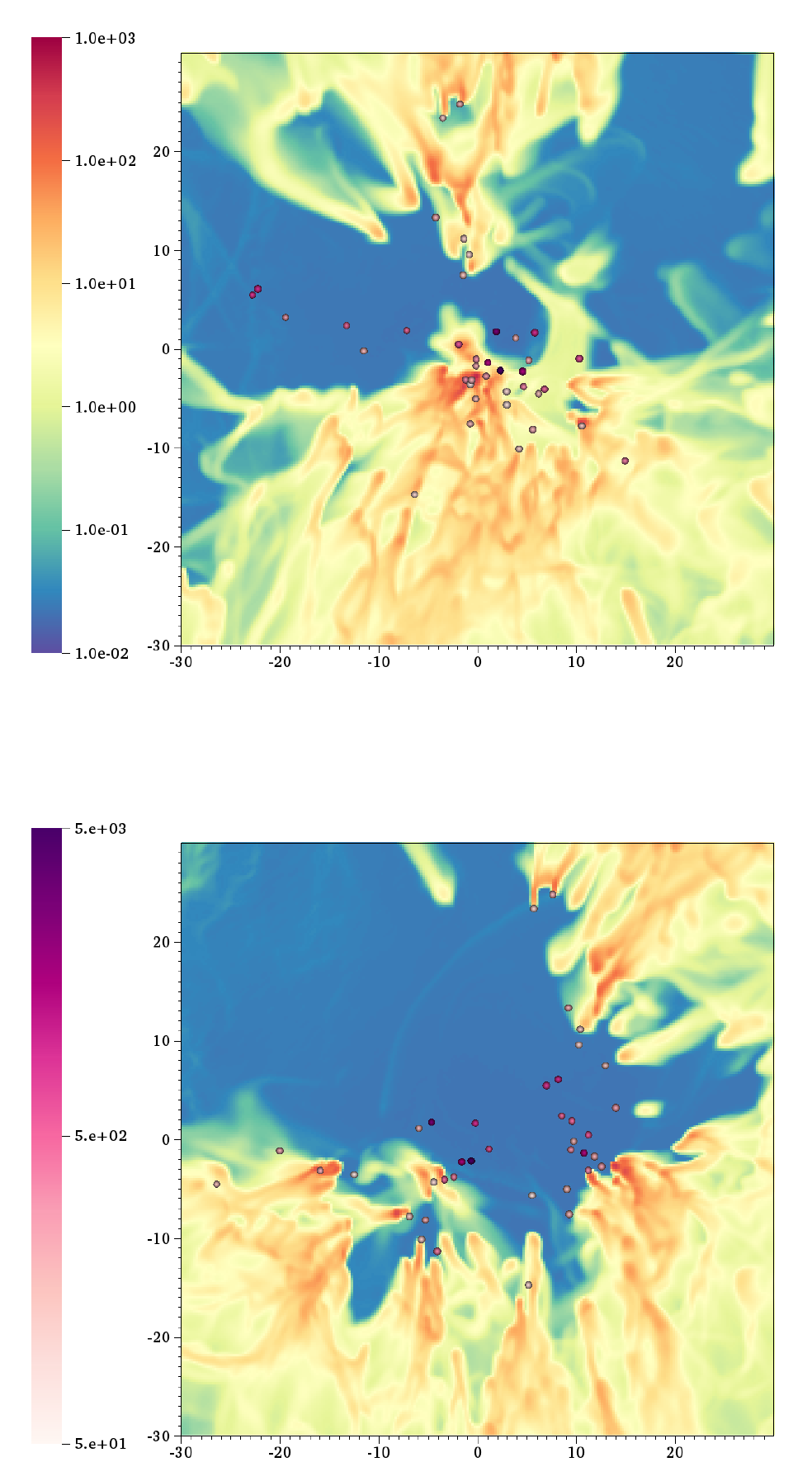}
  \caption{As in Figures~\ref{Fig:FiducialEvolutionColX} and~\ref{Fig:FiducialEvolutionColZ} 
    we show column densities projected in the y-z (top) and x-z (bottom) planes, both at times 
    $t / t_{\rm ff,0} = 2$. The color scale for the gas column density (top) is
    in units of $M_{\odot}~{\rm pc^{-2}}$ and the color scale for the
    particle mass (bottom) is in units of $M_{\odot}$.}
  \label{Fig:FiducialFinalCol}
\end{figure*}

\begin{figure*}
  \centering
  \epsscale{1}
  \includegraphics{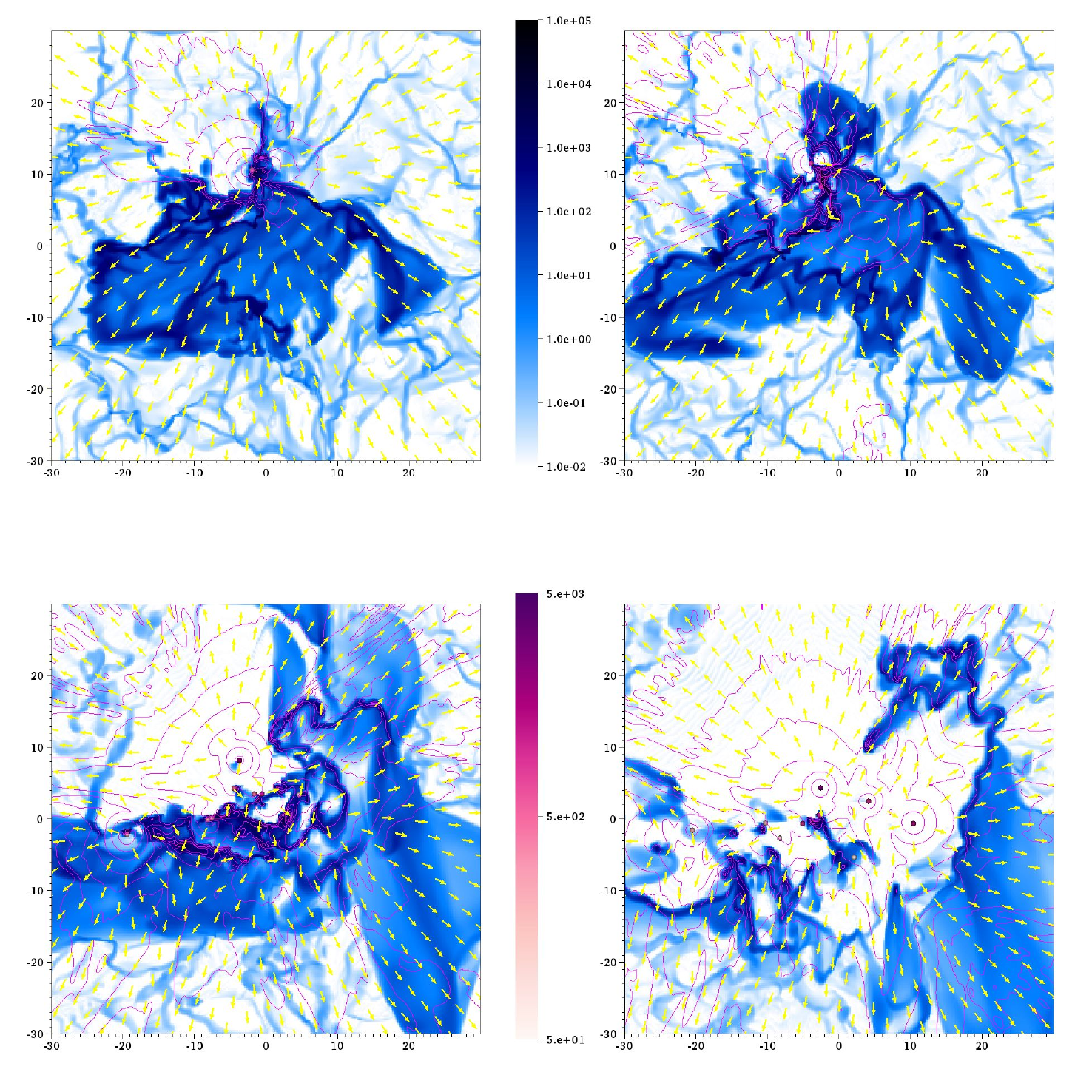}
  \caption{Snapshots of the density in our fiducial model. We show
    slices in the x-y plane passing through the position of the most
    massive star particle at the same times as for
    Figure~\ref{Fig:FiducialEvolutionColX}. The directions of
    radiation flux vectors are overlaid in yellow.  We also show, in pink
    contour lines, the radiation energy density at the same
    times.  Contours represent differences of a decade from the peak
    energy density. Star particles within $\Delta z = \pm 2~{\rm pc}$ of the
    slice are plotted as circles. The color scale for
    the gas density $n_H$ (blue, top) is in units of ${\rm cm^{-3}}$ and
      the color scale for the particle mass (red, bottom) is in units
      of $M_{\odot}$.}
  \label{Fig:FiducialRadiationEvolution}
\end{figure*}

\begin{figure*}
  \centering
  \epsscale{1}
  \includegraphics{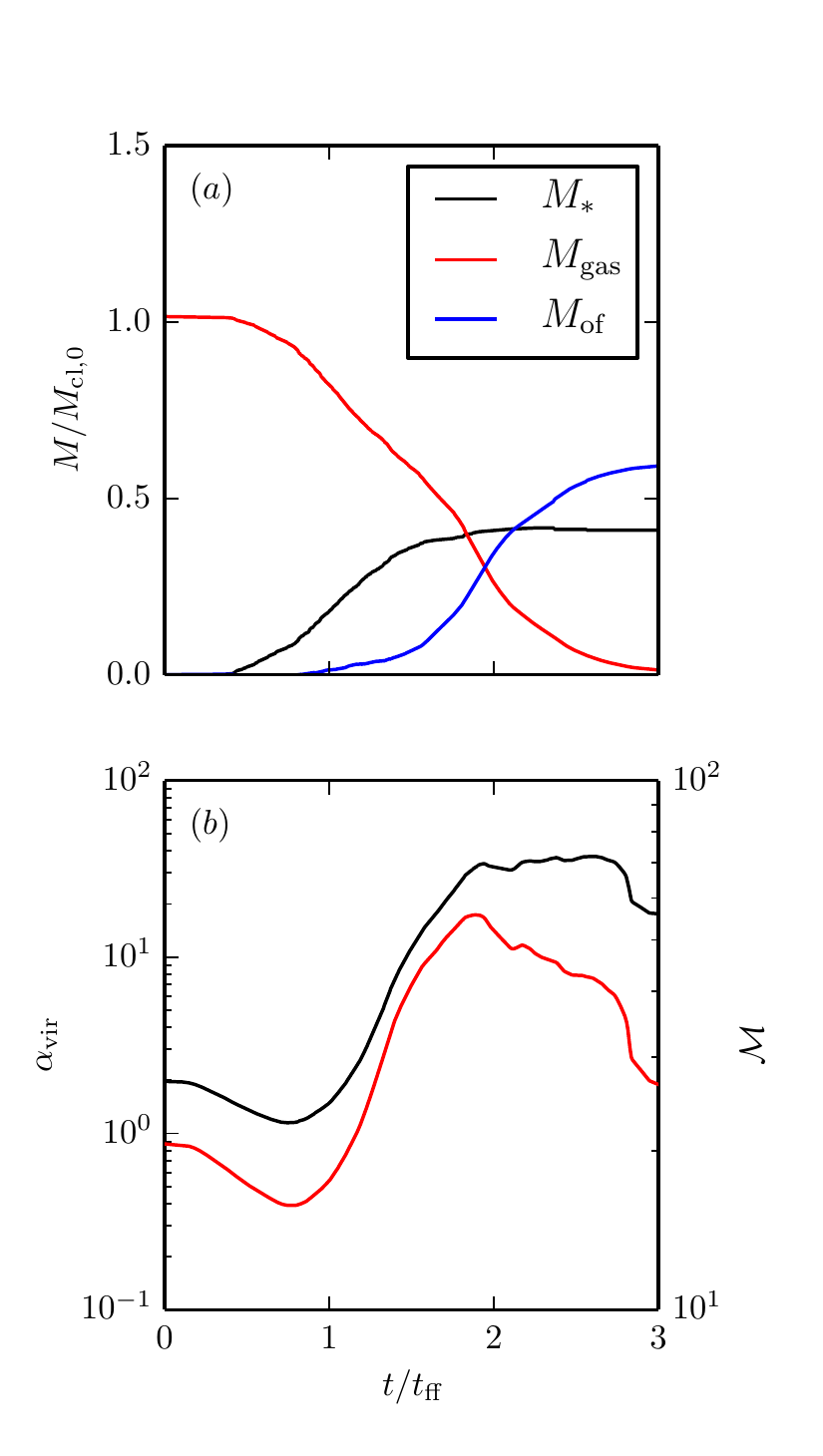}
  \caption{Time evolution of key global characteristics in the
    fiducial model.  We show in (a) the mass history, with
    contributions from gas, stars, and outflows. Evolving cloud
    structure is seen in (b) with the gas virial parameter, defined as
    $\alpha_{\rm vir} \equiv 2E_K/|E_G|$ (black) and the Mach number 
    defined as $\mathcal{M} = \sqrt{2E_K / M_{\rm cl,0}} / c_s$ (red).}
  \label{Fig:FiducialTimeHistory}
\end{figure*}

\begin{figure*}
  \centering
  \epsscale{1}
  \includegraphics{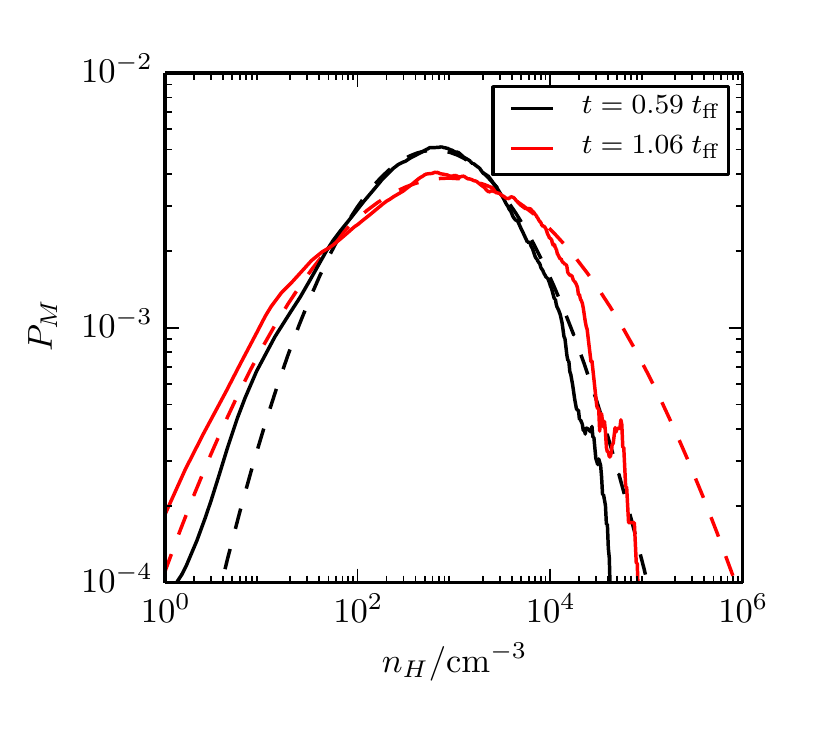}
  \caption{\bf The gas density distribution in our fiducial model at two separate times, $t_{10} = 0.59~t_{\rm ff,0}$ (black solid) and $t_{50} = 1.06~t_{\rm ff,0}$ (red solid). We show the distribution of 
  mass as well as the best-fit lognormal to each distribution (dashed), fitting between 
  the $10^{th}$ and $90^{th}$ percentiles by mass.}
  \label{Fig:PLDensity}
\end{figure*}

We begin by considering the overall time evolution of our fiducial model. 
In Figures~\ref{Fig:FiducialEvolutionColX}
and~\ref{Fig:FiducialEvolutionColZ} we show evolving column density
maps of this cloud in the 
y-z and x-y planes respectively. We
see that turbulence drives the cloud into collapse very rapidly, and
that by a little before half a free-fall time, mass has gathered
preferentially along two perpendicular filaments, roughly coincident with the
x and z-axes. The first star-formation event occurs around
this time, as the density peaks in the cloud continue to collapse
under self-gravity. The filamentary structure of gas in the cloud and
the shape of its density distribution do not however change
significantly. By $\sim t_{10}$, where $t_x$ denotes the time at which 
$x~\%$ of the stellar mass is assembled, the main difference 
from earlier is that the density contrast between the
filaments and surrounding gas has increased.
As star formation progresses, the UV radiation begins to drive gas in
even the densest filaments away from the sites of star formation. By
the time half the stars have formed, much of the gas, even at higher
densities, is already flowing outwards from the center of mass. 
At late stages, shown for example in the surface density projections
at $t=2~t_{\rm ff,0}$ in Figure~\ref{Fig:FiducialFinalCol}, all of the material is streaming away
from the center, which has been cleared of gas.

Figure~\ref{Fig:FiducialRadiationEvolution} shows the same picture
as in Figure~\ref{Fig:FiducialEvolutionColZ} but for 
a single slice through the x-y plane with the evolving radiation
field flux directions and energy density overplotted in vectors and contours
respectively.  Because of the filamentary nature of the cloud, even
though stars form near the center of the original GMC, the radiation
from these stars very quickly blows a hole in the surrounding
gas. Therefore, by $t_{10}$, as shown in the second snapshot, a significant fraction of
the radiation already escapes the cloud through the second quadrant,
where the gas density is small.

In detail, the gas structure surrounding the most massive star
clusters is far from smooth and presents a very different picture 
from the gas expansion seen in outflows driven by reprocessed
radiation \citep{SkinnerOstriker2015}.  At late stages in 
Figures~\ref{Fig:FiducialEvolutionColX} -- \ref{Fig:FiducialRadiationEvolution}
we see a central cavity surrounded by prominent high density fingers
of gas extending inwards towards the most massive stars.  Regions of
low column density, as seen by the central sources, are evacuated
first once the radiative force on them becomes super-Eddington. 
However, regions that are shielded by higher density clumps
of gas only begin to be driven out at late times, giving rise to the
prominent columns of gas.

This picture of filamentary collapse followed by rapid star formation and subsequent gas expulsion 
can also be seen in the time histories shown in Figure~\ref{Fig:FiducialTimeHistory}. 
In Figure~\ref{Fig:FiducialTimeHistory}a, we show the evolution of the total mass in stars, gas, and 
outflows from the simulation volume. While the cloud is initially collapsing
there is relatively little star formation,
although a small number of stars are formed through the effects of
turbulence initiating compression.
Meanwhile, the outflows driven by this initial turbulence
only start leaving the simulation volume at $\sim 0.8t_{\rm ff,0}$,
which is roughly the time taken for gas traveling at $\sim 2$ times
the escape velocity to reach the corner of the box.

At around the same time ($\sim 0.8t_{\rm ff,0}$) there is a break in
the stellar mass evolution as stars begin forming more rapidly; the
majority of the stellar mass is assembled over the next free-fall
time. By $\sim 1.8t_{\rm ff,0}$ the accretion onto star particles is
essentially complete, while radiation from these stars continues to
accelerate the remaining gas so that it becomes unbound from the
central cluster.  By $\sim 3t_{\rm ff,0}$ most of the 
outflowing gas has left the simulation box.

The evolution of the global gas distribution is more
difficult to characterize. One simple measure is the global virial
parameter, which is a rough proxy for the collapse and expansion of
the cloud and is shown in 
Figure~\ref{Fig:FiducialTimeHistory}b. Initially this decreases from
its starting value of $\alpha_{\rm vir} = 2$ as the turbulence decays
slightly and the total potential energy increases due to contraction
along local filaments. However, as for the stellar mass, there is a
break at around $\sim 0.8 t_{\rm ff,0}$, where collapse ceases and radiation 
from the first star particles begins to unbind gas and drive outflows.

The evolution in Mach number parallels that of
  the virial parameter very closely. 
  In fact, the two are well correlated until at least around $\sim 2~t_{\rm ff,0}$, with
$\mathcal{M} \propto \alpha_{\rm vir}^{0.4}$. This roughly corresponds to the 
evolution in Mach number for a shell expanding around a fixed central mass $M$. 
For a shell of fixed mass $M_{\rm sh}$, expanding with velocity $v$, the virial parameter is 
$\alpha_{\rm vir} = v^2 r / G M$, while the mass-weighted Mach number is roughly 
$\mathcal{M}^2 = (M_{\rm sh} / M)(v^2 / c_s^2)$ so that 
$\mathcal{M}^2 = (GM_{\rm sh} / r) (\alpha_{\rm vir} / c_s^2)$. 
The evolution we find is slightly different from   
$\mathcal{M} \propto \alpha_{\rm vir}^{1/2}$  
since the radius of the shell expands. In general though, we see that the evolution of both 
Mach number and virial parameter beyond $\sim 0.8~t_{\rm ff,0}$ is dominated by the 
cloud expansion and not turbulence.

It is for this reason that the increase in Mach number has no
correlated increase in the width of the lognormal distribution (see
below), unlike the case for driven turbulence when the width of the
density distribution is entirely set by the Mach number
\citep{PadoanNordlund2011, Molina2012, Hopkins2013, Krumholz2014,
  ThompsonKrumholz2014}. While the lowest density regions, which are
expanding away at high velocity, dominate the evolution of the virial
parameter and the Mach number, they have little influence on the
density distribution since, until late times, they only represent
around $10$ to $20~\%$ of the mass. Therefore, the lognormal density
distribution, which is fit to the majority of gas remaining in the
cloud, shows no significant change over the bulk of star formation.

A measure quantifying the evolution of the gas is the shape of the gas 
density distribution (PDF). In Figure~\ref{Fig:PLDensity} we show the density distribution at 
two separate times ($t_{10} = 0.59~t_{\rm ff,0}$ in black and $t _{50}= 1.06~t_{\rm ff,0}$
in red). We show only 
the mass distribution, since the distribution by volume comprises more than
$90\%$ empty space. 
We see that by $\sim 0.6~t_{\rm ff,0}$, when the virial parameter 
begins to turn around, the gas density is roughly lognormal in shape, characteristic of 
supersonic turbulence 
\citep{Vazquez-Semadeni1994, Ostriker2001, Vazquez-SemadeniGarcia2001, Federrath2009}. 
Interestingly, even towards the end of star formation, the shape of the distribution is not significantly 
different at the high density end. Certainly, there is an excess of low density gas escaping the 
cloud, but the highest density portion, which represents the star forming regions, remains 
essentially the same.

Our fiducial simulation displays a number of key stages of
evolution. Initially, there is rapid collapse and structure formation
driven by turbulent compression and self-gravity, resulting in a
filamentary gas distribution within around half a free-fall time. This
is followed by on-going collapse under self-gravity, during which the
filamentarity remains while collapse around the density peaks begins
to form stars at an accelerating rate. Starting at $t \sim 0.8 t_{\rm
 ff,0}$ there is a transition from collapse dominated by self-gravity
to cloud expansion driven by radiative feedback, where the lowest
density regions are accelerated out of the cloud first, but where the 
star forming regions remain largely unaffected. Star formation
is largely complete by $\sim 1.5 t_{\rm ff,0}$, and the remnant gas in
the cloud is mostly removed by radiation-driven outflows within the
next free-fall time.

\subsection{Convergence Tests}
\label{SubSec:Convergence}

\begin{figure*}
  \centering
  \epsscale{1}
  \includegraphics{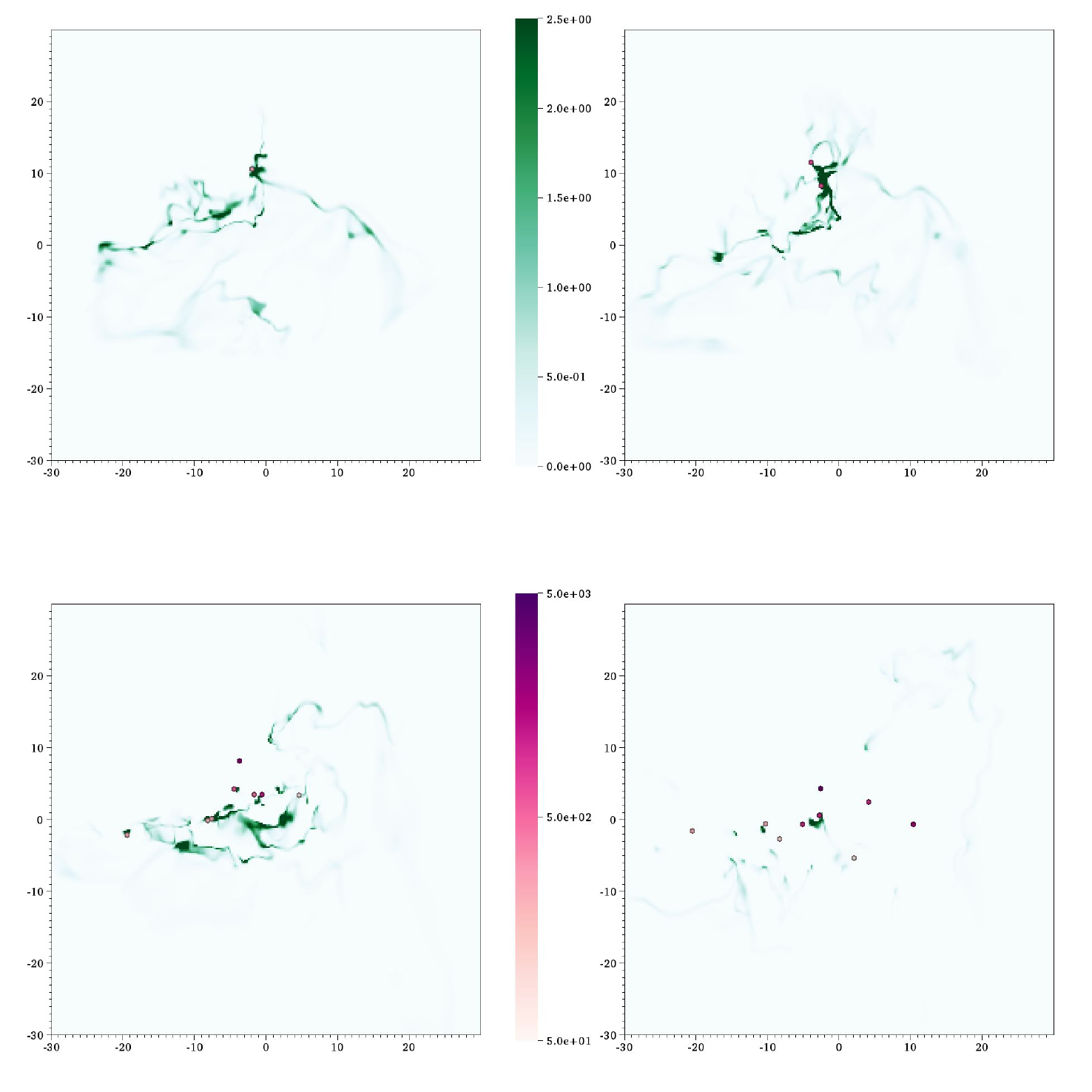}
  \caption{Snapshots of the optical depth $\tau_{\rm cell}$ for our
    fiducial model.  We show slices in the x-y plane, passing through
    the plane of the most massive star particle, at the same four
    times as Figures~\ref{Fig:FiducialEvolutionColX}-
    \ref{Fig:FiducialRadiationEvolution}. As in those figures, we show
    also star particles within $\Delta z = \pm 2~{\rm pc}$ of the slice.}
  \label{Fig:FiducialEvolutionTau}
\end{figure*}

\begin{figure*}
  \centering
  \epsscale{1}
  \includegraphics{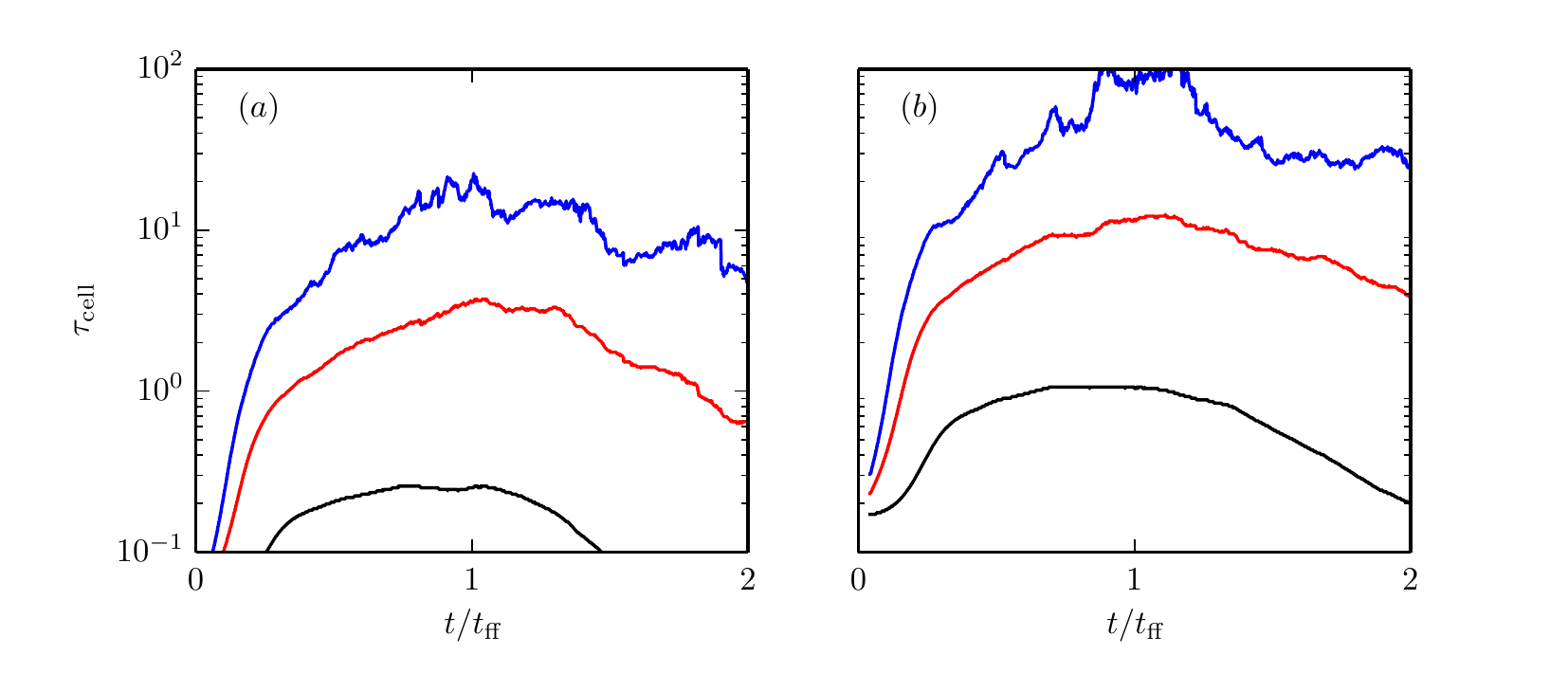}
  \caption{Percentiles of the optical depth distribution as a function
    of simulation time. We show results for both our fiducial cloud
    (left), and a high surface density cloud with $M_{\rm cl, 0} = 2
    \times 10^5~M_{\odot}$ and $r_0 = 15$~pc, corresponding to
    $\Sigma_{\rm cl, 0} = 283~{\rm M_{\odot}~pc^{-2}}$ (right). In
    both cases we show the optical depth at the $50^{\rm th}$ 
    percentile of the mass distribution (black), the $90^{\rm th}$ percentile 
(red), and the $100^{\rm th}$ percentile (blue).}
  \label{Fig:TauDistribution}
\end{figure*}

\begin{figure*}
  \centering
  \epsscale{1}
  \includegraphics{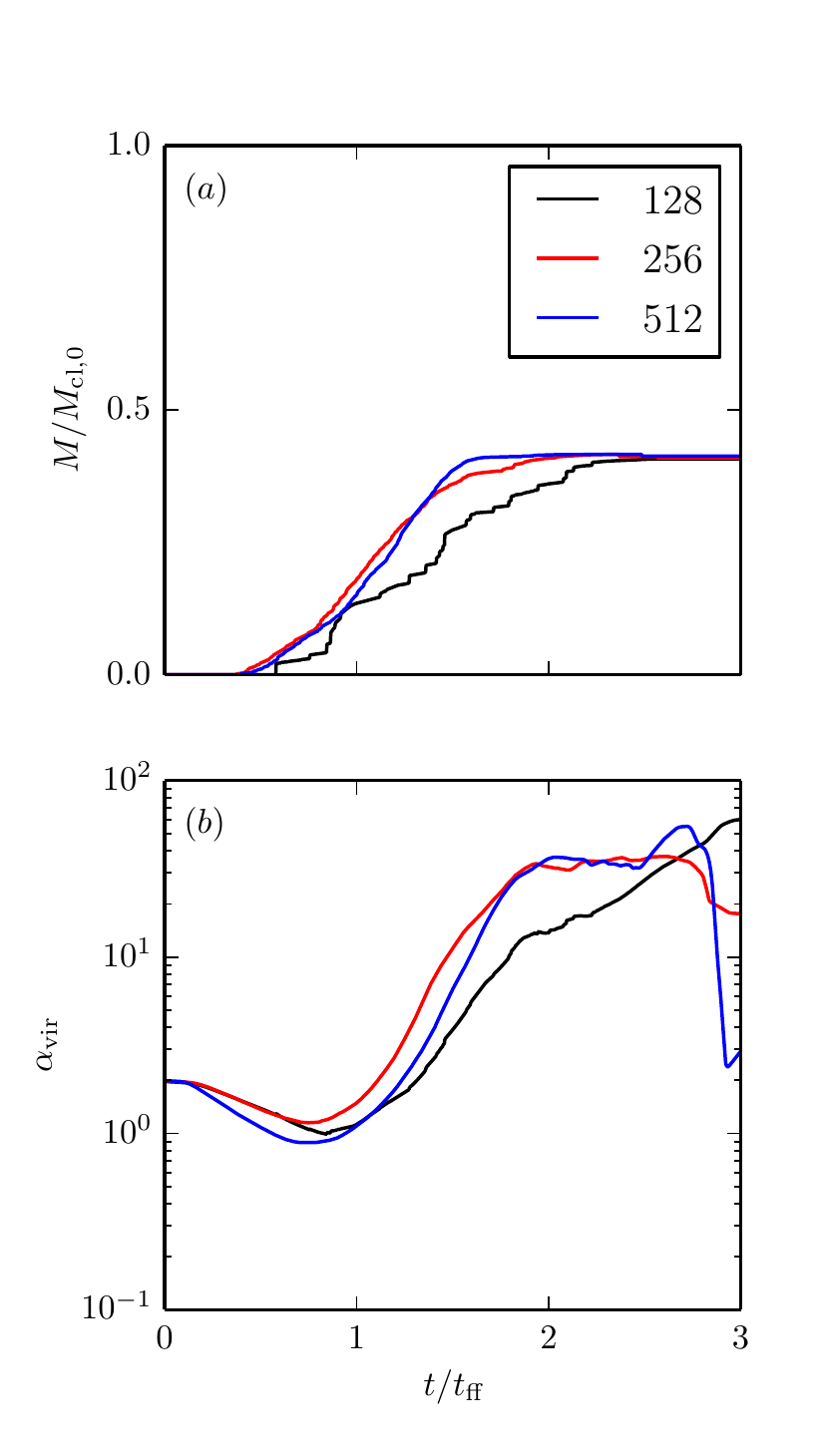}
  \caption{Convergence study for 
    the fiducial cloud model with varying resolution $N$ (shown in the key). 
We show (a) the stellar mass divided by initial cloud mass, 
and (b) the virial parameter of the gas.}
  \label{Fig:Convergence}
\end{figure*}

\begin{figure*}
  \centering
  \epsscale{1}
  \includegraphics{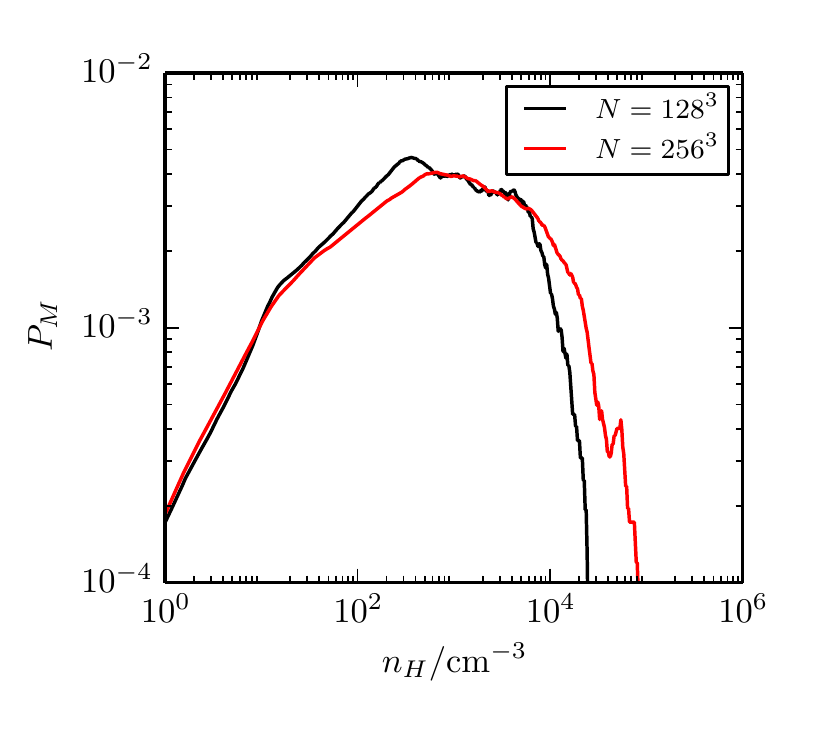}
  \caption{\bf The gas density distribution in our fiducial model at $t_{50} = 1.06~t_{\rm ff,0}$. 
    We show only the distribution in mass.}
    \label{Fig:PLResolution}
\end{figure*}

\begin{figure*}
  \centering
  \epsscale{1}
  \includegraphics{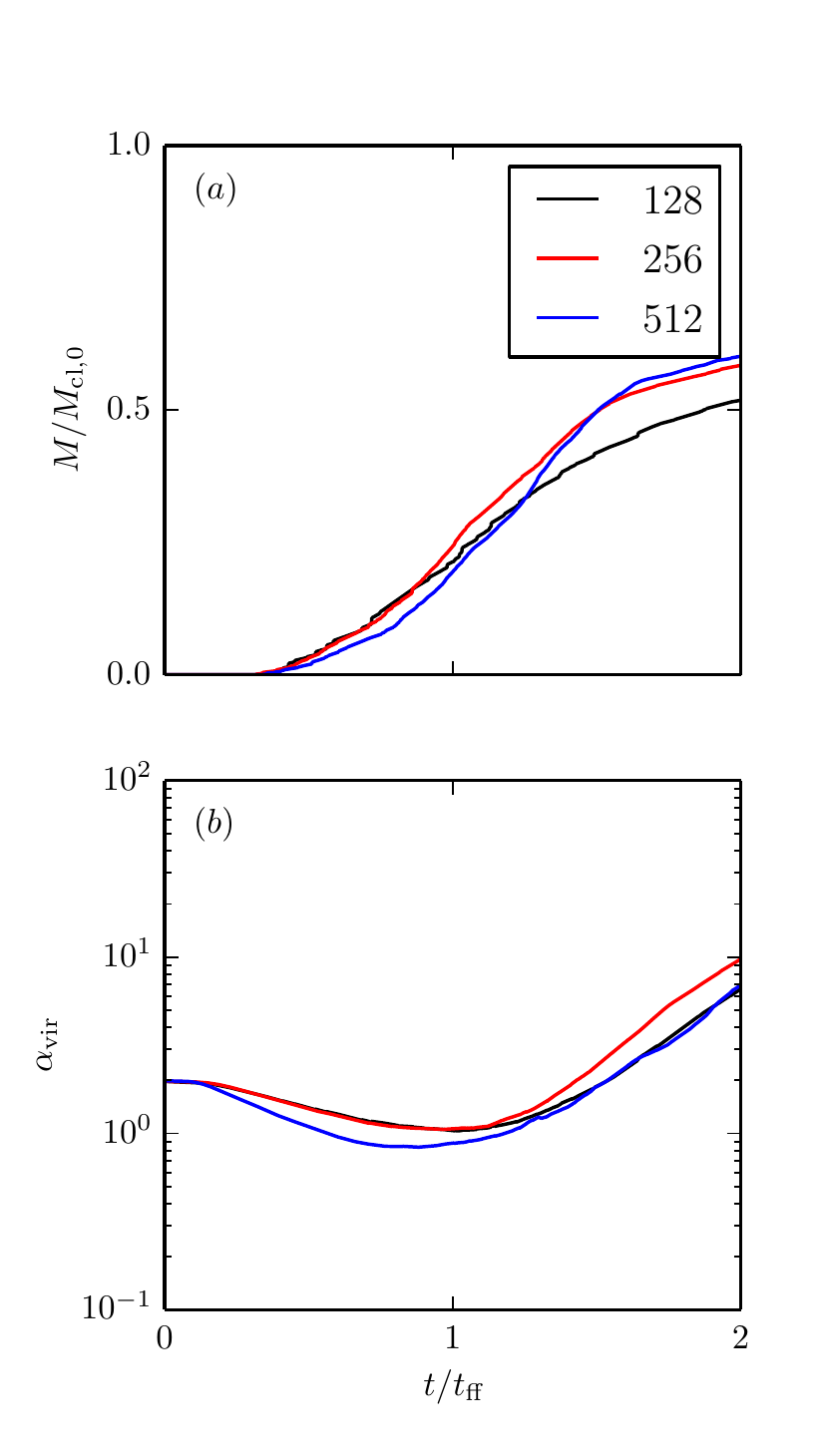}
  \caption{Same as Figure~\ref{Fig:Convergence}, but for a higher mass, higher surface density model 
	with $M_{\rm cl, 0} = 2 \times 10^5~M_{\odot}$ and $r_0 = 15$~pc, corresponding to
    	$\Sigma_{\rm cl, 0} = 283~{\rm M_{\odot}~pc^{-2}}$.}
  \label{Fig:ConvergenceHighMass}
\end{figure*}

In numerically simulating turbulent cloud evolution with 
radiation feedback, we want to ensure that (i) we accurately model the formation of stars driven by
turbulence and gravitational collapse and (ii) we accurately model gas dynamics driven by radiative forces.

The first of these is related to the methodology of sink-particle
creation, which is necessary because 
gas collapse becomes unresolved on the numerical grid.  
To have a sufficiently
``fine-grained'' representation of star formation, we would like to
ensure that a single star particle cannot represent a substantial
fraction of the cloud. The minimum sink particle mass is defined in
terms of the density threshold and the cell size $\Delta x = L/N = 4r_0/N$ as
\begin{eqnarray}
	M_{\rm sink,min} &=& \rho_{\rm th}(\Delta x)^3 \\
	&=& \frac{8.86c_s^2 \Delta x}{G \pi} = \frac{35.4 c_s^2 r_0}{G \pi N} \\ 
	&=& 24.5~{\rm M_{\odot}} 
	\left(\frac{c_s}{0.2~{\rm km~s^{-1}}}\right)^2
	\left(\frac{r_0}{10~{\rm pc}}\right)
	\left(\frac{256}{N}\right).
\end{eqnarray}
As a practical limit, we require that this mass is not more than 
$\sim 0.1 \%$ of the initial cloud mass, so that a 
reasonable number of star particles can be created, even at low efficiencies. This translates 
to around $\Sigma \gtrsim 10~{\rm M_{\odot}~pc^{-2}}$ for the masses and radii that we consider, 
though we will examine this limit in more detail below.

We also need to ensure that we can sufficiently resolve the highest density regions, which 
are converted into star particles. High-resolution simulations of compressible 
turbulence modeling molecular cloud conditions generally show both a lognormal 
component around the mean density, characteristic of supersonic turbulence
\citep{Vazquez-Semadeni1994, Ostriker2001, Vazquez-SemadeniGarcia2001, Federrath2009}, 
and an extended power law tail that arises when self-gravity is included 
\citep{Klessen2000, DibBurkert2005, Vazquez-Semadeni2008, Federrath2008, 
BallesterosParedes2011, Kritsuk2011, Collins2012, FederrathKlessen2013}. 
There is evidence that the power-law component is specifically associated with dense 
prestellar cores (see below). High-resolution observations of 
star forming clouds also generally show power law tails in the surface density whenever there is 
star formation \citep{Kainulainen2009, Schneider2013, Schneider2015}, although there is some debate 
as to whether the power law arises from self-gravity \citep{Brunt2015} and what is the exact relationship 
between the surface density and density distributions \citep{Brunt2010}. 

By contrast, when we consider the mass-weighted density distribution
in our simulations, we find no strong evidence for a power law tail. In
other simulations, this tail is associated with self gravity and, in
particular, with collapsing cores, which themselves have power-law
density profiles \citep{Kritsuk2011,Lee2014}. Thus,  
its absence could indicate that we do not
capture the details of core collapse. In general,
these power law tails are seen to begin at
densities $\sim 100$ times the mean density \citep{Kritsuk2011,
  DibBurkert2005, Vazquez-Semadeni2008, Collins2012,
  FederrathKlessen2013, Lee2014}. In our fiducial simulation, the
threshold for star particle formation is at $\rho \sim 150 \rho_0$ (or
$n_H \sim 1.5 \times 10^4~{\rm cm^{-3}}$). With the peak of the
lognormal in Figure~\ref{Fig:PLDensity} at ${\rm ln}(\rho_{\rm
  peak} / \rho_0) \sim 2.2$ (or $n_H \sim 10^3~{\rm cm^{-3}}$)
and a variance of $\sigma_{{\rm ln}\rho} \sim 2$, the $1\sigma$
density will be at $n_H \sim 10^4~{\rm cm^{-3}}$; hence there is only a
limited range of density over which to sample the power law. 
Effectively, the majority of what would make up the
power law core is hidden in these star particles. Nevertheless, 
tests at higher resolution (see below) appear to indicate that all our models are 
converged in the SFE.

In order to test how well our code captures the physics of gas expansion, we conducted a number of tests 
involving a spherical shell of gas surrounding a single central star particle, and the results of these are 
shown in Appendix~\ref{Sec:Tests}. These tests suggest that the primary numerical limit in
simulating the interaction between radiation and gas with
high accuracy is in ensuring
that $\tau_{\rm cell} = \rho \kappa \Delta x \lesssim 2$ for the optical depth 
within individual cells. This
corresponds to a maximum density that is resolution dependent: 
\begin{equation}
	n_{\rm H, max} = 2.8 \times 10^3~{\rm cm^{-3}}
	\left(\frac{\Delta x}{0.1~{\rm pc}}\right)^{-1}
	\left(\frac{\kappa}{1000~{\rm cm^{2}~g^{-1}}}\right)^{-1}
	\left(\frac{\tau_{\rm cell, max}}{2}\right).
\end{equation}

To gain an idea of how the optical depth evolves in the fiducial model, 
we show in
Figure~\ref{Fig:FiducialEvolutionTau} snapshots of the spatial
distribution of $\tau_{\rm cell}$ at the same four times as
Figures~\ref{Fig:FiducialEvolutionColX}-\ref{Fig:FiducialRadiationEvolution}. 
We see that even in this
model, which has an initial surface density of $\Sigma_{\rm cl, 0} =
71~{\rm M_{\odot}~pc^{-2}}$ and $\tau_{\rm cell, init} = 3 \Sigma_0
\kappa / N = 0.17$, over time the densest regions are enhanced by at
least two orders of magnitude and a portion of the cells have optical depth
$\tau_{\rm cell} \gtrsim 2$.

Even if some localized regions have large single-cell optical depth and do not resolve the radiation
field well, the overall resolution may still be acceptable. In particular,
by around a free-fall time, when stars are beginning to drive gas away from
the sites of star formation, only very small regions have 
$\tau_{\rm cell} \gtrsim 2$. While we may underestimate the radiation
force applied to a small number of cells, for the bulk of the mass we
still follow the physics governing the expulsion of gas and the
propagation of the radiation field.

Quantitatively, the distribution of $\tau_{\rm cell}$ has a long tail,
such that close to $90 \%$ of the mass is found an order of magnitude
below the maximum density.  In Figure~\ref{Fig:TauDistribution}a we
show the $50^{\rm th}$ percentile, $90^{\rm th}$ percentile, and
maximum of the $\tau_{\rm cell}$ distribution as a function of time
for the fiducial model. Even though the maximum $\tau_{\rm cell}$
reaches $\sim 20$ at quite early times, $\tau_{\rm cell, 90} \approx
1$ for the majority of the simulation, which is only a factor of $10$
higher than the initial optical depth.    A similar ratio of
$\tau_{\rm cell, 90} / \tau_{\rm cell, init}$ appears to hold in
higher surface density models.  We see in
Figure~\ref{Fig:TauDistribution}b that a model with $\Sigma \approx
280~{\rm M_{\odot}~pc^{-2}}$, reaches $\tau_{\rm cell, 90} \approx 8$.  
This suggests that up to cloud surface densities $\Sigma \approx 100~{\rm M_{\odot}~pc^{-2}}$, 
$\tau_{\rm cell, 90}$ will remain below $\sim 2$, while for higher surface 
density clouds $\tau_{\rm cell, 90}$ may be higher.

The above considerations suggest that we can satisfactorily simulate
clouds over a range of initial surface densities $10~{\rm M_{\odot}
~pc^{-2}}\lesssim \Sigma \lesssim 100~{\rm M_{\odot}~pc^{-2}}$.
However, in reality we may be able to do better than that, since as
discussed in Appendix~\ref{Sec:Tests}, even at higher cell optical
depths, we only underestimate the velocity of gas expulsion by around
$\sim 10 \%$. We are primarily interested in ensuring that we are able
to accurately capture the net star formation efficiency and evolution
in response to radiation. By looking at how these properties vary as
we change the numerical resolution, we may gain a better idea of what
range of parameters provide converged results.

Figure~\ref{Fig:Convergence}a shows evolution of 
the global efficiency of star formation in the fiducial model 
as a function of time with varying numerical resolution for $N = 128, 256,$ and $512$. 
We see that in the $N = 128$ case, the fiducial
simulation is clearly not converged. It shows substantial differences
from the higher resolution models, including ``stair-stepping'' in the stellar
mass history. This makes sense, since the minimum sink mass for this
simulation is around $M_{\rm sink} \approx 80~{\rm M_{\odot}}$, which
is close to $0.2 \%$ of the global cloud mass.
  Importantly however,
  the $N=256$ and $N=512$ models are converged, and even though the
  $N=128$ model underestimates the SFE at intermediate times, its final 
SFE is the same as for the higher resolution models.

We show also in Figure~\ref{Fig:PLResolution} the density distribution at
$t_{50}$, when 50\% of the final stellar mass has been assembled. 
With resolution reduced by a factor of two in the $128^3$ models, 
the star particle density threshold is at $\sim 40 \rho_0$ or $n_H \sim 4 \times 10^3~{\rm cm^{-3}}$ 
compared to $\rho \sim 150 \rho_0$ or $n_H \sim 1.5 \times 10^4~{\rm cm^{-3}}$ for 
$N=256^3$. We believe that the lack of convergence evident in the low resolution models 
(see Figure~\ref{Fig:Convergence}), which leads to
an underestimation of the SFE, is because this threshold is not sufficiently high compared to the 
lognormal distribution, as evident from the cutoff in Figure~\ref{Fig:PLResolution} here.
However, the convergence of the $256^3$ and $512^3$ models in terms of stellar efficiency
(see Figure~\ref{Fig:Convergence}) would suggest that in our highest resolution models, 
we capture the distribution of gas dominated by turbulence, with only collapsing regions 
assigned to sink particles. Thus, while insufficient resolution can lead to an underestimate of the 
SFE, we believe that increasing resolution beyond a certain point primarily provides improved 
resolution of collapsing cores without changing the SFE. This speculation could be tested with 
AMR simulations.

A similar picture holds for models at the highest
surface densities we simulate. In
Figure~\ref{Fig:ConvergenceHighMass}a we again show evolution of 
the global stellar
efficiency as a function of resolution, but in this case for a cloud
of mass $M_{\rm cl, 0} = 2 \times 10^5~M_{\odot}$. For this model, the
$90^{\rm th}$ percentile in optical depth lies at $\tau_{\rm cell,
  90} \sim 8$, which is a factor of 4 larger than the limit found
in Appendix~\ref{Sec:Tests} for best accuracy in capturing radiation forces. 
However, the larger values of $\tau_{\rm cell}$
do not have a major impact on the evolution of the 
global efficiency, presumably because errors are only $\sim 10\%$ at larger 
$\tau_{\rm cell}$ (see Appendix~\ref{Sec:Tests}), and because only a small fraction 
of the gas has large $\tau_{\rm cell}$.  
As for the
fiducial cloud, stars do begin forming slightly earlier in the higher
resolution run and they do form at a slightly faster rate, but the
final efficiency is the same.

In Figures~\ref{Fig:Convergence}b and ~\ref{Fig:ConvergenceHighMass}b,
we show the gas virial parameter, for our fiducial and highest-$\Sigma$ 
models at different
resolution. As for the stellar efficiency, our simulations are not
well converged at $N =128$, but there are few differences between the
$N = 256$ and $N = 512$ runs, at least until the bulk of star
formation is already complete.

\section{Star Formation Efficiencies}
\label{Sec:SFE}

We now turn to the central question of this paper: what mechanisms are
dominant in setting the star formation efficiency in gaseous clouds,
and how do these depend on cloud parameters? 
As the key elements are turbulence and radiative
feedback, it is interesting to consider these effects separately at
first.

\begin{deluxetable}{l|cccccccccccc}
\tabletypesize{\scriptsize} \tablewidth{0pt}
\def\arraystretch{1.0}
\tablecaption{Model Results}
\tablehead{
 \vspace{-0.2cm} &
 &
 &
\colhead{$t_{\rm *}$} & 
\colhead{$t_{\rm 1/2}$} & 
\colhead{$t_{\rm 90 \%}$} &
\colhead{$t_{\rm unb}$} &
 & 
\colhead{$t_{\rm break}$} & 
 \\ 
\colhead{Model} \vspace{-0.2cm} &
\colhead{$\varepsilon$} &
\colhead{$\varepsilon_{\rm adj}$}  &
 &
 &
 &
 & 
\colhead{$\beta$} & 
 &
\colhead{$\varepsilon_{\rm ff, \bar \rho}$} \\
 &
 &
 &
\colhead{[${t}_{\rm ff}$]}&
\colhead{[${t}_{\rm ff}$]}&
\colhead{[${t}_{\rm ff}$]}&
\colhead{[${t}_{\rm ff}$]}&
 &
\colhead{[${t}_{\rm ff}$]}&
\colhead{}
}

\startdata
$\Sigma$-M2E4-R25 & 0.12 & 0.14 & 0.67 & 0.88 & 1.51 & $1.14^{+0.18}_{-0.24}$ & --$^{a}$ & --$^{a}$ & --$^{a}$ \\
$\Sigma$-M5E4-R35 & 0.18 & 0.20 & 0.63 & 1.07 & 2.05 & $1.17^{+0.22}_{-0.23}$ & --$^{a}$ & --$^{a}$ & --$^{a}$ \\
$\Sigma$-M2E4-R20 & 0.22 & $0.25$ & 0.61 & 1.08 & 1.72 & $1.35^{+0.25}_{0.12}$ & --$^{a}$ & --$^{a}$ & --$^{a}$ \\
$\Sigma$-M5E4-R25 & 0.25 & 0.28 & 0.52 & 1.01 & 1.71 & $1.18^{+0.16}_{-0.20}$ & 1.64 & 0.80 & 0.49 \\
$\Sigma$-M1E5-R35 & 0.23 & 0.26 & 0.49 & 1.00 & 1.53 & $1.22^{+0.19}_{-0.22}$ & 0.77 & 0.85 & 0.24 \\
$\Sigma$-M2E4-R15 & 0.30 & 0.34 & 0.47 & 1.07 & 1.65 & $1.21^{+0.12}_{-0.19}$ & 1.08 & 0.70 & 0.37 \\
$\Sigma$-M1E4-R10 & 0.30 & 0.34 & 0.54 & 1.09 & 1.64 & $1.29^{+0.16}_{-0.20}$ & 1.11 & 0.78 & 0.34 \\
$\Sigma$-M5E4-R20 & 0.32 & 0.36 & 0.46 & 1.02 & 1.51 & $1.19^{+0.14}_{-0.16}$ & 1.17 & 0.73 & 0.45 \\
$\Sigma$-M1E4-R08 & 0.39 & 0.44 & 0.47 & 1.11 & 1.75 & $1.34^{+0.16}_{-0.20}$ & 0.85 & 0.92 & 0.41 \\
$\Sigma$-M1E5-R25 & 0.32 & 0.37 & 0.46 & 1.05 & 1.48 & $1.26^{+0.14}_{-0.19}$ & 1.25 & 1.02 & 0.36 \\
$\Sigma$-M2E5-R35 & 0.31 & 0.35 & 0.47 & 1.08 & 1.51 & $1.32^{+0.16}_{-0.24}$ & 0.87 & 0.74 & 0.29 \\
$\Sigma$-M5E3-R05 & 0.42 & 0.49 & 0.54 & 1.17 & 1.95 & $1.45^{+0.17}_{-0.21}$ & 0.69 & 1.02 & 0.37 \\
$\Sigma$-M2E4-R10 & 0.42 & 0.48 & 0.44 & 1.08 & 1.58 & $1.33^{+0.16}_{-0.22}$ & 0.93 & 0.76 & 0.43 \\
{\bf $\Sigma$-M5E4-R15} & {\bf 0.42} & {\bf $0.48$} & {\bf 0.37} & {\bf 1.06} & {\bf 1.57} & {\bf $1.33^{+0.14}_{-0.21}$} & {\bf 0.95} & {\bf 0.81} & {\bf 0.43} \\
$\Sigma$-M1E5-R20 & 0.41 & 0.47 & 0.32 & 1.02 & 1.56 & $1.31^{+0.18}_{-0.21}$ & 0.88 & 0.84 & 0.45 \\
$\Sigma$-M2E4-R08 & 0.49 & 0.56 & 0.42 & 1.10 & 1.71 & $1.46^{+0.18}_{-0.24}$ & 0.94 & 0.80 & 0.49 \\
$\Sigma$-M2E5-R25 & 0.41 & 0.41 & 0.32 & 1.04 & 1.57 & $1.43^{+0.20}_{-0.24}$ & 0.87 & 0.92 & 0.42 \\
$\Sigma$-M1E4-R05 & 0.52 & 0.60 & 0.44 & 1.16 & 1.80 & $1.59^{+0.23}_{-0.25}$ & 0.83 & 0.91 & 0.51 \\
$\Sigma$-M1E5-R15 & 0.52 & 0.59 & 0.29 & 1.04 & 1.63 & $1.40^{+0.22}_{-0.13}$ & 1.31 & 1.14 & 0.45 \\
$\Sigma$-M5E4-R10 & 0.58 & 0.65 & 0.31 & 1.09 & 1.67 & $1.55^{+0.21}_{-0.28}$ & 0.95 & 0.78 & 0.49 \\
$\Sigma$-M2E5-R20 & 0.48 & 0.54 & 0.28 & 1.21 & 1.86 & $1.43^{+0.31}_{-0.18}$ & 1.05 & 0.63 & 0.51 \\
$\Sigma$-M5E4-R08 & 0.62 & 0.70 & 0.31 & 1.11 & 1.77 & $1.73^{+0.28}_{-0.30}$ & 1.15 & 0.78 & 0.49 \\
$\Sigma$-M2E4-R05 & 0.61 & 0.69 & 0.37 & 1.20 & 1.86 & $1.83^{+0.34}_{-0.32}$ & 1.11 & 0.78 & 0.49 \\
$\Sigma$-M2E5-R15 & 0.61 & 0.69 & 0.26 & 1.08 & 1.63 & $1.75^{+0.25}_{-0.25}$ & 1.30 & 0.56 & 0.47 \\
\hline
$\alpha$-A0.1 & 0.91 & 0.91 & 0.67 & 1.02 & 1.13 & $1.16^{+0.04}_{-0.13}$ & 1.24 & 0.83 & 0.42 \\
$\alpha$-A0.2 & 0.87 & 0.87 & 0.60 & 1.04 & 1.22 & $1.27^{+0.06}_{-0.19}$ & 1.19 & 0.78 & 0.43 \\
$\alpha$-A0.4 & 0.67 & 0.67 & 0.51 & 0.99 & 1.23 & $1.28^{+0.07}_{-0.17}$ & 1.39 & 0.75 & 0.44 \\
$\alpha$-A0.8 & 0.58 & 0.59 & 0.44 & 1.00 & 1.31 & $1.28^{+0.10}_{-0.15}$ & 0.98 & 0.77 & 0.36 \\
$\alpha$-A1.5 & 0.47 & 0.54 & 0.43 & 1.03 & 1.54 & $1.29^{+0.17}_{-0.19}$ & 1.10 & 0.70 & 0.41 \\
{\bf $\alpha$-A2.0} & {\bf 0.42} & {\bf $0.48$} & {\bf 0.37} & {\bf 1.06} & {\bf 1.57} & {\bf $1.33^{+0.14}_{-0.21}$} & {\bf 1.03} & {\bf 0.80} & {\bf 0.43} \\
$\alpha$-A3.0 & 0.28 & 0.35 & 0.36 & 0.99 & 1.47 & $1.33^{+0.18}_{-1.33}$ & 0.92 & 0.80 & 0.36 \\
$\alpha$-A6.0 & 0.12 & 0.35 & 0.27 & 1.03 & 1.66 & --$^{b}$ & 1.57 & 0.87 & 0.16 \\
\vspace{-0.1 cm} $\alpha$-A10.0 & 0.05 & 0.32 & 0.14 & 1.00 & 1.39 & --$^{b}$ & 0.20 & 1.95 & 0.07
\enddata
\tablecomments{Columns display the following information (i) final efficiency of star formation, 
(ii) final efficiency adjusted for the inital turbulence-driven mass outflow,
(iii) time of first star formation,  
(iv) time at which half of the final stellar mass is assembled, 
(v) time at which $90 \%$ of the final stellar mass is assembled,
(vi) time at which the cloud becomes unbound (reaches a virial parameter of 
$\alpha_{\rm vir} = 5^{+5}_{-3}$), 
(vii) post-break power law exponent of the stellar mass evolution using a double power-law fit, 
(viii) break time in double power law fit to the stellar mass evolution, and 
(ix) star formation efficiency per freefall time as defined in Equation~\ref{Eq:epsff_lin}.\\
$^{a}$: In these models, the star formation rate was not fit as there were insufficient numbers of discrete star 
particles to perform a meaningful fit \\
$^{b}$: In these models, $t_{\rm unb}$ was not calculated since the models were initially unbound.}
\end{deluxetable}

\subsection{The Effect of Radiative Feedback}
\label{SubSec:SFERad}

\begin{figure*}
  \centering
  \epsscale{1}
  \includegraphics{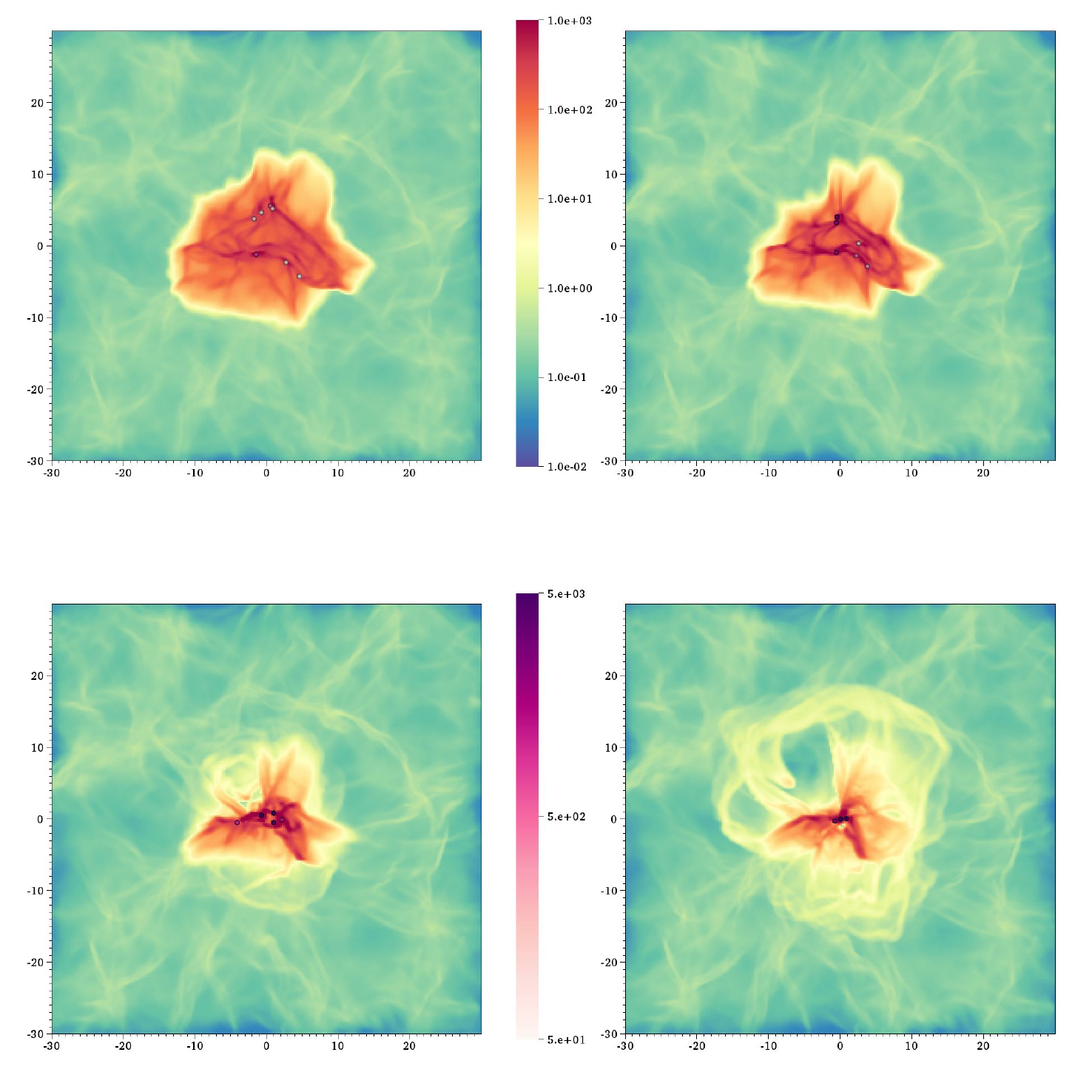}
  \caption{Snapshots of the surface density for the fiducial model
    with reduced turbulence $\alpha_{\rm vir,0} = 0.1$. Surface
    densities are projected in the x-y plane, and snapshots are shown
    for $t / t_{\rm ff,0} = 0.72, 0.83, 1.02$,~ and $1.13$, corresponding to 
    $t_2$, $t_{10}$, $t_{50}$ and $t_{90}$. As for
    Figure~\ref{Fig:FiducialEvolutionColZ}, we show also all star
    particles projected onto the x-y plane. The color scale for the
    gas column density (top) is in units of $M_{\odot}~{\rm pc^{-2}}$
    and the color scale for the particle mass (bottom) is in units of
    $M_{\odot}$.}
  \label{Fig:FiducialLowAlphaEvolutionColZ}
\end{figure*}

We begin by presenting results from a set of models with the initial
turbulent energy a factor of 20 lower than in the fiducial case
(i.e., with $\alpha_{\rm vir,0} = 0.1$). In
Figure~\ref{Fig:FiducialLowAlphaEvolutionColZ}, we show snapshots of
the column density in the x-y plane for a model cloud with the
fiducial initial surface density, but with reduced initial turbulence. The lack of
turbulent support means that for the first $\sim 0.6 t_{\rm ff,0}$, the
cloud undergoes nearly free-fall collapse, contracting to around $40
\%$ of its initial radius and converting potential to kinetic energy.
Star formation only begins once this contraction is complete and the
virial parameter for the cloud is near unity.

After the initial contraction and virialization, the evolution is
quite similiar to the fiducial case as the gas density distribution
becomes similarly filamentary, with preferential directions set by the
seed turbulent field. Low initial turbulence simplifies the picture in
several respects. Firstly, stars tend to form very close to the center
of the cloud, with very little initial momentum away from the center
of mass. This can be seen in the lower two panels of
Figure~\ref{Fig:FiducialLowAlphaEvolutionColZ}, where star particles
close to the cloud center eventually converge to a single, massive,
central star particle.  Secondly, star formation occurs on a relatively
short timescale compared to the initial free-fall time due to the high
densities and small physical length scales involved. Therefore, the
majority of stellar mass is assembled in less than $25 \%$ of a
free-fall time after the initial collapse.

As a very simple model of star formation, one could imagine an initial
roughly spherical collapse, until the cloud reaches a characteristic
radius $x r_0$ at which the virial parameter is of order unity. Once
stars have begun to form close to the center and have evacuated their local
environment, we might expect the remnant cloud's effective radius to evolve
under the competing effects of the inward force of gravity and the
outward force of radiation. Star formation, or at least accretion on to 
the young central star cluster would then continue so long as
gravity dominates over the effects of stellar feedback, and it would stop
once enough stars form for radiation forces to disperse the remaining
gas in the cloud.

This picture is essentially that of a radiation dominated HII region,
with a single, central stellar cluster that stops accreting mass once
the luminosity is sufficient to drive away all surrounding gas.  Such
systems, in different limits, have been analyzed previously in
\cite{Elmegreen1983, Scoville2001, KrumholzMatzner2009, Murray2010,
Fall2010, Kim2016} and considering effects of inhomogeneity by 
\cite{ThompsonKrumholz2014}.  We review the
ideas involved, connecting to the results of our simulations.

We consider a system consisting of a spherical shell of gas with mass
$M_{\rm sh} = (1 - \varepsilon) M_{\rm cl, 0}$ surrounding a point
mass representing a stellar cluster of mass $M_* = \varepsilon M_{\rm
  cl, 0}$.
At any point in time, the circumcluster shell will have a distribution of
surface densities around the mean surface density $\langle
\Sigma^c(\varepsilon, x, \Sigma_{\rm cl, 0}) \rangle$, where
$\Sigma_{\rm cl,0} = M_{\rm cl,0} / \pi r_0^2$ is the initial, observed cloud
column density. The mean surface density would decrease as gas is
converted to stars, but increase under global collapse, so the outcome depends
on the initial cloud reference surface density $\Sigma_{\rm cl, 0}$, the
time-dependent stellar efficiency $\varepsilon$, and the radius to
which the cloud collapses $xr_0$.
We note also that here $\langle\Sigma^c\rangle = M_{\rm sh} / (4 \pi r^2)$
represents the mean surface density seen by the central star cluster, as 
distinct from the cloud column density seen by an external observer 
$M_{\rm sh} / (\pi r^2) = 4 \langle \Sigma^c\rangle$. 

In this simple model, assuming all the flux is absorbed, 
the radiative force per unit area in the shell
due to the central luminous source is $F_{\rm rad} = L/(4 \pi r^2 c) =
\Psi M_{*} / (4 \pi r^2 c)$. Meanwhile, there are two components
to the gravitational force applied to a local patch of surface
density $\Sigma^c$. Firstly, there is the force per unit area due to the
central point mass $F_1 = GM_{*}\Sigma^c / r^2$. Secondly, unlike
previous studies \citep{Fall2010}, we also include a term for the
self-gravity of the shell per unit area, which is
$F_2 = GM_{\rm  sh}\Sigma^c / (2 r^2)$. The exact normalization of this latter force
is derived for a uniform density shell and may change with surface
density variations. However, with the basic argument that gas at the
radial center of the shell will feel the force of all interior mass,
or half the total shell mass, this is at least
approximately correct even when the shell surface density is not
uniform. 

For a region with surface density $\Sigma^c$, the Eddington ratio
$F_{\rm rad}/F_{\rm grav}=F_{\rm rad}/(F_1+F_2)$ is equal to
$\Psi M_*/[2\pi c G \Sigma^c(2 M_* + M_{\rm sh})]$.
At a time when the net star formation efficiency is $\varepsilon$, we
can evaluate the Eddington surface density $\Sigma_E$ such that the
inward and outward forces balance and the Eddington ratio is unity:
\begin{eqnarray}
	\Sigma_E &=& \frac{\varepsilon}{\varepsilon + 1}\frac{\Psi}{2 \pi c G} \nonumber \\
	&=& 759~{\rm M_{\odot}~pc^{-2}} \frac{\varepsilon}{1 + \varepsilon}
	\left(\frac{\Psi}{2000~{\rm erg~s^{-1}~g^{-1}}}\right);
	\label{Eq:FinalSigma}
\end{eqnarray}
here we have used $M_{*} = \varepsilon M_{\rm cl, 0}$ and $M_{\rm
  sh} = (1 - \varepsilon) M_{\rm cl, 0}$. In local patches where surface densities
within the shell exceed the Eddington level 
$\Sigma^c>\Sigma_E$, gas will be able to continue
collapsing, and where $\Sigma^c<\Sigma_E$, the gas can can be
driven outwards.
We note that even for $\varepsilon \rightarrow 1$, $\Sigma_E < 400
{\rm M_{\odot}~pc^{-2}}$, implying that UV radiation feedback by itself is not
expected to be effective in expelling gas from very high surface density GMCs
\citep[see][for a study of reprocessed radiation effects in this regime]{SkinnerOstriker2015}

We now consider a very simple hypothesis, similar in spirit to
\cite{Fall2010}, in which star formation, and accretion halts completely once the
mean circumcluster
surface density reaches the Eddington value,
$\langle \Sigma^c \rangle = \Sigma_E$. For a given $\varepsilon(t)$, the
mean circumcluster surface density can be related to the initial cloud
mass and surface density by
\begin{eqnarray}
	\langle \Sigma^c \rangle &=& \frac{(1 - \varepsilon)M_{\rm cl,0}}{4 \pi (xr_0)^2} \nonumber \\
	&=& \frac{(1 - \varepsilon)}{4x^2}\Sigma_{\rm cl, 0},
	\label{Eq:MeanSurfaceDensity}
\end{eqnarray}
where the $(1 - \varepsilon)$ comes from the conversion of gas to
stars, the factor of $x$ from cloud contraction, and the additional factor of $4$ from the 
fact that $\Sigma_{\rm cl, 0} = M_{\rm cl,0} / (\pi r_0^2)$ is an observed column 
density. If we substitute $\langle \Sigma^c \rangle$ from Equation~(\ref{Eq:MeanSurfaceDensity})
for $\Sigma_E$ in Equation~(\ref{Eq:FinalSigma}), we obtain a
simple relationship between the star formation efficiency and the
initial cloud surface density:
\begin{eqnarray}
	\frac{1}{\varepsilon} - \varepsilon &=& \frac{2 \Psi x^2}
	{\pi c G {\Sigma_{\rm cl, 0}}} \nonumber \\
	&=& 30.5 x^2 \left(\frac{\Sigma_{\rm cl, 0}}{100~{\rm M_{\odot}~pc^{-2}}}\right)^{-1}
	\left(\frac{\Psi}{2000~{\rm erg~s^{-1}~g^{-1}}}\right).
	\label{Eq:FinalEfficiency}
\end{eqnarray}
Similar to Equation (6) of \cite{Fall2010}, this predicts a low
efficiency ($\varepsilon \sim 3\%$) of star formation in GMCs with $\Sigma \sim 100~{\rm
  M_{\odot}~pc^{-2}}$, typical of the Milky Way.

\begin{figure}
  \centering
  \epsscale{1}
  \includegraphics{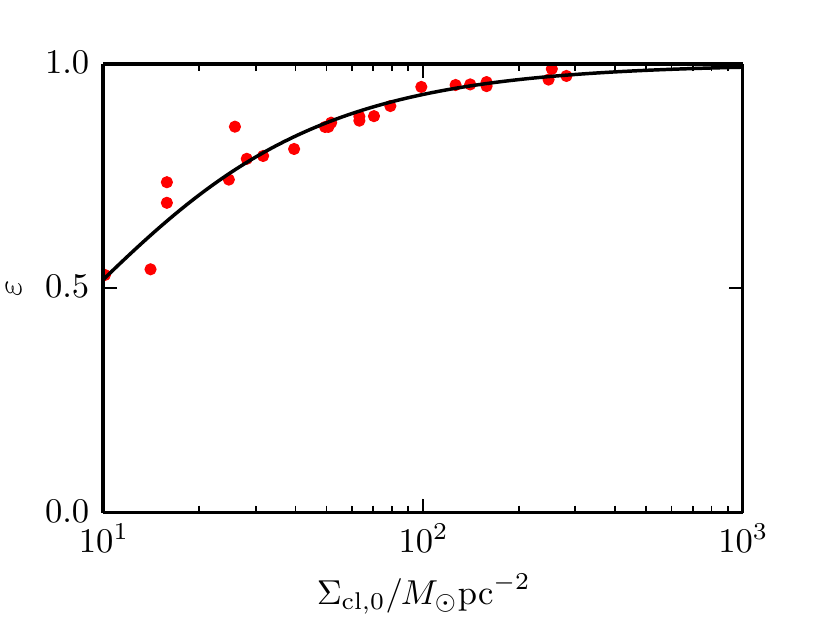}
  \caption{Final star formation efficiency $\varepsilon$ as a function
    of surface density (in units $M_\odot~{\rm pc^{-2}}$) for the 
    $\Sigma$-series simulations with reduced turbulence $\alpha_{\rm
      vir,0} = 0.1$. We show both simulation outputs (red circles) as
    well as the best-fit to Equation~(\ref{Eq:FinalEfficiency})
    assuming a uniform density shell (black line).}
  \label{Fig:a0.1Efficiencies}
\end{figure}

To test whether this simple radiative force balance argument can
capture simulated cloud behavior, we first consider low turbulence
models, $\alpha_{\rm vir,0} = 0.1$, for a range of surface densities.
In Figure~\ref{Fig:a0.1Efficiencies} we show the final star formation
efficiency for this set of simulations, which have initial masses and radii
the same as for our $\Sigma$-series. 

In Figure~\ref{Fig:a0.1Efficiencies} we also show the fit of the
simple model represented by Equation~(\ref{Eq:FinalEfficiency}) to our
cloud simulations. In the low turbulence case, the shape of
Equation~(\ref{Eq:FinalEfficiency}) fits relatively well, matching the
drop in efficiencies in our low surface density
clouds.
However, the best fit requires a value of $x^2 = 0.005$, or
linear contraction of the cloud radius by $x \sim 0.07$. Modeling the
cloud contraction is not straightforward due to the non-spherical gas
distribution. Even so, it is clear from
looking at Figure~\ref{Fig:FiducialLowAlphaEvolutionColZ} that the
low-turbulence clouds collapse to around $x\sim 1/2$ of their initial
radii. This is a factor of more than five larger than the nominal best fit.  We
conclude that although Equation~(\ref{Eq:FinalEfficiency}) can fit the
numerical results for the reduced turbulence models, the parameters
required are not consistent with the true evolution and structure of
the cloud.

The source of this discrepancy is likely that the surface density seen
by the radiation field is not uniform, but is instead distributed over
at least an order of magnitude. As a consequence, at a given
$\varepsilon$, even if the mean surface density in a cloud is
comparable to $\Sigma_E$ as given in Equation~(\ref{Eq:FinalSigma}),
there will be many higher surface density regions that can still
collapse, leading to an increase in $\varepsilon$. Conversely, as
noted by \cite{ThompsonKrumholz2014}, this also means that low 
surface density regions can be driven away even when the stellar
luminosity is not sufficient to drive off the bulk of a cloud's
mass. To assess the impact of this, we now turn to the effect that
turbulence has on changing the gas density distribution and therefore
also this simple picture of star formation regulation.

\subsection{The Effect of Turbulence}
\label{SubSec:SFETurb}

Qualitatively, we expect turbulence to affect cloud evolution in two
separate ways. Firstly, since our $\Sigma$-series clouds are only
marginally bound, the initial turbulence will unbind a certain amount
of the gas on the edges of the cloud immediately. This effectively
lowers the initial surface density, and sets a maximum on the star
formation efficiency. Secondly and more fundamentally, turbulence dramatically alters the
density and surface density distributions.

We may gain a more concrete idea of what turbulence means for star
formation by considering a simplified version of our $\Sigma$- and
$\alpha$-series in which radiative feedback is turned off, so that
only turbulence and self-gravity set the star formation rate. In these
simulations, star formation continues until all the gas is consumed,
and the only gas that is not converted to stars is that which is
initially unbound by the turbulence.

\begin{figure*}
  \centering
  \epsscale{1}
  \includegraphics{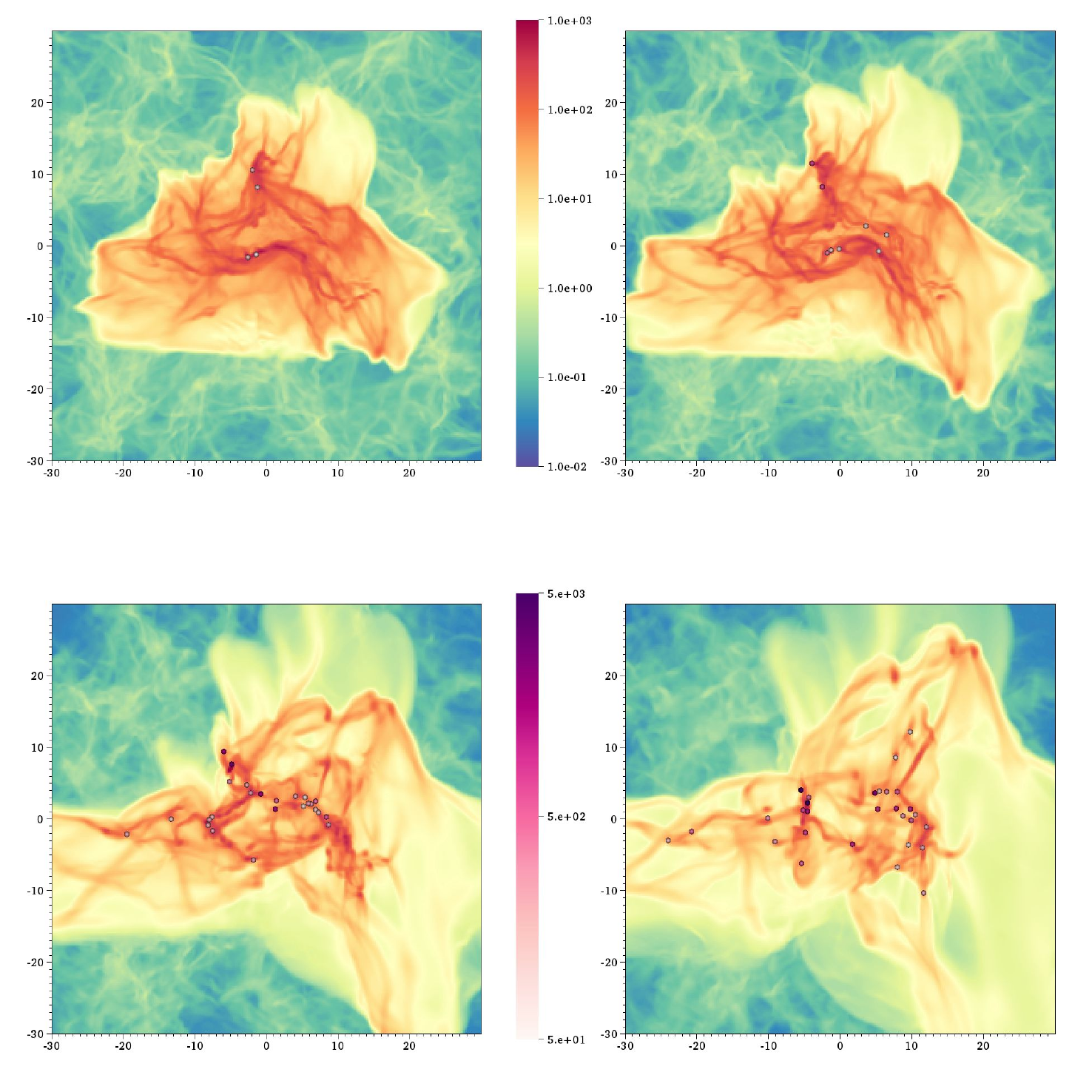}
  \caption{Snapshots of the surface density for our fiducial model
    with no radiative feedback. Snapshots are shown for $t / t_{\rm ff,0} =
    0.43, 0.59, 1.06$ and $1.57$, i.e., $t_2$, $t_{10}$, $t_{50}$ and $t_{90}$ in 
    the fiducial model {\it with feedback}.  We
    also show all star particles projected onto the x-y plane. The
    color scale for the gas column density (top) is in units of
    $M_{\odot}~{\rm pc^{-2}}$, and the color scale for the particle mass
    (bottom) is in units of $M_{\odot}$.}
  \label{Fig:FiducialNFEvolutionColZ}
\end{figure*}

Figure~\ref{Fig:FiducialNFEvolutionColZ} shows the resulting column density evolution for our fiducial 
model with no radiation. Until more than a freefall time the no-feedback and 
feedback cases are very similar. Gas collapses along 
the same filaments, with the same initial density structure. Moreover, stars appear to form at roughly 
the same rate despite the lack of radiative feedback to stir up additional turbulence. This is likely because 
even though the no-feedback model does not drive out low-density gas, and so does not show the same 
prominent columns seen in Figures~\ref{Fig:FiducialEvolutionColZ} -- \ref{Fig:FiducialRadiationEvolution}, 
the high density sites of star formation remain similar. Differences between the two models appear only 
at late times, since in the model without feedback, star formation continues at the same rate for close to 
an extra free-fall time. Eventually, the majority of the no-feedback cloud is converted into stars. Only the gas driven away by the initial turbulence does not contribute to star formation.

\begin{figure*}
  \centering
  \epsscale{1}
  \includegraphics{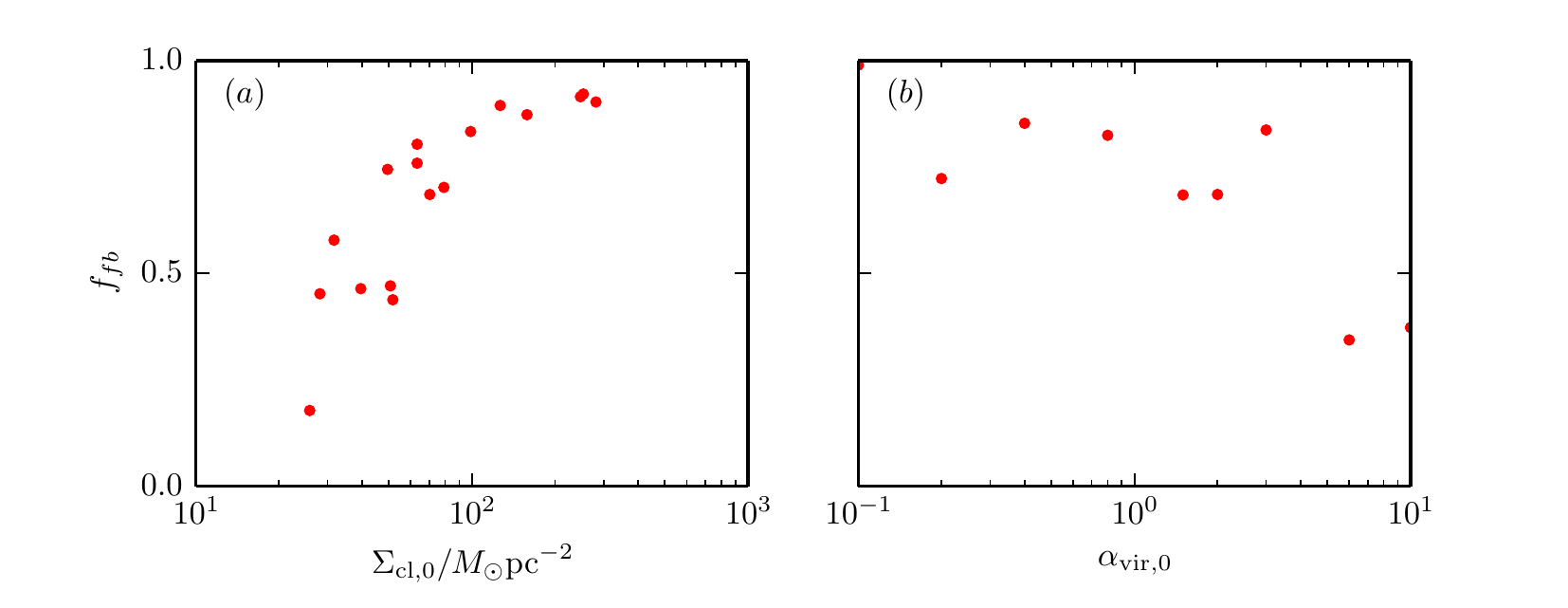}
  \caption{Ratio $f_{\rm fb}$ of the stellar mass at time $t_{\rm fb}$ to the final stellar mass 
	for the (a) $\Sigma$- series, and (b) $\alpha$-series
    simulations.}
  \label{Fig:TurbRatios}
\end{figure*}

In fact, if we compare the evolution of stellar mass in all of our
$\Sigma$ and $\alpha$-series models (discussed in detail below) 
with and without
feedback, we find that they are essentially identical 
up until the point that radiation begins to disperse the gas.
We may define a time
$t_{\rm fb}$, at which there is a $10 \%$ difference in stellar mass
between simulations with and without radiation feedback. 
We then define $f_{\rm fb}$, the fraction of the final stellar
mass in the feedback model that has been assembled by time $t_{\rm fb}$.  In
Figure~\ref{Fig:TurbRatios} we show 
$f_{\rm fb}$ as a function of
both initial surface density and virial parameter. We see that for
almost all models, except for those at low surface density or high virial
parameter (where there is little stellar mass formed), between $70$ and
$90 \%$ of the stellar mass is assembled before radiative feedback
can significantly affect star formation.

This suggests that the star formation rate in our clouds is determined
primarily by the initial conditions, via the density structure that is
imposed by the interplay between gravity and turbulence. The majority
of stellar mass growth is through stars formed early on before
radiative feedback becomes important.  As will be discussed further in
Section~\ref{SubSec:SFERates}, radiation does little to suppress this
early star formation. Instead, radiative feedback is primarily
important in rapidly truncating star formation by driving gas from the
cloud, once sufficiently many stars have formed in the
turbulent density field.

\subsubsection{Turbulent Outflows}
\label{SubSubSec:Outflows}

\begin{figure}
  \centering
  \epsscale{1}
  \includegraphics{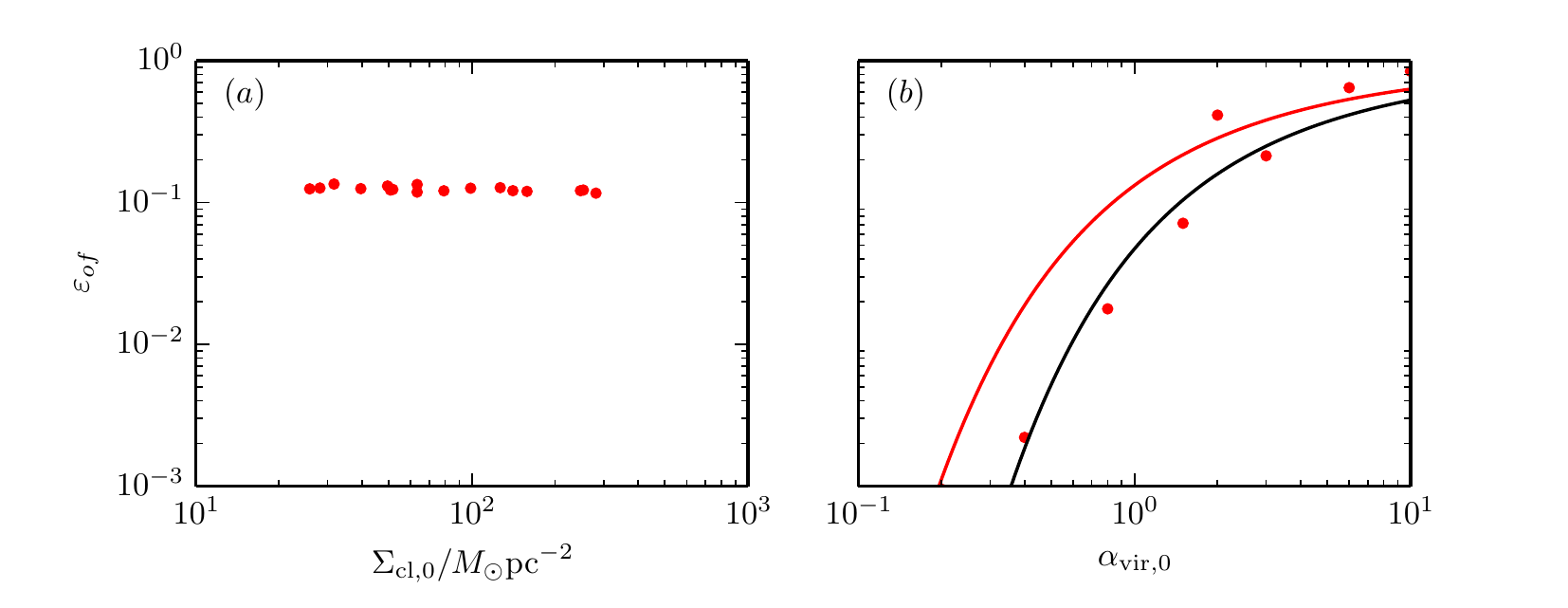}
  \caption{Outflow efficiency $\varepsilon_{\rm of}$ for the
    no-feedback versions (i.e., with radiation turned off) of the (a)
    $\Sigma$-series and (b) $\alpha$-series simulations. The outflow
    mass is calculated as the total mass flow out of the box over the
    4 free-fall times during which our simulations are run.  See text
    for explanation of black and red curves.}
  \label{Fig:Outflow}
\end{figure}

The most obvious initial effect of turbulence is the unbinding of small
fractions of mass with high velocities in the tail of the distribution.
We may calculate the total mass in outflows due to the initial turbulence alone by running
simulations with no radiative feedback until all of the gas in the
box is converted to stars. In these models, any gas that leaves the
cloud is unbound by the initial turbulence.

In Figure~\ref{Fig:Outflow}a, we show the
corresponding outflow mass fractions $\varepsilon_{\rm of}$ for our
no-feedback 
$\Sigma$-series models. The outflowing mass amounts to between $10$
and $15 \%$ of the initial cloud mass, with little variation between
low and high surface density. Without radiation, the primary
dimensionless parameter that varies for different values of $R$ and
$M$ at fixed $\alpha_{\rm vir,0} = 2.0$ is the Mach number $\mathcal{M}$. However,
there is little direct dependence of the outflow mass on $\mathcal{M}$
at fixed $\alpha_{\rm vir,0}$, since the escape velocity from the edge
of the cloud increases proportionally to the Mach number at fixed
virial parameter.

As shown in Figure~\ref{Fig:Outflow}b, 
the outflow mass depends strongly only on the initial virial parameter
$\alpha_{\rm vir,0} \propto \mathcal{M}^2 / (GM/r)$, i.e., the strength
of the turbulence relative to gravity.  Below $\alpha_{\rm vir,0} \sim
1$, there is effectively no outflow, as seen for the case of
$\alpha_{\rm vir,0} = 0.1$. Meanwhile, above $\alpha_{\rm vir,0}
\sim 5$, the outflow mass fraction approaches unity.

The dependence on $\alpha_{\rm vir,0}$ can be understood in terms of the fraction of gas
that escapes due to an initial turbulent velocity above the local escape velocity. 
For a uniform density cloud, the
escape velocity at radius $r$ is $v_e^2(r) = (GM/r_0)(3 - r^2 /
r_0^2)$, while in our models the initial velocity field is Gaussian
with dispersion $v_{RMS}^2 = 3 GM \alpha_{\rm vir,0} / (5r_0)$. Therefore, the
total  outflow mass fraction due to the initial turbulence
is just the fraction of mass, at any given radius, above $v_e(r)$:
\begin{equation}
	\varepsilon_{\rm of,init} = \int_0^1 dx~3 x^2 \left[1 - {\rm erf}\sqrt{\frac{5(3 - x^2)}{6 \alpha_{\rm vir}}}\right].
\label{Eq:ofinitpred}
\end{equation}
This quantity is indicated by the black curve in Figure~\ref{Fig:Outflow}b. For $\alpha_{\rm vir,0}
\le 3$,
Equation~(\ref{Eq:ofinitpred})
somewhat overestimates the total outflow mass, since the
initial turbulence is damped, and gas in the cloud interior, which may
have an initial velocity higher than the escape velocity, will
nevertheless collapse to form stars. However, it does represent a
reasonable estimate, since the majority of outflowing mass is on the
outer edges of the cloud, where the escape velocity is lowest.

Above $\alpha_{\rm vir,0} \sim 3$, we no longer predict the outflow mass
well since we cannot distinguish between gas that leaves the box and
gas that becomes truly unbound. Even accounting for an escape velocity
to $2r_0$ rather than infinity, $v_2(r)^2 = (GM/r_0)(2 - r^2 /
r_0^2)$, indicated by the red line in Figure~\ref{Fig:Outflow}b, we still
underestimate the outflow mass at large $\alpha_{\rm vir,0}$, since the
remaining gas mass is at too low density to form a significant mass in
stars, and so may drift outside the box while still bound. However,
without dramatically increasing the box size, with associated high
computational cost, this degeneracy is unavoidable. Therefore, at
values of $\alpha_{\rm vir,0} \gtrsim 3$,
Equation~(\ref{Eq:ofinitpred})
may underestimate
the net star formation efficiency if, in reality, some gas that is
expelled from our box were to ultimately recollapse.

\subsubsection{Surface Density PDF}
\label{SubSubSec:PDF}

\begin{figure}
  \centering
  \epsscale{1}
  \includegraphics{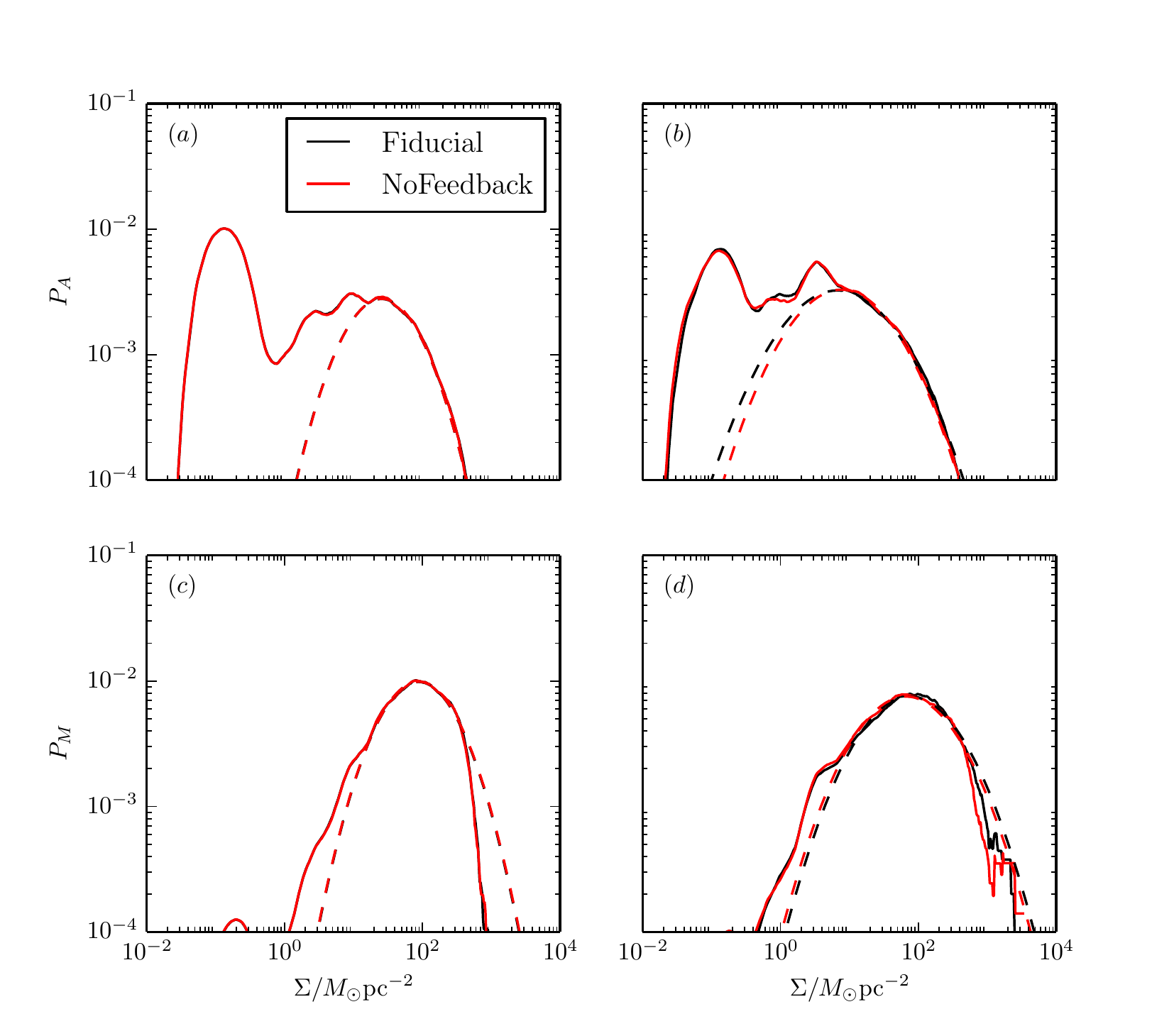}
  \caption{Surface density distributions in (top row) area and (bottom row)
    mass.  Times in (a) and (c) are at
    $t_{10} = 0.59 t_{\rm ff,0}$, and in (b) and (d) are  at 
    $t_{50} =1.06 t_{\rm ff,0}$.
    We show results for both the fiducial
    and no-feedback models. In each case, we show both the simulated
    surface density distributions (solid lines) as well as the
    best-fit lognormal curves (dashed lines).}
  \label{Fig:LogNormal}
\end{figure}

In addition to driving outflows, turbulence also has an effect on the
gas density structure. In particular, the non-uniformity of the cloud means that it has
significant variations in surface density. To quantify this, we
calculate the surface densities projected in the x-y, x-z, and y-z
planes, and we consider the probability distribution function over all
three.

For our fiducial
$\alpha_{\rm vir,0} = 2$ model, the resultant surface density distributions by
both area $P_A(\Sigma)$ and mass $P_M(\Sigma)$ are shown in
Figure~\ref{Fig:LogNormal}a,c at $t_{10}$. By this time, both $P_A$ and $P_M$ are roughly
log-normal in shape at high $\Sigma$, with similar variances of
$\sigma_{\rm ln \Sigma} \sim 1$. The area distribution has a long tail
at low surface density due to the fact that we are sampling over the
whole simulation volume rather than just the cloud volume. Meanwhile,
as there is little cumulative mass in these low density regions, the
mass distribution is much more obviously lognormal, consistent with
previous simulations of GMCs with supersonic turbulence
\citep[see e.g.][]{Ostriker2001, Vazquez-SemadeniGarcia2001,
  Federrath2010}.

We may therefore parameterize the ``cloud'' portion of 
both $P_A(\Sigma)$ and $P_M(\Sigma)$ as
lognormal distributions with mean $\mu_{(A/M)}$ and standard deviation
$\sigma_{\rm ln \Sigma}$:
\begin{equation}
	P_{(A/M)}(\Sigma)~d\ln\Sigma = \frac{1}{\sigma_{\rm ln \Sigma} \sqrt{2\pi}}
	{\rm exp}\left[-\frac{\left({\rm ln}\Sigma - \mu_{(A/M)}\right)^2}{2\sigma_{\rm ln \Sigma}^2}\right]~d\ln\Sigma.
\end{equation}
At any given time, the total mass in the system is conserved so that
$P_M(\Sigma) \propto \Sigma P_A(\Sigma) $. This implies
that both distributions must have the same standard deviation and
gives a relation between the mean surface density and the means
of the two lognormals:
\begin{equation}
	{\rm ln}\langle \Sigma \rangle_{\rm cloud} = \mu_A + \frac{1}{2}\sigma_{\rm ln \Sigma}^2
	 = \mu_M - \frac{1}{2}\sigma_{\rm ln \Sigma}^2 .
	\label{Eq:murel}
\end{equation}
Thus, half of the cloud's mass is at surface densities above/below 
$\langle \Sigma \rangle_{\rm cloud} \exp(\sigma_{\rm ln \Sigma}^2/2)$, and half of its
area is at surface 
densities above/below $\langle \Sigma \rangle_{\rm cloud} \exp(-\sigma_{\rm ln \Sigma}^2/2)$.
Here, we emphasize the difference between the mean cloud surface density 
  $\langle \Sigma \rangle_{\rm cloud} \sim M_{\rm cl} / A_{\rm cl}$, which can be
  obtained using Equation \ref{Eq:murel} after fitting lognormals to obtain
  $\mu_{(A/M)}$ and $\sigma_{\ln \Sigma}$, and the mean surface density over the whole simulation
  volume $\langle \Sigma \rangle_{\rm box} \sim M_{\rm cl} / A_{\rm box}$.

Comparing the fitted lognormal functional form to the PDFs in
Figures~\ref{Fig:LogNormal} we see that the lognormal approximation is
reasonable, particularly at the high-density end
for $P_A$ and around the peak for $P_M$.
The fits we show are only calculated between the $10^{\rm th}$ and $90^{\rm th}$ percentile
in mass, so as to avoid sampling the low surface density regions outside the
cloud. We also assume measurement errors proportional to $\sqrt{N}$ where
$N$ is the number of grid cells in each density bin.
At the time shown in Figures~\ref{Fig:LogNormal}c, the median
or peak surface density of the mass distribution for the fiducial
model is around $\Sigma \sim 100~{\rm M_{\odot}~pc^{-2}}$,
but part of the mass lies beyond even $\Sigma \sim 300~{\rm M_{\odot} ~pc^{-2}}$.
The peak of the distribution by area shown in Figures~\ref{Fig:LogNormal}a
is lower, $\Sigma \sim 25~{\rm M_{\odot}~pc^{-2}}$.
As we shall discuss below, the wide variation in $\Sigma$ implies that at
early times, radiation forces can be much more effective in some regions
than in others.  

The distribution is not just a broad
lognormal at the onset of star formation, but remains broad as it
progresses. For example, Figures~\ref{Fig:LogNormal}b,d show the area and mass PDFs at
$t_{50}=1.06  t_{\rm ff,0}$.
At this time, the best-fit parameters
are $\sigma_{\rm ln \Sigma} = 1.38$ (where this is taken as the
average of fits to the area and mass distributions), $\mu_A = 2.53$, and $\mu_M = 4.45$, so that the 
mass conservation relations are approximately followed and 
$\langle \Sigma \rangle_{\rm cloud} \sim 33~{\rm M_{\odot}~pc^{-2}}$. 

\begin{figure}
  \centering
  \epsscale{1}
  \includegraphics{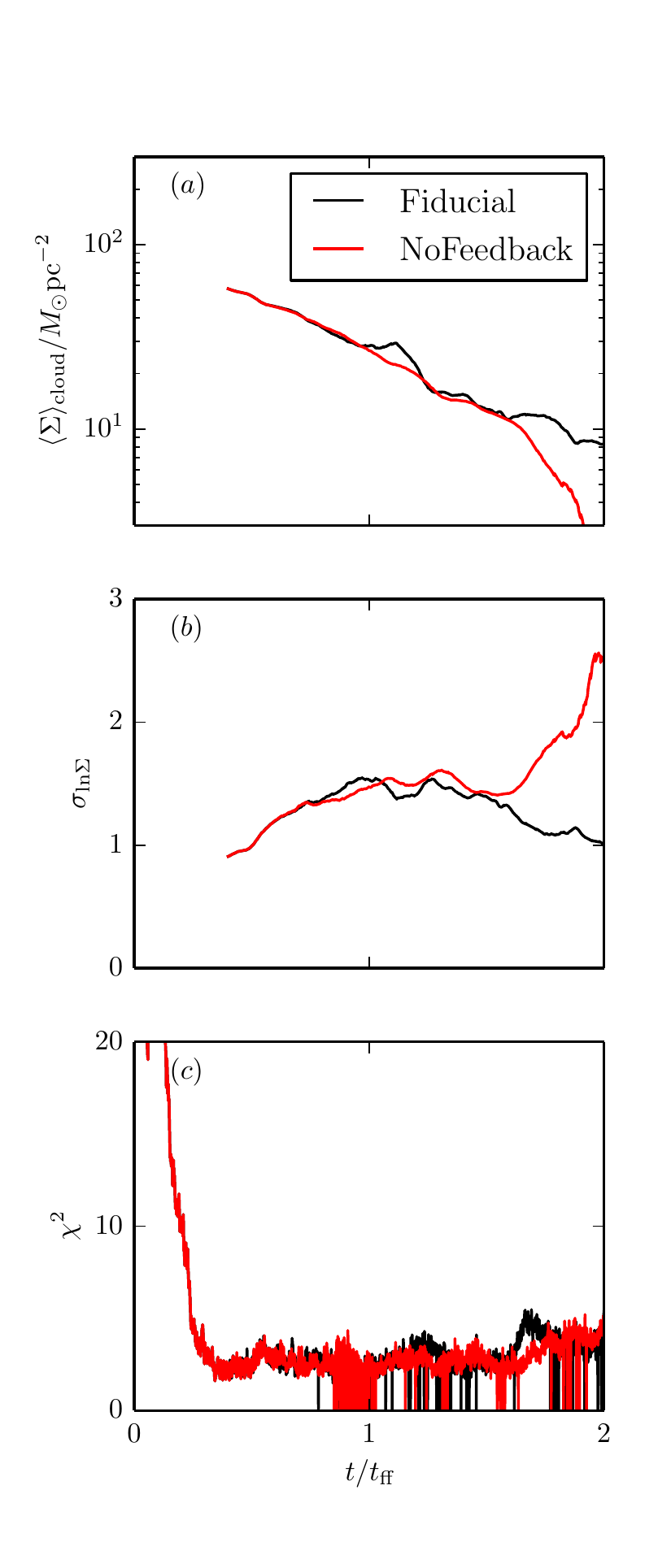}
 \caption{Best fit values as a function of time of lognormal fits to
   our fiducial (black) and no-feedback (red) models.  We show (a)
   the fitted mean surface density $\langle \Sigma\rangle_{\rm cloud}$,
   (b) the standard deviation to the mass distribution, and (c) the reduced $\chi^2$ of the 
  best fit.}
  \label{Fig:LogNormalFits}
\end{figure}

\begin{figure*}
  \centering
  \epsscale{1}
  \includegraphics{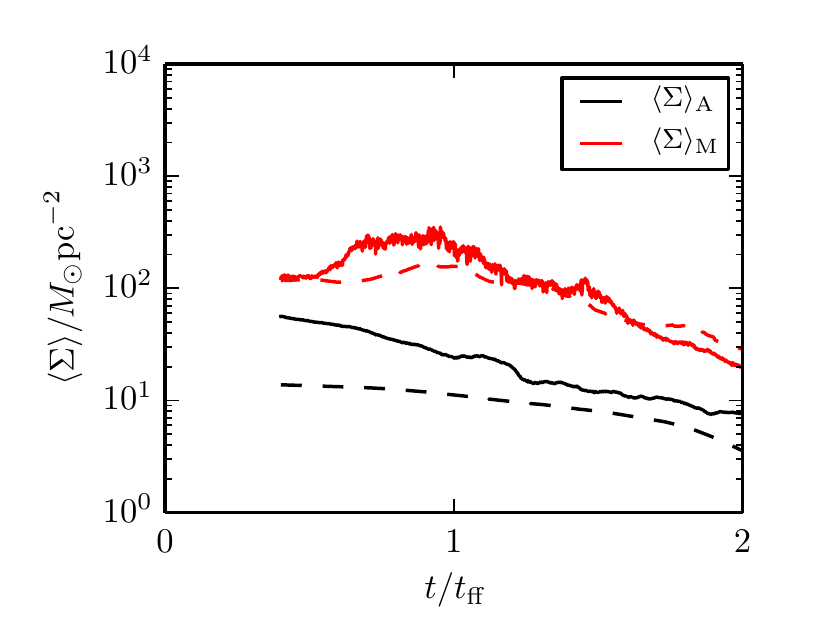}
 \caption{ Mean surface density as a function of time for our fiducial simulation. We show both the 
 area-weighted mean $\langle \Sigma \rangle_{\rm A}$ (black) as well as the mass-weighted mean 
 $\langle \Sigma \rangle_{\rm M}$ (red). For comparison, we show both the values 
 calculated directly by averaging over the whole simulation box (dashed) as well as the values 
 for the cloud alone found by fitting a lognormal distribution to the surface
 density (solid).}
  \label{Fig:SDMean}
\end{figure*}

In Figures~\ref{Fig:LogNormalFits}a,b, we show the best fit
mean and standard deviation parameters as a function of time for
lognormal fits to the surface density distribution of our fiducial
cloud.
A drop in reduced  $\chi^2$ values 
in Figure~\ref{Fig:LogNormalFits}c only really develops after $\sim 0.4 t_{\rm ff,0}$ indicating 
that it is only by this time that turbulence has erased the initial conditions. This is 
expected, since the cloud is initially uniform and the development of density structure has 
a timescale set by the turbulent crossing time 
$t_{\rm cross}=r_0/v_{\rm rms} = 1.2 t_{\rm ff,0}$.  
While the surface density distribution gradually broadens over the first $t_{\rm ff,0}$, 
after this time it effectively reaches a steady state in both the feedback and no-feedback models, 
with $\sigma_{\rm ln \Sigma} \sim 1-1.5$. Even though the surface density distributions 
are visibly different at $t = 1.06 t_{\rm ff,0}$ for the feedback 
(Figure~\ref{Fig:FiducialEvolutionColZ}) and 
no-feedback (Figure~\ref{Fig:FiducialNFEvolutionColZ}) models, they are statistically similar, since it 
is only the lower column density regions that are affected by radiative feedback. This holds true until 
around $1.5 t_{\rm ff,0}$, by which time the majority of stars have formed and the majority of remaining gas in the fiducial model is outflowing. As gas is accreted by star particles or is driven from the cloud, the 
mean and peak of the distribution naturally shift downward. However, the
width of the distribution remains similar throughout star formation.

As noted above, we cannot obtain $\langle \Sigma \rangle_{\rm
    cloud}$ directly from our simulations, since our simulation volume
  comprises both the ``cloud'' and the surrounding empty region (or larger
  scale lower-density ISM in a real system).  However, if we assume that the
  ``cloud'' portion of the distribution is a log-normal, then we can
  fit to the PDFs and use Equation \ref{Eq:murel} to obtain $\langle
  \Sigma\rangle_{\rm cloud}$.  Our fits to obtain $\mu_A$, $\mu_M$, and
  $\sigma_{{\rm ln}\Sigma}$ extend from the $10^{\rm th}$ to the
  $90^{\rm th}$ percentile by mass, which effectively excises the low
  surface density regions external to the cloud.

  We can define the mass-weighted mean surface density over the whole
  box as $\langle \Sigma\rangle_{\rm M, box} = \langle \Sigma^2\rangle_{\rm box} /
  \langle \Sigma\rangle_{\rm box}$, where $\langle \Sigma\rangle_{\rm box}$ is
  the area-weighted mean.  Under the assumption that the ``cloud''
  portion is a lognormal,
  $\langle \Sigma\rangle_{\rm M, cloud} =
  \langle \Sigma\rangle_{\rm cloud} \exp(\sigma_{\ln\Sigma}^2)$, where 
  $\langle \Sigma\rangle_{\rm A, cloud} \equiv   \langle \Sigma\rangle_{\rm cloud}$.   
  Figure~\ref{Fig:SDMean} shows, for the fiducial model, the evolution of the
  area- and mass-weighted surface densities computed in two different ways:
  taking direct averages over the box, and using the fitted lognormals to identify
  just the ``cloud'' material.  

  From Figure~\ref{Fig:SDMean},  $\langle \Sigma \rangle_{\rm
    box}$ starts a factor $\sim 5$ below $\langle \Sigma \rangle_{\rm
    cloud}$, because the projected surface area of the cloud is $\sim
  1/5$ of the box surface area.  Over time, as the cloud
  disperses and fills more of the simulation area and volume, these measures
  become more similar.  In contrast, $\langle \Sigma \rangle_{\rm M, 
    box}$ and $\langle \Sigma \rangle_{\rm M, cloud}$ are much closer to each
  other over the whole evolution, and are each an order of magnitude larger
  than $\langle \Sigma \rangle_{\rm box}$.  
We conclude that both the definitions 
$\langle \Sigma \rangle_{\rm A,cloud} \equiv {\rm exp}(\mu_M - (1/2)\sigma_{\rm ln \Sigma}^2)$ and 
$\langle \Sigma \rangle_{\rm M,cloud} \equiv {\rm exp}(\mu_M + (1/2)\sigma_{\rm ln \Sigma}^2)$ 
well represent the area- and mass-weighted mean values for the cloud material,
and that the latter is also similar to the mass-weighted mean over the
whole box.

\begin{figure*}
  \centering
  \epsscale{1}
 \includegraphics{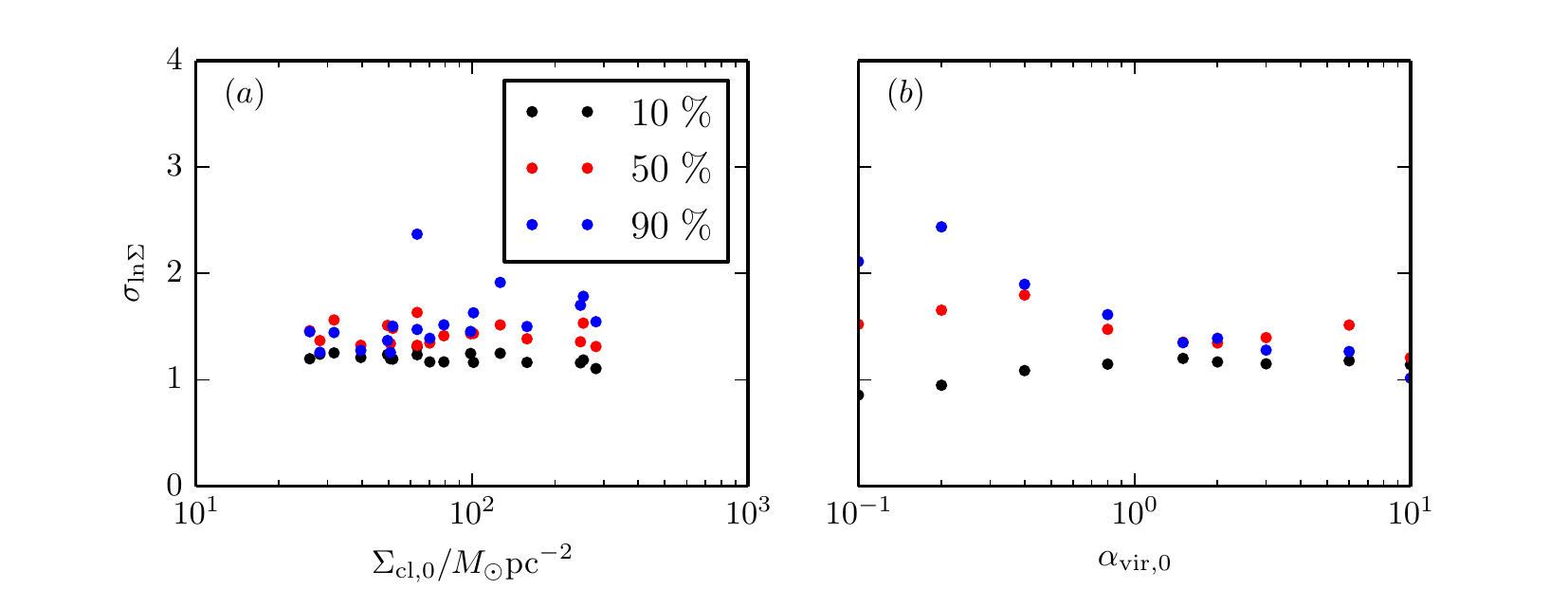}
  \caption{Best fit lognormal standard deviation of ${\rm ln \Sigma}$
    for the (a) $\Sigma$-series models and (b) $\alpha$-series 
    models. We show the best fits at three different times: $t_{10}$, $t_{50}$ and $t_{90}$.}
  \label{Fig:LogNormalAlpha}
\end{figure*}

We note also that there is no significant change in the width of the
surface density distribution with differing  initial cloud mass and
radius. For a number of different models in our $\Sigma$-series, we
compute best-fit values of $\sigma_{\rm ln \Sigma}$ at times $t_{10}, t_{50},$ and 
$t_{90}$, with the results shown in Figure~\ref{Fig:LogNormalAlpha}a.
In most models, there appears to be a modest increase in the lognormal
width with time from the width at $t_{10}$, since this occurs at around
$\sim 0.6 t_{\rm ff,0}$, and as discussed earlier, the lognormal has not
quite reached a steady state by this stage. However, after
$t \sim 1 - 1.2 t_{\rm ff,0}$, the distribution widths remain roughly
constant for the remainder of star formation.  
Moreover, the distribution width is relatively independent of cloud
surface density at $\sigma_{\rm ln \Sigma} \sim 1.3 - 1.5$ (slightly
decreasing towards the low-$\Sigma$ end, which has lower Mach number).

This makes it clear why the simple hypothesis that
stars will form until $\langle \Sigma^c \rangle_{\rm cloud} = \Sigma_E$ is seriously
flawed: even the peak by mass of the surface density,
$\langle \Sigma\rangle_{\rm M, cloud}$, is a factor of 
${\rm exp}(\sigma_{\rm ln \Sigma}^2 / 2) \sim 3.1$ higher than
$\langle \Sigma \rangle_{\rm cloud}$. Thus, half of the gas is at surface density 
more than three times $\langle \Sigma \rangle_{\rm cloud}$, and a large fraction
is in regions at even higher surface density. Forcing these
high-$\Sigma$ regions out of the cloud would demand $\Sigma_E \gg
\langle \Sigma \rangle_{\rm cloud}$, requiring a higher luminosity and
hence (from Equation~\ref{Eq:FinalSigma})
larger $\varepsilon$ than predicted by
Equation~(\ref{Eq:FinalEfficiency}).

In the above, we have analyzed the distribution of surface densities
as would be seen by an external observer.  However, for the purposes
of gauging the effects of radiation forces from a cluster on the
surrounding gas in a cloud, what matters is actually the {\it circumcluster}
distribution of densities.  As we shall show in a separate publication
(\citealt{RaskuttiOstriker2016b}, in preparation [Paper II]), this
distribution is in fact quite similar to the lognormal PDFs shown and 
discussed above. 
For example, for the fiducial model, we find 
$\langle \Sigma^c \rangle_{\rm cloud} \sim 12~M_\odot~{\rm pc^{-2}}$ and 
$\sigma_{\ln \Sigma}=1.42$ at $t = 1.06~t_{\rm ff,0}$, 
which can be compared to the
variance and mean values seen by an external observer
$\sigma_{\ln \Sigma}=1.38$ and
$\langle \Sigma \rangle_{\rm cloud} \sim 33~M_\odot~{\rm pc^{-2}}$
(see also 
Figure~\ref{Fig:LogNormalFits} and \ref{Fig:SDMean}).

The mean value of the cloud surface density at any time is related
to its initial mean surface density via the efficiency and a radial contraction factor $x$ (cf. 
Equation \ref{Eq:MeanSurfaceDensity}), albeit adjusted for the initial turbulent outflows
\begin{equation}
	\langle \Sigma^{c} \rangle = \frac{(1 - \varepsilon)}{4x^2}\Sigma_{\rm adj}
	\label{Eq:MeanAdjustedSurfaceDensity}
\end{equation}
where $\Sigma_{\rm adj} = \Sigma_{\rm cl,0} (1.0 - \varepsilon_{\rm of,init})$. 
In addition, fitting a lognormal form gives a value of $\langle \Sigma \rangle_{\rm cloud}$ from
Equation \ref{Eq:murel}.  Putting these relations together yields $x$.
We shall show in Paper II that for models
with $\alpha_{\rm vir,0}\sim 1-2$, this yields $x\approx 1$.
That is, the overall size of clouds does not vary much over the star-forming period. 

\subsubsection{Dependence of the PDF on Seed Field}
\label{SubSubSec:PDF Seed}

\begin{figure*}
  \centering
  \epsscale{1}
  \includegraphics{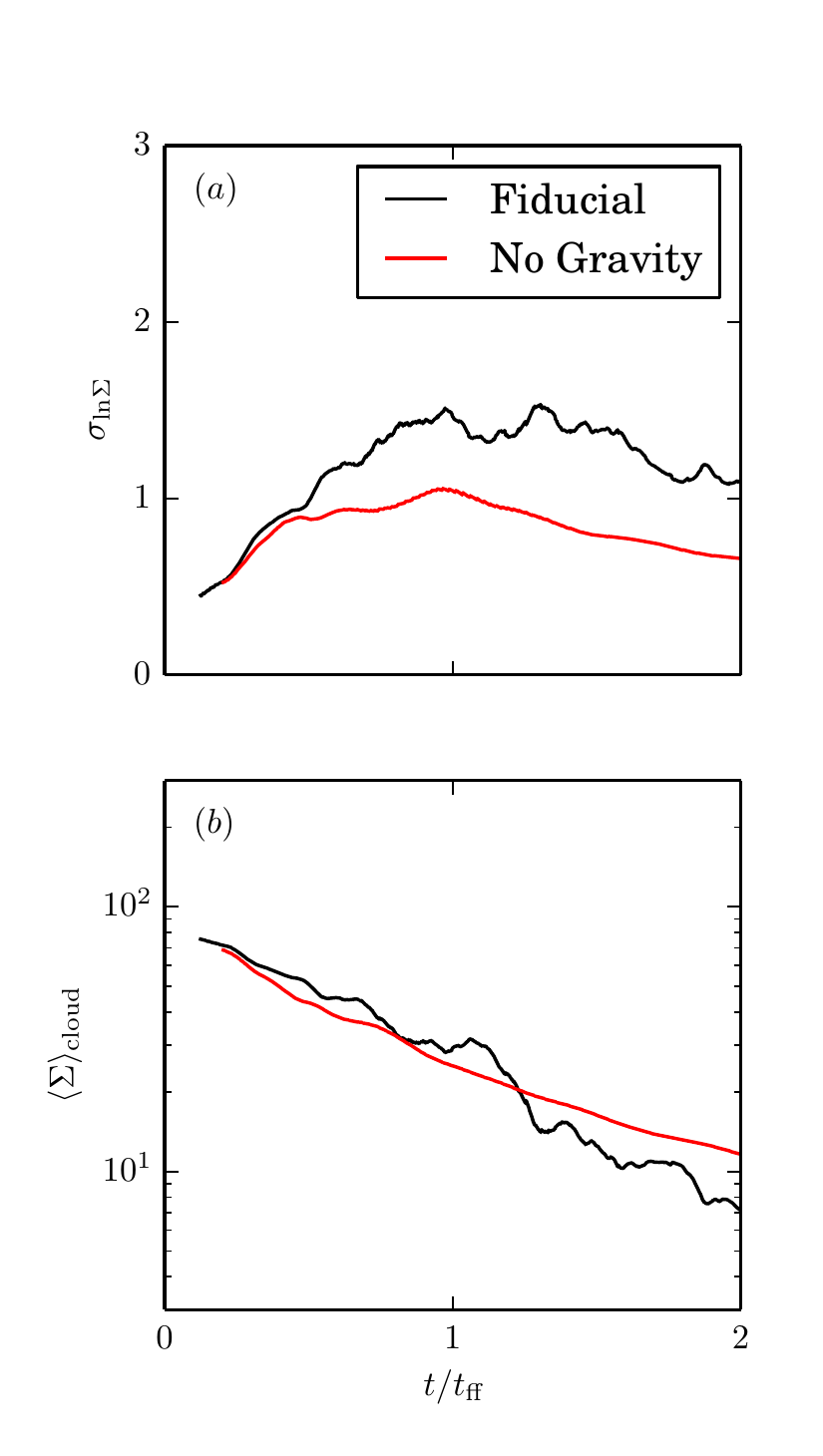} 
  \caption{ Time evolution of the best fit lognormal mean (bottom) and standard deviation (top) to
    the surface density distribution in our fiducial (black) and no-gravity (red) models.}
  \label{Fig:NoSGDensity}
\end{figure*}

Before we turn to how the surface density field affects the final
stellar efficiency, we briefly consider what factors are important in
determining its shape. To do this, we run a variation of the fiducial
model in which we initially allow our cloud -- with the same seed
turbulent velocity field -- to evolve without self-gravity. Without star
formation and feedback, the gas flows steadily out from the simulation
volume due to the initial turbulence, while the Mach number decreases
as the turbulence decays and the highest velocity material leaves the
box.

However, after half a freefall time, when stars are beginning to form
in the fiducial model, the density distribution has reached a rough
steady state (as shown from the variance of the column density PDF in
Figure~\ref{Fig:NoSGDensity}).  From this point forward, the column density
distribution remains roughly lognormal, with a fairly steady, though
slowly declining width, and a steadily declining mean as the gas flows
out of the box. It is also relatively similar to the fiducial model
with gravity, suggesting that the initial density field is set almost
entirely by the initial turbulence, with gravity only being important
on the smallest scales (or highest densities).  The only differences
are a reduced width, due to the absence of self-gravity, which
flattens the distribution at the high density end, and a reduced mean
density due to the increased gas outflows. At the onset of star
formation, therefore, the bulk of gas mass (i.e., up to the highest
densities) has a lognormal shape set almost entirely by the initial
turbulent field.

Figure \ref{Fig:LogNormalAlpha}b shows the dependence of $\sigma_{\rm ln \Sigma}$ 
on the initial virial parameter. Before the majority of star formation
begins at $t_{10}$, the 
lower $\alpha_{\rm vir,0}$ models have slightly narrower distributions 
(lower $\sigma_{\rm ln \Sigma}$) due to their lower Mach number. However,
this trend is reversed over time, with the lower
virial parameter models tending to have much broader distributions by the end of
star formation. Largely this is because these clouds have very high
efficiencies, so that only a small fraction of the cloud mass remains
once $90 \%$ of stars have formed, hence the lognormal fit is
considerably worse.  

\begin{figure*}
  \centering
  \epsscale{1}
 \includegraphics{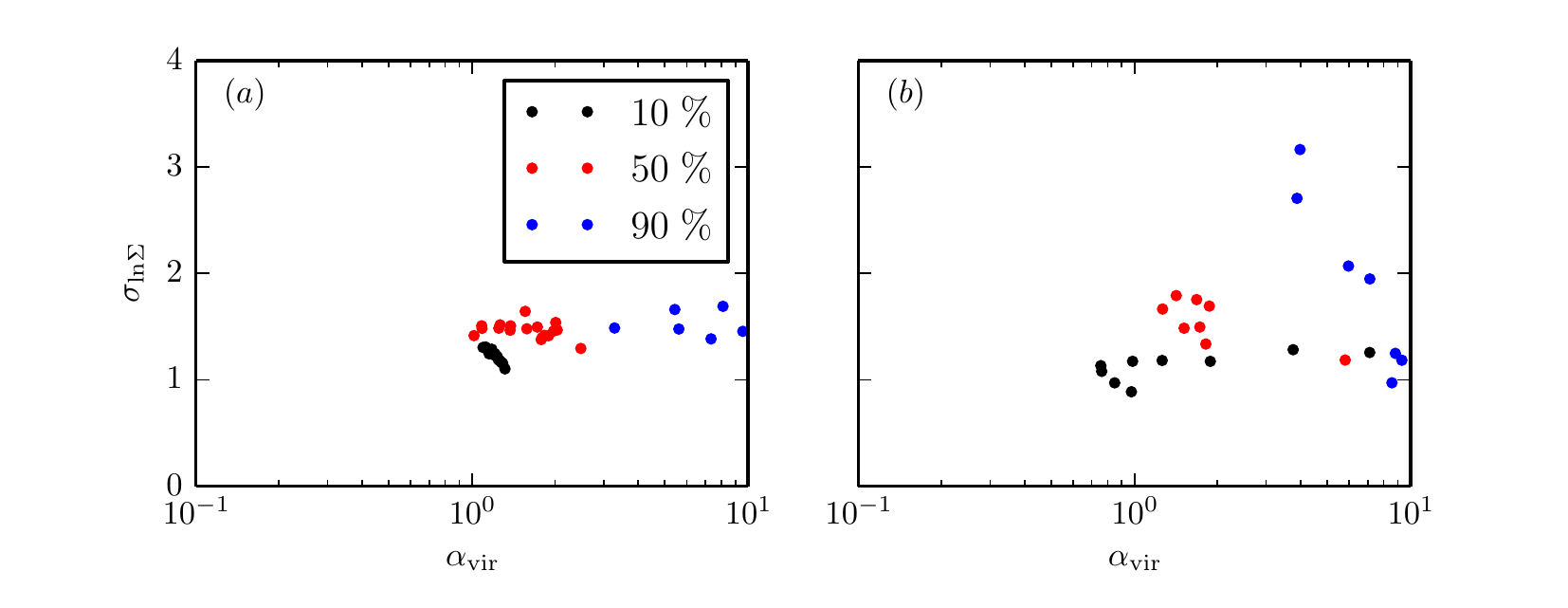}
  \caption{Best fit lognormal standard deviation of ${\rm ln \Sigma}$
    for (a) $\Sigma$-series models and (b) $\alpha$-series 
    models, this time plotted against the instantaneous gas virial parameter in each simulation.}
  \label{Fig:LogNormalAlphaDyn}
\end{figure*}

It must also be kept in mind that $\alpha_{\rm vir}$ changes in time for most of the $\alpha$ series. At 
high initial virial parameter, $\alpha_{\rm vir,0} \gtrsim 5$, the initial
turbulence rapidly drives the highest velocity regions from the box, so
that the virial parameter rapidly decays. Similarly, at low initial
$\alpha_{\rm vir,0}$, including the reduced-turbulence models considered
in Section~\ref{SubSec:SFERad}, the cloud contracts until the virial
parameter is close to unity in all cases. This is evident if we
consider the variation of  $\sigma_{\rm ln \Sigma}$ with instantaneous
rather than initial virial parameter, shown in Figure
\ref{Fig:LogNormalAlphaDyn}b. Clouds in the range $\alpha_{\rm vir,0}
= 0.1 - 3.0$ all converge to a much smaller range of $\alpha_{\rm
  vir} = 0.5 - 1.0$ by the time star formation begins.  Then, as
radiative feedback becomes important, $\sigma_{\rm ln \Sigma}$ increases slightly with 
increasing virial parameter.

\subsection{Final Efficiencies}
\label{SubSec:SFEFinal}

\begin{figure}
  \centering
  \epsscale{1}
  \includegraphics{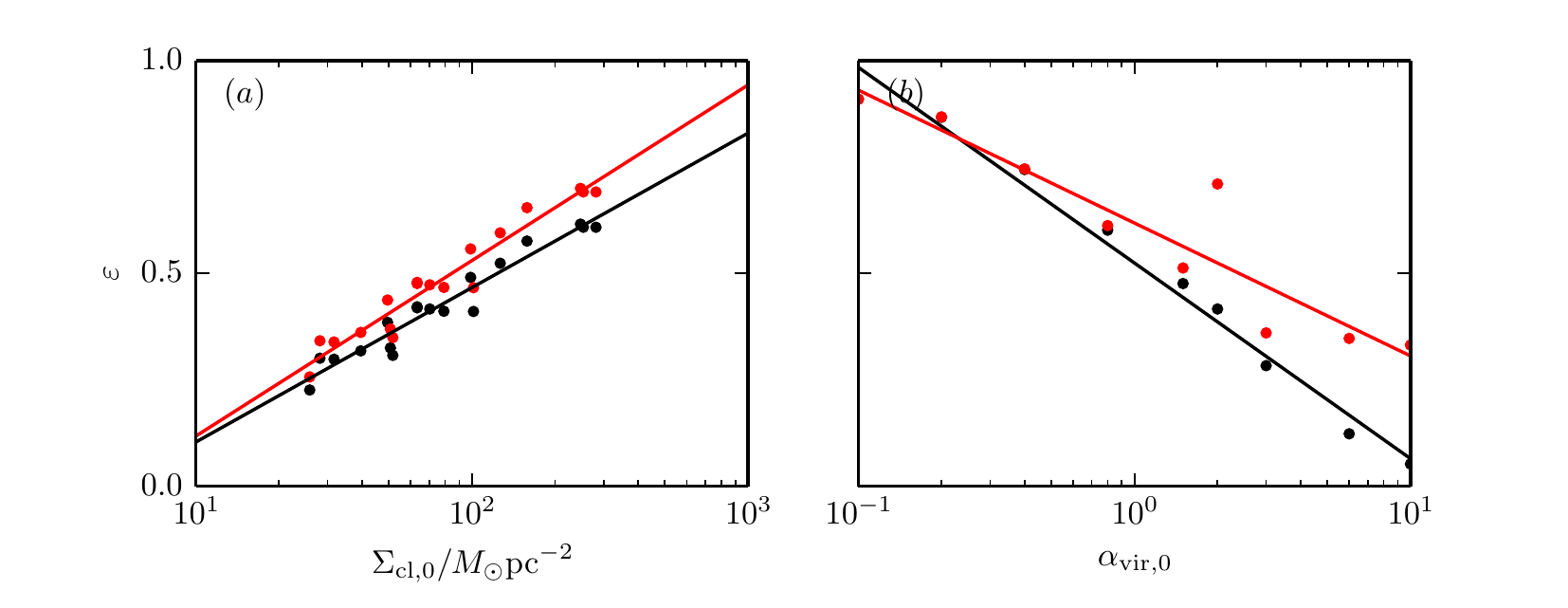}
\caption{Final star formation efficiency $\varepsilon$ 
for (a) $\Sigma$-series simulations, and (b) $\alpha$-series simulations. 
We show both star formation efficiencies normalized
    to the initial cloud mass (black circles) and normalized to the
    cloud mass accounting for initial turbulence-driven outflows (red
    circles). In all cases we show best-fit logarithmic relations as
    solid lines.}
  \label{Fig:Efficiencies}
\end{figure}

In this subsection, we present and analyze results for the full set of 
turbulent cloud collapse models with radiation feedback. 
In Figure~\ref{Fig:Efficiencies} we show the final star
formation efficiencies for our $\Sigma$- and $\alpha$-series cloud
models. We have normalized by both the initial cloud mass and the
cloud mass corrected for turbulence driven outflows.  Interestingly,
the efficiency appears to show a logarithmic dependence on $\Sigma$
across almost two orders of magnitude of variation in the initial surface
density. In fact, Figure~\ref{Fig:Efficiencies} shows a
remarkably good fit to the relation  
$\varepsilon = 0.37{\rm log}\Sigma - 0.26$ (black line). 
This is true even if we account for the mass loss due to
turbulent outflows from the simulation box by setting
$\varepsilon \rightarrow \varepsilon_{\rm adj}
\equiv \varepsilon / (1.0 - \varepsilon_{\rm of,init})$. The
normalization of the relation changes slightly, but we 
still obtain a logarithmic relationship between surface density and final
efficiency, given by $\varepsilon = 0.41{\rm log}\Sigma - 0.26$ (red line).

We can also fit to a logarithmic dependence on the virial parameter,
with $\varepsilon = -0.45{\rm log}\alpha + 0.51$, although in this
case it is less clear whether this is significant, as there are
systematic errors at both low and high virial parameter. Moreover, our
estimate of the net efficiency is uncertain at high virial parameter,
since we overestimate turbulent outflows and correspondingly
underestimate star formation.

\begin{figure}
  \centering
  \epsscale{1}
  \includegraphics{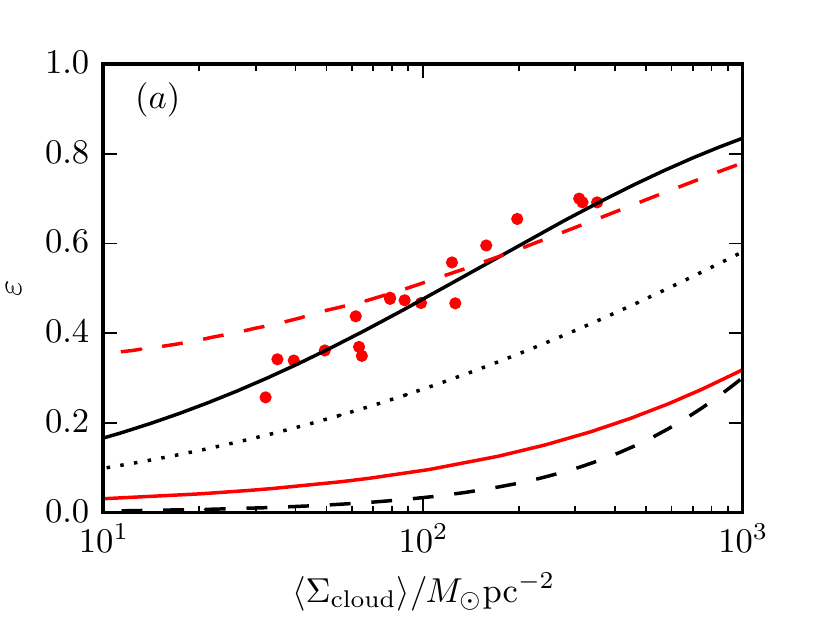}
  \caption{Final star formation efficiency $\varepsilon$ as a function
    of surface density for $\Sigma$-series simulations. Points show
    star formation efficiencies ($\varepsilon_{\rm adj}$)
    normalized to the cloud mass
    accounting for initial turbulence-driven outflows (red
    circles). The black solid line shows the prediction of 
      $\varepsilon_{\rm max}$ and the dotted line shows
      $\varepsilon_{\rm min}$, from 
    Equation~(\ref{Eq:LognormalEff}) with
    $x=0.84$ and $\sigma_{\rm ln  \Sigma} = 1.42$
    taken from the circumcluster surface density
    distribution. For comparison, the simple model of Equation
    \ref{Eq:FinalEfficiency} with $x=1$ is shown (black dashed line).  
    We also show the predictions from the model of \cite{ThompsonKrumholz2014}, with their 
    fiducial parameter values (red solid), and with values of $\varepsilon_{\rm ff}=0.44$ and 
    $\sigma_{\rm ln \Sigma} = 1.42$ based on the results of our simulations (red dashed).}
  \label{Fig:ModelEfficiencies}
\end{figure}

Although the logarithmic form fits the $\Sigma$ series well, it is quite 
different from simple predictions. 
For example, we can compare the numerical results 
to the uniform shell force balance model
(with $x = 1$) of Equation~(\ref{Eq:FinalEfficiency}), as shown with the
red solid line in
Figure~\ref{Fig:ModelEfficiencies}. Evidently, both the shape and 
magnitude of the curve
from Equation~(\ref{Eq:FinalEfficiency}) compare poorly with the simulation 
results. 

As discussed earlier, the key difference between the simulated clouds
and the simple Eddington limit prediction of
Equation~(\ref{Eq:FinalEfficiency}) is that the cloud does not have
uniform surface density. Stars do not keep forming inside a uniform shell 
and then instantaneously drive that shell
away once the efficiency $\varepsilon$ and
luminosity are large enough for $\Sigma_E$ to match
$\langle \Sigma^c \rangle_{\rm cloud}$. Instead, as will be discussed in more 
detail in \cite{RaskuttiOstriker2016b}, early star formation begins to drive
away the lowest surface density regions (those that have have 
$\Sigma^c < \Sigma_E$). Stars continue forming from the remaining mass until the
radiative force is sufficient to drive away much higher surface
density gas (as, from Equation~\ref{Eq:FinalSigma}, an increase in
$\varepsilon$ raises $\Sigma_E$). For example, in our fiducial model, close to the end of star
formation at $t = 1.45 t_{\rm ff,0}$ when $\varepsilon \sim 0.35$, the
mean surface density is $\langle \Sigma \rangle_{\rm cloud} \sim 12~M_{\odot}~{\rm
  pc^{-2}}$, with the circumcluster surface density around a factor of $2$ or $3$ lower again, 
  while the Eddington surface density in
Equation~(\ref{Eq:FinalSigma}) is $\Sigma_E \sim 200~M_{\odot}~{\rm
  pc^{-2}}$, implying that enough stars have formed to drive away gas
at close to 50 times the mean surface density of the
cloud.
Figure~\ref{Fig:LogNormalFits} shows that at this time
$\sigma_{\rm ln \Sigma} = 1.49$ which yields $\mu_M = 2.9$ such that
${\rm ln}\Sigma_E = 5.3$ is roughly 1.5-$\sigma$ above the peak surface
density. This would suggest that star formation is halted when enough
stars form to drive away not the mean surface density, but instead
something closer to the $90^{\rm th}$ percentile of surface density.

Recently, \cite{ThompsonKrumholz2014} have argued that understanding how
radiative feedback limits star formation in GMCs requires an
accounting of the full surface density distribution set by
turbulence. They propose a model in which the instantaneous mass loss
rate from a cloud is set by the cloud's freefall time and fraction of
mass in a lognormal PDF at surface densities below a critical
value. At the same time, they assume the stellar mass and luminosity
increases as $\dot{M_{*}} = \varepsilon_{\rm ff} M_{\rm gas}/t_{\rm ff}$.
Here, we develop a related model by considering the
mass eligible to be expelled from the cloud at any time, for a given
  lognormal PDF and star formation efficiency.

From Equation~(\ref{Eq:FinalSigma}), surface densities above the
Eddington surface density $\Sigma_E$ have a net inward force and can
continue to either collapse and form stars, or be accreted on to the growing 
star clusters. Correspondingly, surface densities below 
$\Sigma_E$ have a net outward force, and we assume here that such 
regions become instantaneously unbound without further mixing with other gas.
Although this is clearly an oversimplification, we have found
  \citep[see][for details]{RaskuttiOstriker2016b} that the distribution
  of outflowing velocities is consistent with essentially ballistic
  outflow of super-Eddington structures.

If the cloud has a circumcluster surface
density distribution $P_M(\Sigma^c)$ by mass, the
fraction of mass eligible to be expelled will be
$\int_{-\infty}^{\ln\Sigma_E}P_M(\Sigma^c) d\ln\Sigma^c$.
If the stellar mass in
the cloud at a given instant during its evolution is $M_{*} =
\varepsilon M_{\rm cl,0}$, then the gas mass remaining is
$(1-\varepsilon) M_{\rm cl,0}$, and the 
fraction $\varepsilon_{\rm of}$ of the original gas that is eligible
  for outflow is:
\begin{equation}
	\varepsilon_{\rm of} = (1 - \varepsilon)
          \int_{-\infty}^{\ln\Sigma_E} P_M(\Sigma^c) d\ln\Sigma^c.
	\label{Eq:Mstarmax}
\end{equation}

In Section~\ref{SubSubSec:PDF}, we showed that the column 
density distribution is well approximated by a lognormal
distribution, and here, we extend that approximation to the 
circumcluster surface density
\citep[see][for details]{RaskuttiOstriker2016b}.
In this case, Equation~(\ref{Eq:Mstarmax}) may be
evaluated as
\begin{equation}
	\varepsilon_{\rm of} = \frac{1}{2}\left(1 - \varepsilon \right)\left(1 + {\rm erf}\left(y_E\right)\right),
	\label{Eq:EpsOF}
\end{equation}
where
\begin{equation}
y_E \equiv \frac{{\rm ln}\Sigma_E - \mu_M}
	{\sqrt{2}\sigma_{\rm ln \Sigma}}.
\label{Eq:yE}
\end{equation}
The quantity $\mu_M = \ln \langle \Sigma^c\rangle + \sigma_{\ln \Sigma}^2/2$
(see Equation~\ref{Eq:murel}) 
is the mean of the mass distribution of ln$\Sigma^c$ and $\sigma_{\ln \Sigma}$
is the variance.  Using  Equation~(\ref{Eq:MeanSurfaceDensity}),
\begin{equation}
  \mu_M = {\rm ln}\left[\Sigma_{\rm cl,0}(1 - \varepsilon)\right] + \frac{\sigma_{\rm ln \Sigma}^2}{2} - {\rm ln}(4x^2)
\end{equation}
if the circumcluster gas is concentrated in a thin shell.  
Substituting into Equation~(\ref{Eq:yE}) and using
Equation~(\ref{Eq:FinalSigma}), we obtain: 
\begin{eqnarray}
	y_E &=& \frac{1}{\sqrt{2}\sigma_{\rm ln \Sigma}}{\rm ln}\left(\frac{4 \Sigma_E x^2}
	{\Sigma_{\rm cl, 0}(1 - \varepsilon)}\right) - 
	\frac{\sigma_{\rm ln \Sigma}}{\sqrt{8}} \nonumber \\
	     &=& \frac{1}{\sqrt{2}\sigma_{\rm ln \Sigma}}\left[
	{\rm ln}\left[\frac{2\Psi}{\pi c G \Sigma_{\rm cl, 0}}\right]
	+ 2 {\rm ln}x - \frac{\sigma_{\rm ln \Sigma}^2}{2} + 
	{\rm ln}\frac{\varepsilon}{1 - \varepsilon^2}\right].
	\label{Eq:xbfinal}
\end{eqnarray}
With Equation~(\ref{Eq:xbfinal}), 
Equation~(\ref{Eq:EpsOF}) gives the unbound or outflowing fraction
$\varepsilon_{\rm of}$ of a cloud with initial
surface density $\Sigma_{\rm cl,0} \equiv M_{\rm cl,0}/(\pi r_{\rm cl,0}^2)$
in terms of the current efficiency $\varepsilon$ and the two free 
parameters $x$ and $\sigma_{\rm ln \Sigma}$.

\begin{figure}
  \centering
  \epsscale{1}
  \includegraphics{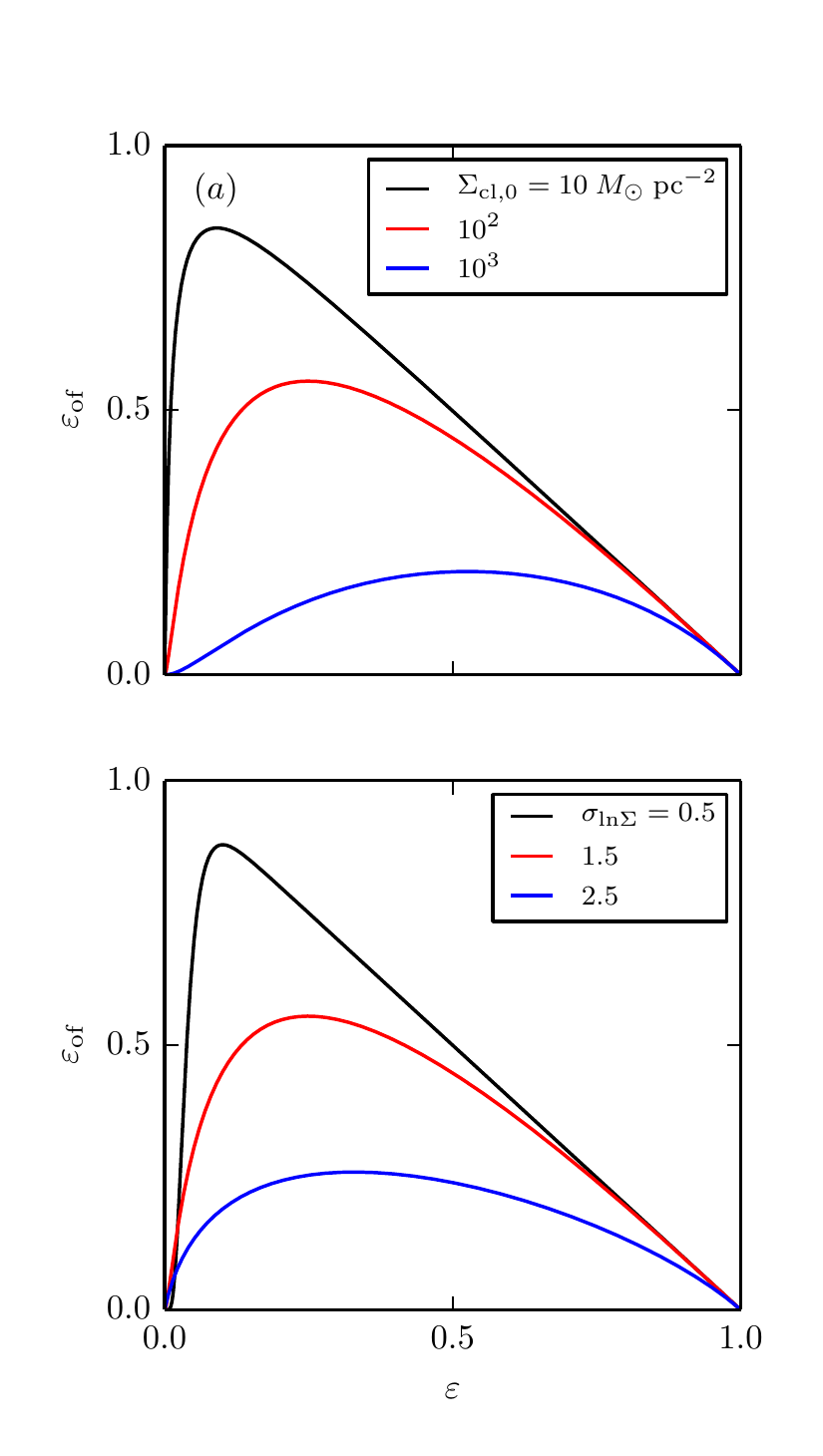}  
  \caption{Maximum possible stellar efficiency $\varepsilon_{\rm max}$
    (Equation~\ref{Eq:EpsOF}) as a function of present stellar
    efficiency $\varepsilon$ for (a) clouds of varying initial surface
    density $\Sigma_{\rm cl, 0}$, and (b) varying lognormal surface density
    distribution width $\sigma_{\rm ln \Sigma}$. For both
    cases we use $x = 1$ in Equation~(\ref{Eq:xbfinal}).  In (a), 
    we set $\sigma_{\rm ln \Sigma} = 1.5$, and the key shows
    $\Sigma_{\rm cl, 0}$ in units of $M_{\odot}~{\rm pc^{-2}}$. For
    (b), we set $\Sigma_{\rm cl, 0} = 100~M_{\odot}~{\rm
      pc^{-2}}$ and the key shows values of $\sigma_{\rm ln \Sigma}$.}
  \label{Fig:EffEvolution}
\end{figure}

\begin{figure}
  \centering
  \epsscale{1}
  \includegraphics{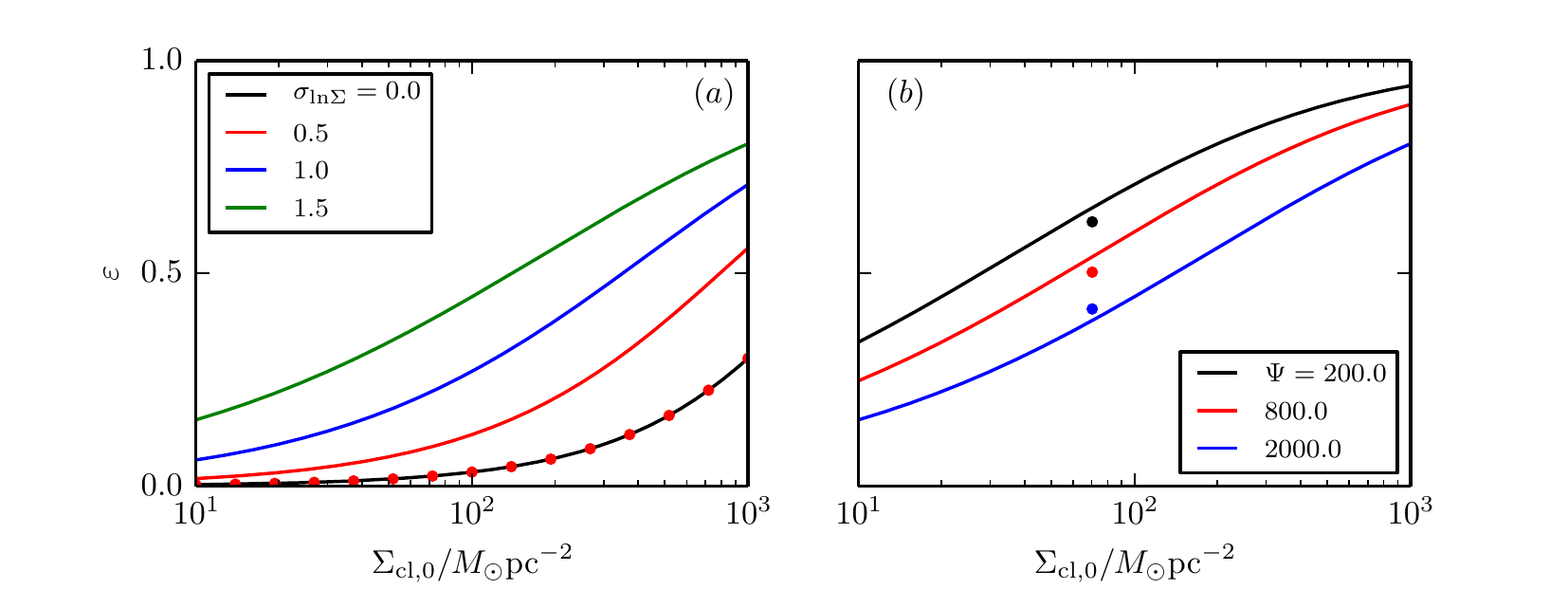}

  \caption{
    Maximum
    stellar efficiency $\varepsilon$ predicted by 
    Equation~(\ref{Eq:LognormalEff}a) as a function of initial surface
    density for clouds with varying lognormal distribution width
    $\sigma_{\rm ln \Sigma}$ (left) and varying $\Psi$ (right). In all cases, we use $x = 1$ and the keys
    shows appropriate values. For comparison, on the left, we also
    show values for the simple model of Equation
    (\ref{Eq:FinalEfficiency}) (red circles). As expected,
    Equation~(\ref{Eq:LognormalEff}a) converges to 
    Equation~(\ref{Eq:FinalEfficiency}) for small $\sigma_{\rm ln \Sigma}$, since
    this is the case of a uniform shell. Meanwhile, for varying $\Psi$ and 
    fixed width $\sigma_{\rm ln \Sigma} = 1.5$, 
    we also show values of the final efficiency 
    taken from simulations with varying $\Psi$ (circles).}
  \label{Fig:EffVarSigma}
\end{figure}

Figure~\ref{Fig:EffEvolution} shows the behaviour of $\varepsilon_{\rm
  of}$ as a function of input stellar efficiency $\varepsilon$ for
varying $\Sigma_{\rm cl, 0}$ and $\sigma_{\rm ln \Sigma}$.  When
  $\varepsilon=0$, $\varepsilon_{\rm of}=0$ since with no stars, the
  Eddington surface density is zero. As the mass in stars
increases, $\varepsilon_{\rm of}$ initially also increases under two
competing influences.  More stars increase the radiative force and
$\Sigma_E$ so that a larger fraction of the cloud is eligible to
  become unbound; this increases the factor $1 + {\rm erf}(y_E)$ in
Equation~(\ref{Eq:EpsOF}). However, a larger stellar mass also
decreases the gas mass since a larger fraction of the cloud is already
bound up in stars; this decreases the factor $1 - \varepsilon$.
  In the limit $\varepsilon \rightarrow 1$, there is no gas reservoir and
  therefore $\varepsilon_{\rm of} \rightarrow 0$.  For some value
  intermediate value of $\varepsilon$ between 0 and 1, an
  infinitesimal increase in $\varepsilon$ would decrease
  $\varepsilon_{\rm of}$ in Equation~\ref{Eq:EpsOF}, since the overall
  decrease in the gas mass reservoir (lower $1-\varepsilon$) outweighs
  the increase in the fraction of gas that is super-Eddington (higher
  $y_E$).  This point represents the maximum value of
  $\varepsilon_{\rm of}$ for any given set of initial cloud
  parameters and variance in the PDF.

We consider a cloud such that, for a given PDF variance, 
the maximum in $\varepsilon_{\rm of}$ has been reached; its efficiency is
then $\varepsilon = {\rm arg~ max~ }\varepsilon_{\rm of}$.
At this point, the outflowing gas mass cannot decrease, because
this material has already become super-Eddington. The remaining
gas reservoir that is neither stars nor outflowing gas is a fraction 
$(1-{\rm max }~\varepsilon_{\rm of} - {\rm arg~ max~ }\varepsilon_{\rm of})$
of the original cloud.  If this material collapses faster than the PDF in 
larger-scale cloud can adjust,
it will all be added to the existing stars 
and the final star formation
efficiency will be $\varepsilon_{\rm final}= 1-{\rm max }~\varepsilon_{\rm of}$.
The luminosity from these additional stars would also
increase the radiation pressure on the outflowing gas.  
Alternatively, if the collapse is slower, there may be time
for the log-normal density distribution of the remaining gas
to adjust, and the final
star formation efficiency may rise to a level between 
${\rm arg~ max~ }\varepsilon_{\rm of}$ and $1-{\rm max }~\varepsilon_{\rm of}$.
This suggests that for a given PDF variance, there are
upper and lower bounds on $\varepsilon_{\rm final}$:
\begin{subequations}
\begin{eqnarray}
	\varepsilon_{\rm max} &=& (1 - \max_{0 < \varepsilon < 1} \varepsilon_{\rm of}) \nonumber \\
	&=& \frac{1}{2} \min_{0 < \varepsilon < 1} \left[1 + \varepsilon + 
	(\varepsilon - 1){\rm erf}\left(y_E\right)\right] \\
	\varepsilon_{\rm min} &=& {\rm arg}\max_{0 < \varepsilon < 1} \varepsilon_{\rm of}.
\end{eqnarray}
\label{Eq:LognormalEff}
\end{subequations}
From Figure~\ref{Fig:EffEvolution}, the maximum of the function
$\varepsilon_{\rm of}$ increases with decreasing $\Sigma_{\rm cl, 0}$,
since at lower surface densities the gas reservoir
can be driven away more easily.
Also, broader surface density distributions (larger
$\sigma_{\rm ln \Sigma}$) tend to decrease ${\rm max}~\varepsilon_{\rm of}$
since more gas is at the highest density, which
is more difficult to unbind.

  In Figure
\ref{Fig:ModelEfficiencies}, we compare the predictions of
Equation~(\ref{Eq:LognormalEff})
to our numerical results
for net star formation efficiencies as a function of initial surface
density. We note that we are primarily interested in the efficiency
relative to the cloud mass adjusted for initial 
turbulent outflows
$\varepsilon_{\rm adj} = \varepsilon / (1 - \varepsilon_{\rm of,init})$, 
so we substitute the adjusted surface density
$\Sigma_{\rm cl,0} 
\rightarrow \Sigma_{\rm adj} = \Sigma_{\rm cl,0} (1 - \varepsilon_{\rm of,init})$ 
in Equation~(\ref{Eq:xbfinal}). For the parameters entering Equation~(\ref{Eq:xbfinal}),
we use $x = 0.84$ and $\sigma_{\rm ln \Sigma} = 1.42$,
based on the time average of the best-fit values for the circumcluster 
surface density in the fiducial model (see Paper II for details).
We use these same values of $x$ and $\sigma_{\rm ln \Sigma}$ at all
$\Sigma_{\rm cl, 0}$. Additionally, since we are interested in comparing 
to the physical observed cloud, rather than the artificial initial conditions,
we show the adjusted efficiency as a function of the adjusted cloud
surface density $\langle \Sigma_{\rm cloud} \rangle = \Sigma_{\rm adj} / x^2$ rather 
than $\Sigma_{\rm cl, 0}$.

Figure \ref{Fig:ModelEfficiencies} shows that 
$\varepsilon_{\rm max}$ from Equation~(\ref{Eq:LognormalEff}a)
represents a reasonable estimate of the actual SFE found in the simulations, 
both in normalization and in the shape of the dependence on
$\Sigma_{\rm cloud}$.
In principle, however, the final efficiency might be closer to
$\varepsilon_{\rm min}$ if conditions were such that
the star formation rate were lower.

Equation~(\ref{Eq:LognormalEff}) predicts much higher efficiencies
than the fiducial model of \cite{ThompsonKrumholz2014}, shown for
comparison in Figure~\ref{Fig:ModelEfficiencies}.
Their model has similar ingredients to 
ours (a lognormal surface density
distribution, with both sub-Eddington and super-Eddington regions),
with the principal difference being that the \cite{ThompsonKrumholz2014}
formalism assumes a fixed star formation rate per unit gas mass,
whereas there is no assumption about the star formation rate in our
model.  Specifically, they assume that stars form at a rate
$\dot{M_*} = \varepsilon_{\rm ff} M_g / t_{\rm ff}$ from the total
remaining mass (including super-Eddington $\Sigma < \Sigma_E$
regions), and that winds are driven out at a rate proportional to
$1/t_{\rm ff}$ from the super-Eddington gas.  It is their adoption of a
very small fiducial value $\varepsilon_{\rm ff} = 0.01$ (implying a
vast discrepancy between star formation and wind expulsion rates) that
leads to a highly suppressed final efficiency in their fiducial model.
If instead we adopt $\varepsilon_{\rm ff} = 0.44$ (similar to our
simulation results in Section \ref{SubSec:SFERates}) and apply their
formula, their model prediction is somewhat closer to ours, albeit with a
lower normalization and shallower dependence on $\Sigma_{\rm cloud}$
(see Figure~\ref{Fig:ModelEfficiencies}).

The correspondence between the prediction of $\varepsilon_{\rm max}$ in 
Equation~(\ref{Eq:LognormalEff}a) and our $\Sigma-$series numerical
model results is close enough to suggest that the dominant effect in
suppressing star formation is radiative feedback driving out
structures at successively higher surface densities
  until a maximum mass of outflowing material is reached.
However, there
are a number of issues, or at least questions, surrounding this model. 
First among these is whether the correlation timescale of the
lognormal surface density distribution in the cloud is long enough to
allow persistent acceleration by the central stars.
This is because  
the dynamical evolution of any given fluid element depends on the
coherence time of the (Lagrangian) evolution for the surface density
region surrounding it.
In principle, the surface 
density distribution could remain statistically lognormal at all
times, while individual regions fluctuate rapidly. If these
fluctuations in time are much shorter than the time taken for
radiation to accelerate gas from the cloud, then any fluid element would 
fully sample the distribution of surface densities, and only the mean 
cloud surface density would be relevant.  

The question of the column density correlation 
timescales in comparison to the cloud destruction timescale has
already been discussed in a heuristic manner in
\cite{ThompsonKrumholz2014}. They compare the turbulence crossing time
to the acceleration timescale $t_{\rm acc} \propto r_0/v_{\rm esc}(r_0)$ 
and argue that so long as the radiation force is several times
stronger than the force of gravity, densities will fluctuate on longer
timescales than it takes for the cloud to unbind.

In our simulations, we can measure the temporal correlations 
of the column in a given area of the sky.  We find 
a correlation time $\sim 0.5 t_{\rm ff,0}$ once star
formation has begun. This is roughly comparable to the timescales on
which gas is accelerated out of the cloud, suggesting that outflowing
low surface density regions might merge with collapsing higher surface
density regions before they have a chance to escape the
cloud. However, this overall Eulerian correlation time is not necessarily 
representative for low density regions. 
In addition, we do find an outflowing velocity distribution consistent with low 
density regions remaining correlated until they escape the cloud
\citep[see][for details]{RaskuttiOstriker2016b} and so conclude that this interpretation is not 
unreasonable. 
In future tests, to answer this question realistically, we would need to use tracer
particles in the gas to track the flow of individual fluid elements.

We note also that the values of $\sigma_{\rm ln \Sigma}$ from our 
simulations (see Figures \ref{Fig:LogNormalAlpha} and 
\ref{Fig:LogNormalAlphaDyn}) are somewhat larger than current estimates 
from observations, which typically find $\sigma_{\rm ln \Sigma}<1$.  This
may owe in part to line-of-sight contamination, which tends to 
reduce the observed $\sigma_{\rm ln \Sigma}$ \citep[e.g.,][]{Schneider2015}, 
and in part to the absence of magnetic fields in the present models, 
as magnetization reduces shock compression and therefore density 
variance \citep[e.g.,][]{Ostriker2001,Molina2012}.  Figure \ref{Fig:EffVarSigma}a shows the 
prediction of Equation (\ref{Eq:LognormalEff}a) for 
$\varepsilon_\mathrm{final}$ as a function of $\Sigma_{\rm cl}$ 
for a range of $\sigma_{\rm ln \Sigma}$, demonstrating that the 
predicted net SFE in a cloud could be considerably lower at low 
$\sigma_{\rm ln \Sigma}$.  

\subsection{Effect of varying $\Psi$}
\label{SubSec:VaryPsi}

\begin{figure*}
  \centering
  \epsscale{1}
  \includegraphics{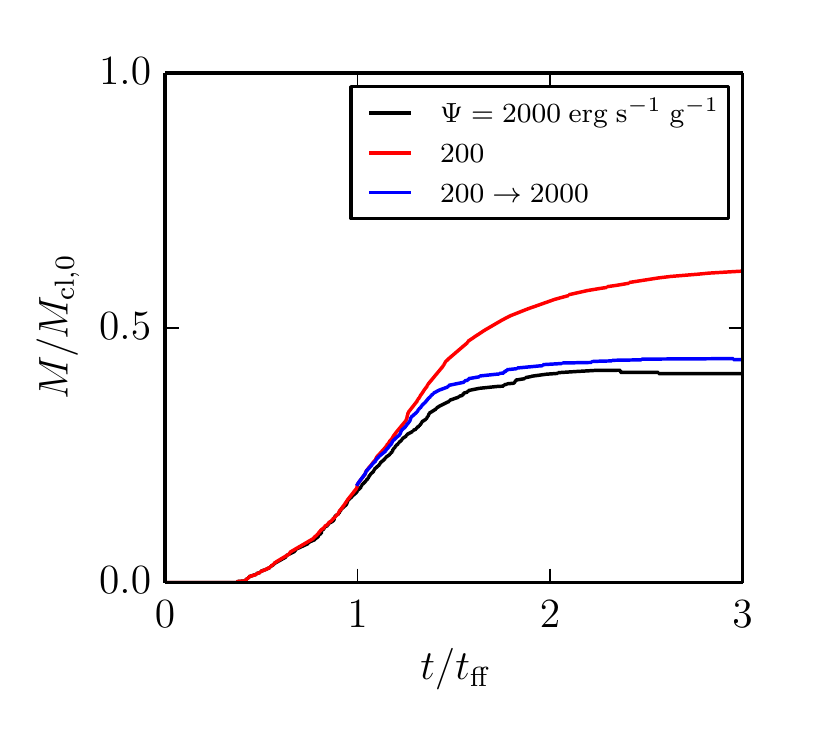} 
  \caption{Evolution with time of the star formation efficiency for
    varying values of the luminosity per unit mass $\Psi$ (shown in
    the legend). We show the fiducial model (black), a low luminosity
    model (red), and a mixed model for which $\Psi = 200~{\rm
      erg~s^{-1}~g^{-1}}$ until a freefall time (at which point $M_* =
    8.5 \times 10^3~M_{\odot}$), which is then set to the fiducial value
    $\Psi = 2000~{\rm erg~s^{-1}~g^{-1}}$ beyond that point.}
  \label{Fig:VarPsi}
\end{figure*}

As a final point, it is worth considering the effects on our model of a varying
$\Psi$ due to undersampling of the IMF. \cite{Kim2016} recently
studied the effects of an undersampled \citet{Chabrier2003} IMF on the
value of $\Psi$, both stochastically and on average, by using the SLUG
code \citep{Krumholz2015} to simulate the photon output from stellar
clusters as a function of mass. They found that although
$100~M_{\odot}$~clusters produce significantly fewer photons, with
$\Psi \sim 200~{\rm erg~s^{-1}~g^{-1}}$, by the time $M_{\star} \sim 2
\times 10^3~M_{\odot}$, the cluster luminosity per unit mass has
settled to the final fully sampled value of $\Psi \sim 2000~{\rm
  erg~s^{-1}~g^{-1}}$, with a variation of around $\pm 0.3$~dex around
this median. For even larger clusters of $M_{\star} \sim
10^4~M_{\odot}$, this variance drops to only $\pm 0.1$~dex. We may
therefore expect the stellar IMF to be fully sampled for $M_{\star}
\gtrsim 2 \times 10^3~M_{\odot}$.

In all of our simulations, the 
stellar mass exceeds $2 \times 10^3~M_{\odot}$ by the end of the
star formation epoch. Our lowest mass cloud has a mass of only 
$5 \times 10^3~M_{\odot}$ but an efficiency of more than $40\%$, so that more than 
$2 \times 10^3~M_{\odot}$ is in stars. When star particles first form,
realistically  the IMF would be undersampled.  By setting $\Psi$ to
a constant, we overestimate the radiative force at early times,
while correspondingly underestimating the 
star formation rate. However, the final efficiencies are largely set by the balance between 
the radiative force and the gas self-gravity at late times, so these are not changed significantly 
by our overestimation of $\Psi$ at early times.

The smallest of our star particles are 
less than $M_{\star} \sim 10^2~M_{\odot}$, which would imply
correspondingly low median values of 
$\Psi \sim 200~{\rm erg~s^{-1}~g^{-1}}$. Therefore, when these star particles first form, we are 
overestimating the radiative force by a factor of almost 10. We may
bracket the effect of this by 
simulating our fiducial model over a range of possible $\Psi$ values.
In Figure~\ref{Fig:VarPsi} we 
show the stellar efficiency for three different models with $\Psi$ varying between 
$200$ and $2000~{\rm erg~s^{-1}~g^{-1}}$. The models with lower $\Psi$
form stars at a 
slightly faster rate and end up with a higher efficiency, since less gas is driven from the cloud by 
the effects of UV radiation. 

However, we note that the variation in efficiency is only
$\sim 50~\%$ despite an order of magnitude variation in the luminosity per unit mass. In 
simple models of radiative feedback that assume a uniform shell of gas driven 
away by radiative pressure, $1 / \varepsilon - \varepsilon$ is linear with $\Psi$. Therefore, variation 
by an order of magnitude in $\Psi$ would  correspond
to almost an order of magnitude variation in 
$\varepsilon$, or at least a saturation at unity. By contrast, the model
we develop, which takes into account the lognormal 
surface density distribution with best-fit values of $x$ and
$\sigma_{{\rm ln}\Sigma}$, does a much better job of matching the final stellar
efficiency, as shown in Figure~\ref{Fig:EffVarSigma}b. 

The above shows that the final SFE is much less sensitive to
undersampling of the IMF than simple models would predict.
More importantly, by the time star formation stops, all of our clouds, even the lowest mass ones, 
have enough stellar mass to sample the IMF. Furthermore, the gas surface density structure of our clouds does not change significantly for varying $\Psi$, since it is largely set by the initial turbulence. 
Therefore, by the time the IMF becomes fully sampled, the cloud appears roughly 
the same regardless of the strength of radiative feedback prior to that time. 
We might then expect the final efficiency to only depend on the final value of $\Psi$, which is 
$\Psi \sim 2000~{\rm erg~s^{-1}~g^{-1}}$ in all cases. In fact, when we 
run simulations in which the value of $\Psi$ changes from 
$\Psi \sim 200~{\rm erg~s^{-1}~g^{-1}}$ at early times to 
$\Psi \sim 2000~{\rm erg~s^{-1}~g^{-1}}$ after a freefall time, as shown in Figure~\ref{Fig:VarPsi}, 
the final stellar efficiency is largely unchanged.

\subsection{Evolution of Star Formation Rate}
\label{SubSec:SFEEvol}

\begin{figure}
  \centering
  \epsscale{1}
  \includegraphics{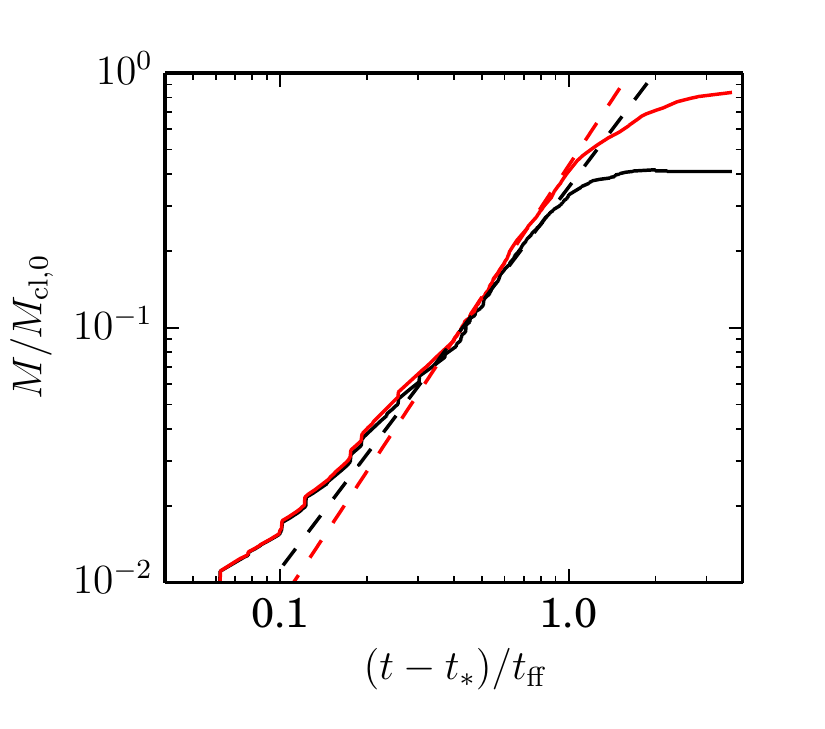}
  \caption{Stellar mass as a function of time after the first star was formed for both our fiducial (black) and 
	no-feedback (red) models. For comparison we also show the fits obtained using a single power law in 
	each case (dashed lines).}
  \label{Fig:FiducialEvolLog}
\end{figure}

\begin{figure}
  \centering
  \epsscale{1}
  \includegraphics{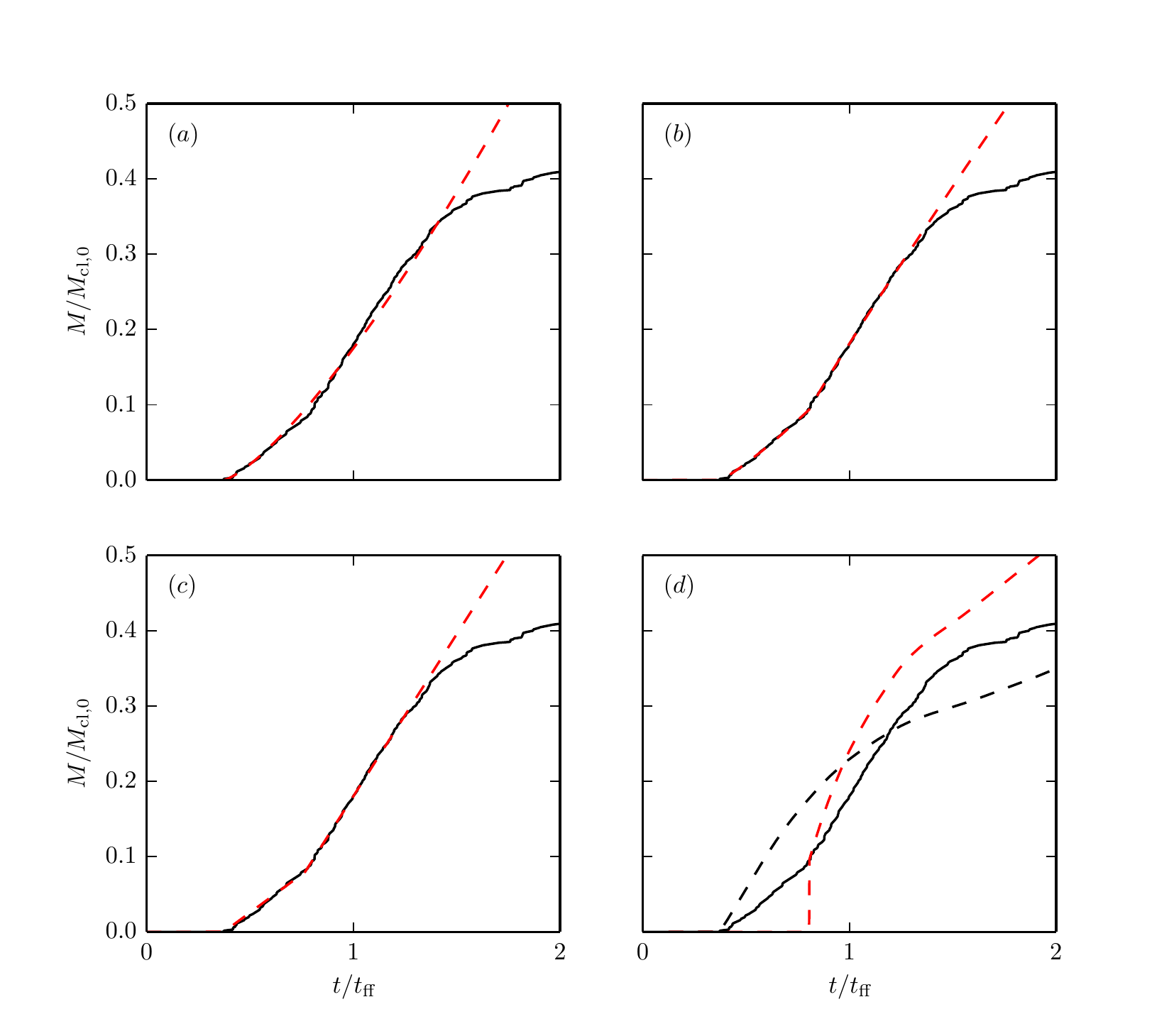}
  \caption{Stellar mass as a function of time for our fiducial model
    in comparison to several fits.  In all cases, we show the
    simulated stellar mass as a solid line. Panel (a) compares to a
    single power law, panel (b) to a broken power law, and panel (c) a
    piecewise linear model. In panel (d), we show a comparison to
    Equation \ref{Eq:Mstardot} with constant best-fit
    $\varepsilon_{\rm ff}$ starting either at $t_*$ (black dashed), or
    starting at $t_{\rm break}$ (red dashed). For (d) we compute $\rho$ and
    $t_{\rm ff}$ allowing for evolving gas mass and
    evolving cloud size based on a log-normal fit.}
  \label{Fig:FiducialEvol}
\end{figure}

\begin{figure*}
  \centering
  \epsscale{1}
  \includegraphics{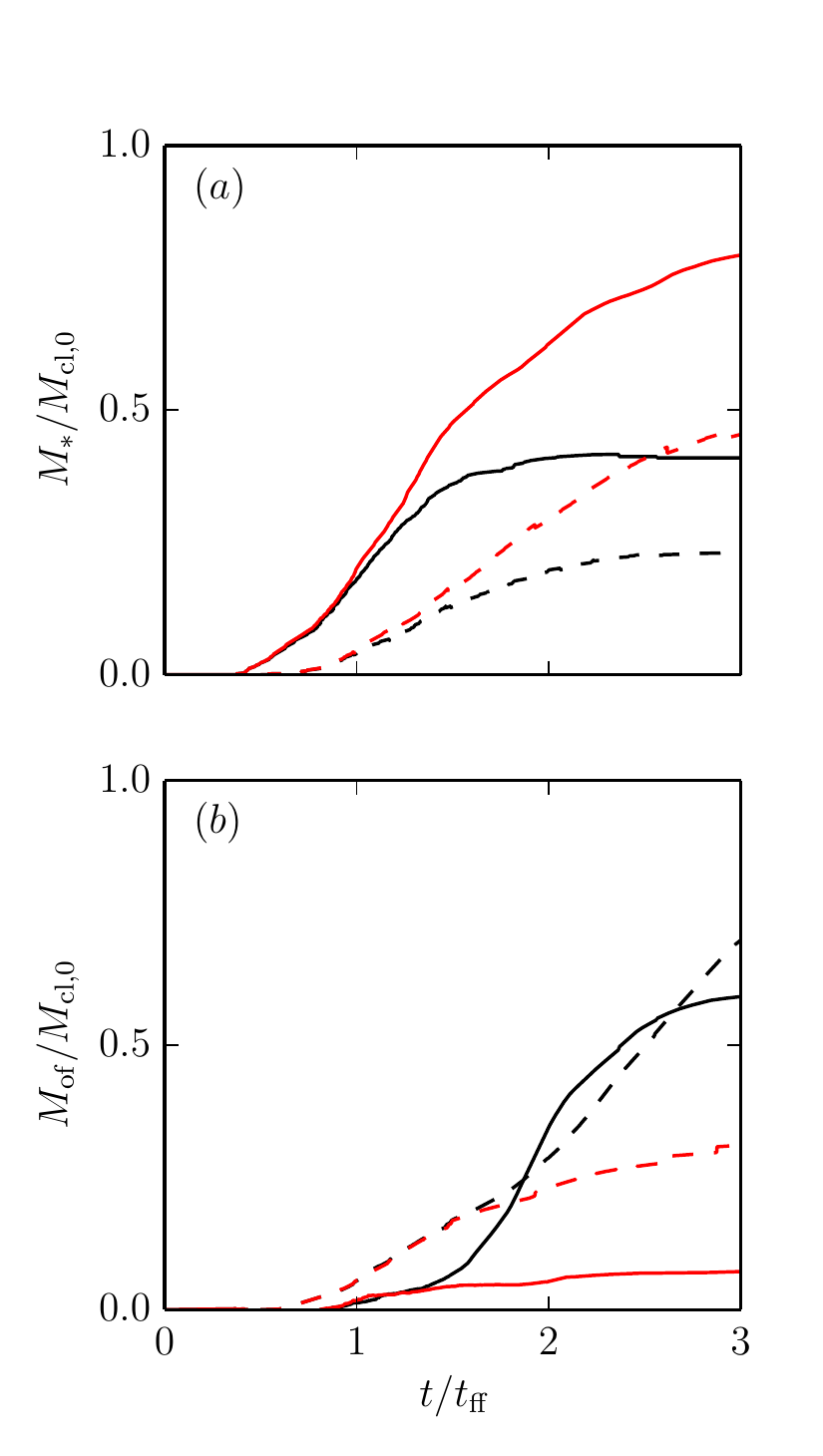} 
  \caption{Evolution of the star formation efficiency (top) and outflowing mass (bottom) with time for 
  varying initial conditions. We show results for both our fiducial model (solid), as well as a model 
  initialized from conditions in which the density relaxes in response to
  turbulence before gravity is turned on (dashed). In both cases, 
  we show both models with radiative feedback (black) and without (red).}
  \label{Fig:VarInit}
\end{figure*}

So far, we have only been considering the final efficiencies of star
forming clouds as a means of assessing when radiative feedback
becomes important in unbinding GMCs. However, of just as much interest
is the rate of star formation, which is often parameterized by the
star formation rate per freefall time:
\begin{equation}
\dot M_{*}(t) \equiv \varepsilon_{\rm ff}(t)\frac{M_g(t)}{t_{\rm ff}(t)}.
 \end{equation}
We now turn to an analysis of how this star formation rate varies in our models.

The majority of previous studies have tended to focus only on the mean
SFR per freefall time $\langle \varepsilon_{\rm ff}(t) \rangle$
averaged over the whole epoch of star formation and with a fixed
$t_{\rm ff}=t_{\rm ff,0}$ set by the cloud's initial mean density
\citep[e.g.,][]{Wang2010, PadoanNordlund2011, Padoan2012,Bate2012,
  Krumholz2012, FederrathKlessen2012, Myers2014}. Here, we adopt the
methodology of \cite{Lee2014} and fit the stellar mass history of our
clouds to determine if there is any systematic evolution in time. We
fit with a power law defined in terms of $t - t_{*}$, where $t_{*}$ is
the time at which the first star particle is formed.  We also
initially tested the fitting region of \cite{Lee2014} ranging from
$M_{*} = 0.015 M_{\rm cl,0}$ to $M_{*} = 0.3 M_{\rm cl,0}$. Adopting
this choice for our fiducial model, similar to \cite{Lee2014} we can
confirm a super-linear, though slightly less than quadratic power-law
evolution of the star formation efficiency, $M_{*} \propto t^{\beta}$
with $\beta \sim 1.5$, as shown in Figure~\ref{Fig:FiducialEvolLog}.

However, there are a number of issues with fitting a power law to the
SFR. Firstly, there is no reason a priori to assume that the gas
density distribution or the SFR have reached some sort of steady state
at $t_{*}$. This is certainly a concern for our simulations, since the
cloud begins with uniform density, which is very far from a self-consistent
quasi-steady state.  It will undergo turbulent collapse, and the
highest density regions may form stars while the shape of the density
distribution is still changing considerably. Therefore, a power law
starting at $t_{*}$ will not necessarily capture the physics of star
formation in realistic clouds as it may still be affected by the
artificial initial conditions.

This can be seen very clearly if we consider the stellar mass
evolution in our fiducial model as a function of $t - t_{*}$, as shown in 
Figure~\ref{Fig:FiducialEvolLog}.  There is an obvious break in
the power law at around $M_{*} \approx 0.1 M_{\rm cl,0}$ so 
that a single power law underestimates the stellar mass at
both early and late times.  This break is not the result of feedback;
the model without radiative feedback also shows a similiar break, and
an almost identical evolution up to at least $\sim 0.3 M_{\rm cl,0}$.

A much better fit can be achieved if we allow for a broken power law 
$M_* \propto (t - t_{\rm break})^\beta$, with a transition time $t_{\rm break}$. 
For the fiducial model, $t_{\rm break} / t_{\rm ff,0} = 0.80$ when $M_{*}/M_{\rm cl,0} = 0.094$. 
In this case, as shown in Figure~\ref{Fig:FiducialEvol}b, we have a much smaller least squares
error for the broken power law model ($\chi^2_{\rm broken}$) as opposed to the simple 
power law ($\chi^2_{\rm pl}$) with $\chi^2_{\rm broken} / \chi^2_{\rm pl} = 0.042$ 
over the whole fitting range and $\chi^2_{\rm broken} / \chi^2_{\rm pl} = 0.24$ just above the 
break mass.  More importantly, the broken power law does not systematically depart from the evolution 
at either low or high stellar mass, and thus more accurately captures the cloud behaviour below 
the transition.

Interestingly, when we fit to a broken power-law, both regimes show roughly linear growth in the stellar 
mass with time, albeit at significantly different rates. In fact, if we restrict both regimes to have a 
constant star formation rate (as shown in Figure~\ref{Fig:FiducialEvol}c) we still get a much better fit 
than the simple power law above the break time, with $\chi^2_{\rm lin} / \chi^2_{\rm pl} = 0.38$. 
This suggests two things. Firstly, our fiducial cloud simulation exhibits some transient behaviour even 
after star formation has begun; until close to a freefall time it seems to be adjusting from the artificial 
initial state. Secondly, the cloud appears to emerge from this transient state at approximately 
$t_{10}$. It then exhibits roughly linear growth in stellar mass 
until radiative feedback becomes important.

We may test the presence of this transient state by considering a
  ``pre-relaxed'' model, in which the cloud 
  is allowed to evolve without self-gravity for the first half of 
a freefall time and then evolves with gravity beyond this point. When this is done, the stellar 
and outflowing mass evolve as shown in Figure~\ref{Fig:VarInit}. We
see that the pre-relaxed model has both a lower final efficiency and
lower SFR. It is however difficult to compare this result to our fiducial
model, since it is at lower surface density due to both expansion of the
cloud and the much larger fraction of gas unbound by the initial
turbulence (around $30~\%$ compared to $10~\%$). Moreover, it is at a
lower virial parameter since the turbulence decays away over the first
half a freefall time of relaxation. What we can see is that there is
no evidence of a break in the SFR for this model, suggesting that this
break is an artificial one born of our initial
conditions.

The origin of a constant SFR in later stages
is not trivial to explain. The simplest star formation law, and one assumed 
in a number of previous studies, adopts the form
\begin{equation}
	\dot M_{*}(t) = \varepsilon_{\rm ff}\frac{M_g(t)}{t_{\rm ff}(t)},
	\label{Eq:Mstardot}
\end{equation}
where $\varepsilon_{\rm ff}$ is a constant and
$t_{\rm ff} \propto \rho_{g}^{-1/2}$ is the 
instantaneous freefall time for $\rho_{g}(t)$ the (time-dependent) 
volume-averaged gas density. 
A focus of both numerical and observational studies has been to fit to this form and estimate 
$\varepsilon_{\rm ff}$. In this picture, star formation halts through a combination of gas 
depletion by star formation and gas expulsion through cloud expansion and a subsequent increase in
$t_{\rm ff}$. 
However, the form of Equation~(\ref{Eq:Mstardot}) does not seem to apply in our simulations. 
In our turbulent clouds, the cloud radius remains roughly constant while the majority of stars are forming so
that $\rho_g \propto M_g$ and
$t_{\rm ff} \propto M_g^{-1/2}$, which would yield 
$\dot{M_{*}} \propto M_g^{3/2}$ in Equation~(\ref{Eq:Mstardot}). Therefore, as the gas mass is depleted,
Equation~(\ref{Eq:Mstardot}) would predict a decrease in the SFR until cloud expansion from feedback 
drives a rapid increase in the freefall time. In fact, our power law or broken power law fits show that the
SFR is constant or increasing until gas is expelled by feedback.

Fits to the simple star formation law of Equation~(\ref{Eq:Mstardot})
are shown in Figure~\ref{Fig:FiducialEvol}d.
The gas mass $M_g(t)$ is taken directly 
from each simulation, and
$t_{\rm ff}(t)$ is found by fitting for the mean cloud
density through the density PDF.  If we fit this form with star
formation beginning at $t = t_{*}$, the best-fit value is 
$\varepsilon_{\rm ff}=0.45$, and there is a huge
discrepancy between the numerical and analytic results. 
Even if we fit only after $t = t_{\rm break}$, the fit is not good; 
the best-fit efficiency $\varepsilon_{\rm ff} = 1.21$ is high 
since the early mass growth must be large to compensate for the steady
decrease in SFR. 

\begin{figure}
  \centering
  \epsscale{1}
  \includegraphics{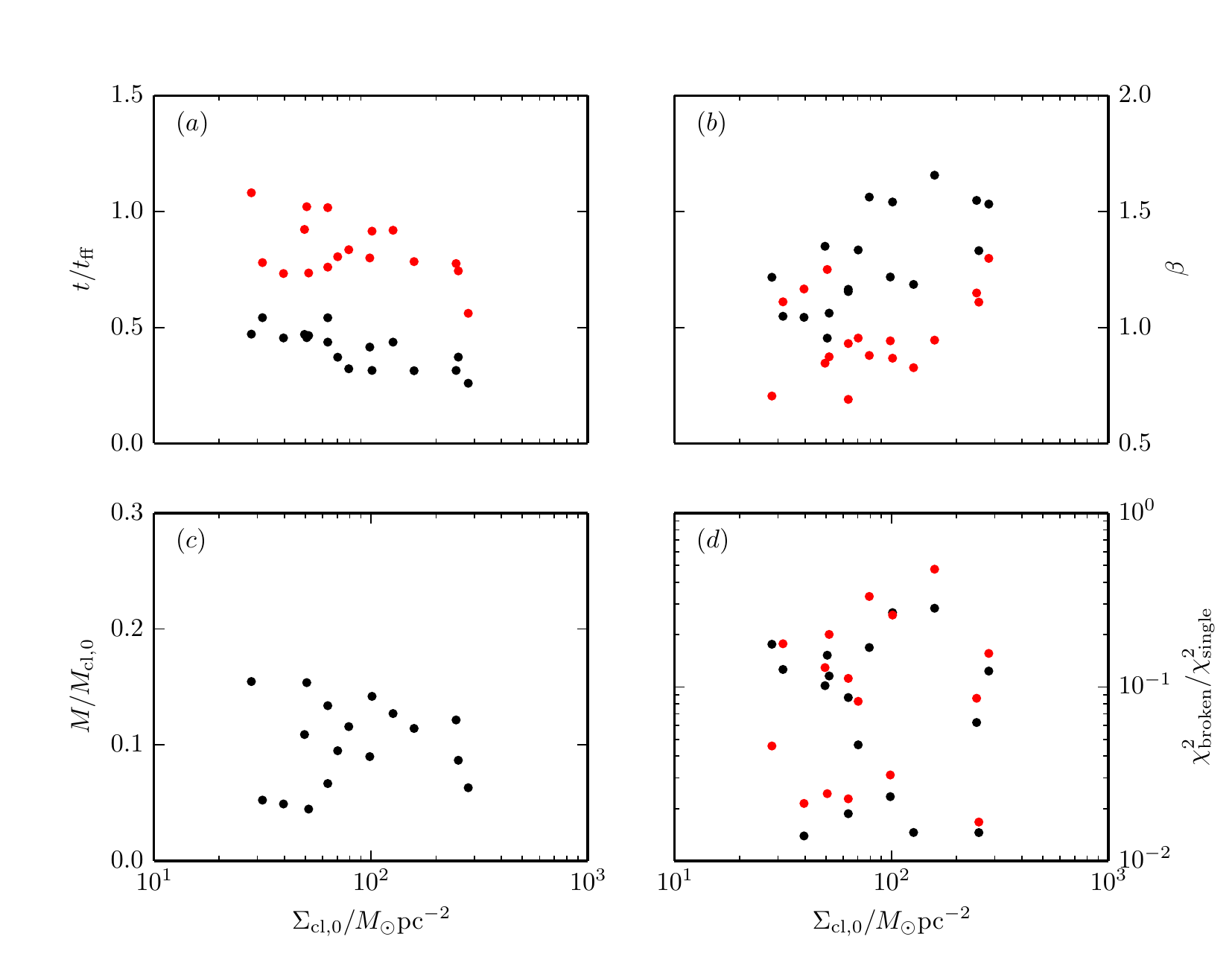}
  \caption{Characteristics of our best fit single power law and broken power law fits for the 
	$\Sigma$-series. Panel (a) shows the break time $t_{\rm break}$ for the broken power law (red) 
	and $t_{*}$ (black), while Panel (c) shows the corresponding break masses. Panel (b) shows the 
	post-break exponent in the broken power law (red) or single exponent in the single power law (black). 
	Finally, in Panel (d), we show the relative values of $\chi^2$ for the broken and single power law models, both over the whole fitting range (black) and just after $t_{\rm break}$ (red).}
  \label{Fig:FiducialEvolExponent}
\end{figure}

We may test the generality of these conclusions by applying the same methodology to other members of 
our $\Sigma$-series simulations. As a caveat, we saw in Section~\ref{SubSec:Convergence} that 
lower resolution simulations will capture the final SFE, but may underestimate the SFR if they are
not converged. This is particularly evident in our low surface density and high virial parameter simulations 
in which the number of individual star particles formed is small. With this in mind, we omit
simulations with $\varepsilon < 0.15$ or $\Sigma_{\rm cl, 0} < 20~M_{\odot}~{\rm pc}^{-2}$ from our 
studies of the SFR.

In Figure~\ref{Fig:FiducialEvolExponent} we show the best-fit characteristics for both the single 
power-law fit and the high mass portion of our broken power-law fit, as a function of surface
density. For the $\Sigma$-series, we observe little variation in either the power law exponent or the break 
time and mass across the sequence. As for our fiducial model, for the broken power law, the fitted 
exponent is close to $\beta = 1$ for most cases. Meanwhile, a single power law almost always shows 
super-linear behavior, with $\beta \sim 1.5$. 

At the break time, roughly $0.1~M_{\rm cl,0}$ in stellar mass has
formed across all models.  The break times also show little variation
with surface density, being close to $t_{\rm break} \sim 0.8 t_{\rm
  ff,0}$ for the broken power law, with $t_{*} \sim 0.5 t_{\rm
  ff,0}$.  The most massive clouds begin star formation a little
earlier, since they are already denser, hence regions reach critical
densities high enough to undergo local collapse at earlier times. This
slight surface density dependence also persists in the break time for
the broken power law, suggesting that it may be as arbitrary as
$t_{*}$. Potentially, the initial conditions still affect the state
of the system at $t_{\rm break}$, although our simulations do not show
any strong evidence for a continuous acceleration in the star
formation rate. Nevertheless, simulations with more realistic cloud
initial conditions are needed before any definitive statements can be
made about varying star formation rates in turbulent clouds.

\subsection{Star Formation Per Freefall Time}
\label{SubSec:SFERates}

\begin{figure}
  \centering
  \epsscale{1}
  \includegraphics{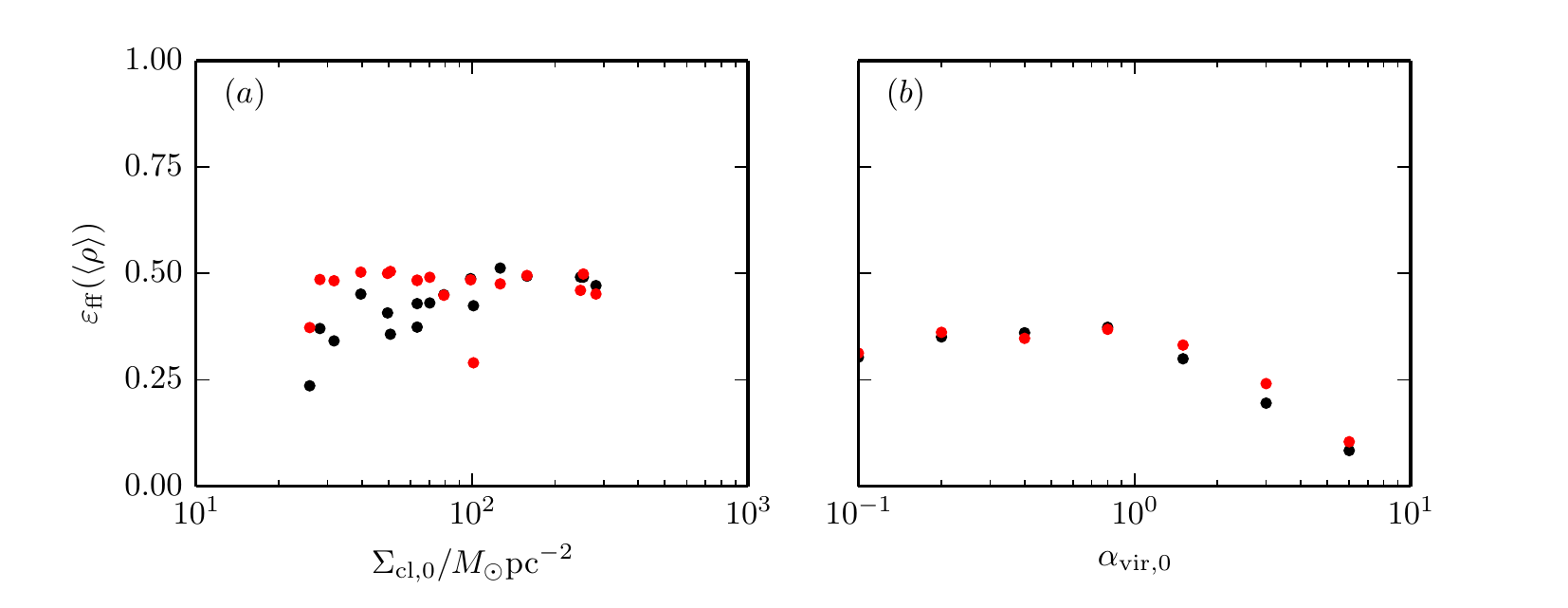}
  \caption{Star formation rate coefficient $\varepsilon_{\rm ff,\bar\rho}$ 
(defined in Equation \ref{Eq:epsff_lin}) for the (a) $\Sigma$-series
    simulations and (b) $\alpha$-series simulations. 
    We show results from simulation with radiation feedback (black)
    and without (red).}
  \label{Fig:SFERates}
\end{figure}

The fact that our star formation efficiencies grow roughly linearly with time allows us to quantify these 
star formation rates quite easily. We replicate the method of \cite{PadoanNordlund2011, 
Krumholz2012, Myers2014} and fit a straight line to the stellar mass vs. time above the transition mass 
discussed in Section~\ref{SubSec:SFEEvol}. We then calculate the efficiency per freefall time defined as
\begin{equation}
	\varepsilon_{\rm ff, \bar{\rho}} \equiv \frac{
\langle \dot{M_*}\rangle t_{\rm ff, \bar{\rho}}}{M_{\rm cl,0}}
\label{Eq:epsff_lin}
\end{equation}
where the mean density used in $t_{\rm ff, \bar{\rho}}$ is calculated from 
directly fitting a lognormal form to the density PDF at the start of star formation, defined here as 
$t_{\rm break}$, and extracting the mean density $\langle \rho \rangle$. For the $\Sigma$-series, this 
is very close to the initial cloud density since only around $10 \%$ of the mass is lost in early 
turbulent outflows and there is very little global cloud contraction or expansion. However, the 
mean density differs significantly from the initial cloud value for the models with low virial parameter, 
where the cloud contracts significantly.

In Figure~\ref{Fig:SFERates} we show the resulting star formation rate
coefficients as a function of both surface density and virial
parameter. Almost irrespective of both virial parameter and surface
density, we find a rate coefficient 
$\varepsilon_{\rm ff,\bar\rho} \sim 0.25 - 0.5$. 
Notably, the inclusion of radiative feedback only mildly decreases the
star formation rate, with no systematic surface density dependence. 
This is not surprising as we had already noted
that turbulence dominates the star formation at early times.  
These results again suggest that UV feedback in clouds
can be effective as a means of limiting star formation and unbinding
clouds, but does little to suppress the instantaneous star formation
rate.

\subsection{Cloud Lifetimes}
\label{SubSec:Lifetimes}

\begin{figure}
  \centering
  \epsscale{1}
  \includegraphics{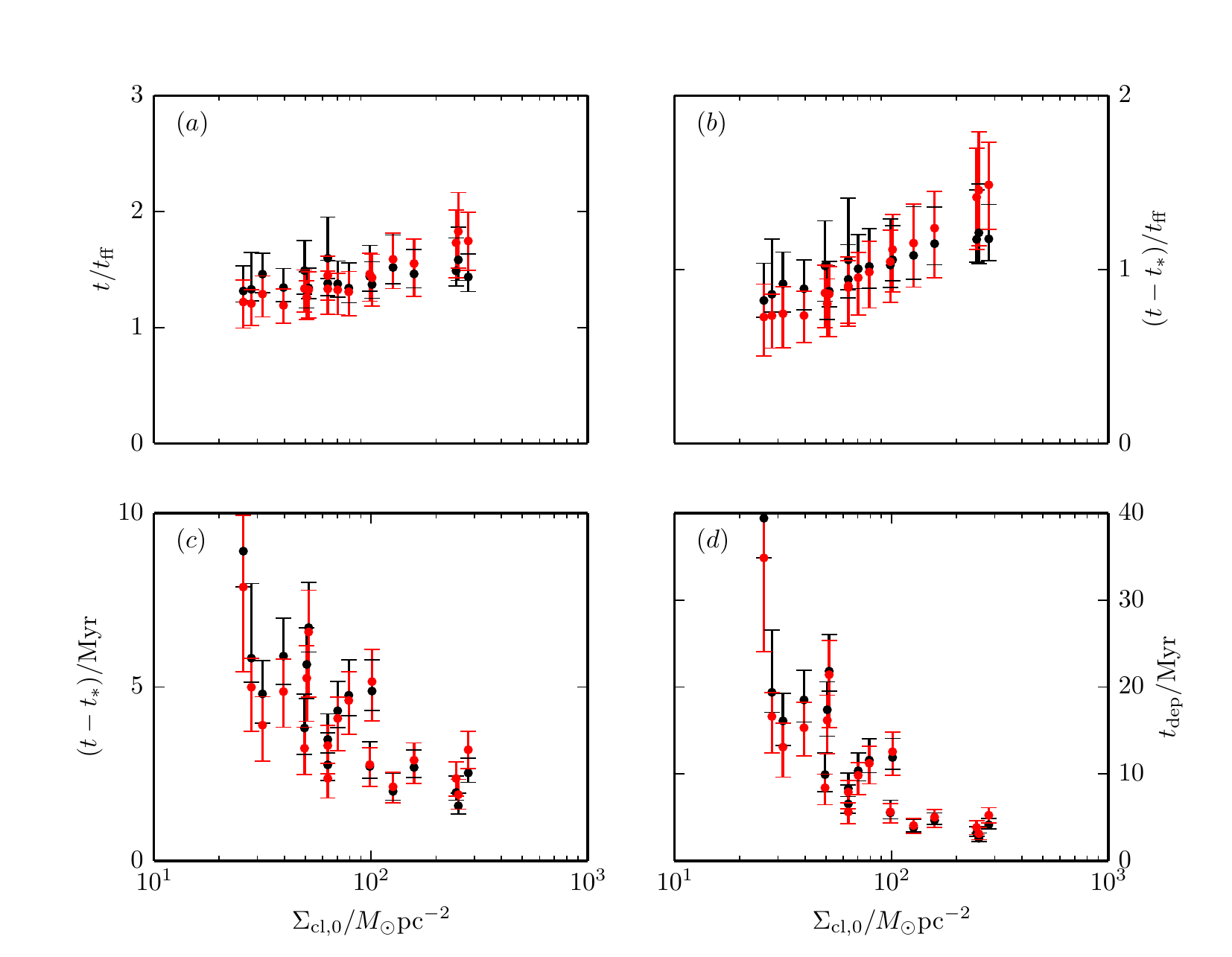}
 \caption{Cloud lifetimes as a function of surface density for our
   $\Sigma$-series of models.  We show lifetimes calculated using the
   time $t_{80}$ when $80^{+10}_{-10} \%$ of stars are formed (black) and the time
   $t_{\rm unb}$ when the virial parameter reaches $5^{+5}_{-3}$ (red). Error bars denote 
	the limits on these respective values so that the uppermost limit shows $t_{90}$ and the 
	time when $\alpha_{\rm vir} = 10$. In both cases, we show (a) the simulation time when this 
	occurs in units of $t_{\rm ff,0}$, (b) the time after the first star is formed
   ($t-t_*$) in units of $t_{\rm ff,0}$, (c) $t-t_*$ in Myr, and (d)
   the depletion time $t_{\rm dep} \equiv (t-t_*) / \varepsilon_{\rm final}$.}
  \label{Fig:Lifetimes}
\end{figure}

The consequence of such high star formation rates is that our model
clouds either convert most of their gas mass to stars, or become
unbound on short timescales. While it is difficult to characterize
exactly when the bulk of gas mass from a filamentary cloud becomes
unbound, the proxy we use is the virial parameter. Similar to
\cite{Colin2013}, we find that in all our $\Sigma$-series simulations,
the virial parameter remains close to unity when the dynamics are set
by gravitational collapse and turbulence, but then quickly expand to
$\alpha_{\rm vir} \sim 50$ once radiative feedback begins to
dominate. Therefore, we arbitrarily take $\alpha_{\rm vir} = 5$ as our
criterion for unboundedness, noting that the expansion from being
formally unbound at $\alpha_{\rm vir} = 2$ to $\alpha_{\rm vir}
\gtrsim 10$ takes $\sim 0.3 t_{\rm ff,0}$.

In Figure~\ref{Fig:Lifetimes}a, we show the time $t_{\rm unb}$ when
clouds become unbound ($\alpha_{\rm vir} = 5$) as a function of
surface density. We find that clouds last between $1.2$ and
$1.9~t_{\rm ff,0}$ at most and that radiative feedback acts very rapidly
to unbind the clouds, with the transition from $\alpha_{\rm vir} = 2$
to $\alpha_{\rm vir} = 10$ never taking longer than half a freefall
time. Star formation still continues slowly as clouds continue to
expand beyond $\alpha_{\rm vir} = 5$, but if we look at the time
$t_{80}$ as shown in
Figure~\ref{Fig:Lifetimes}a, it is generally similar to $t_{\rm unb}$,
while $t_{90}$ roughly corresponds to clouds reaching a virial
parameter of $\sim 10$.

Given that the first stars only begin forming at around $\sim 0.5 t_{\rm
  ff,0}$, and clouds only reach a state where the artificial initial
conditions are erased at $\sim 0.8 t_{\rm ff,0}$ (where this number is
taken from both the minimum in the virial parameter, and the break
times in fits to the SFR), this means that the majority of clouds form
stars over a period shorter than a global freefall time.

This can be seen if we consider the cloud star-forming time, defined
as $t_{\rm unb} - t_*$, in Figure~\ref{Fig:Lifetimes}b. Since
$\varepsilon_{\rm ff,\bar\rho}$ varies mildly 
but $\varepsilon_{\rm final}$ increases more strongly with $\Sigma$, 
there is an increase in cloud lifetime in units of $t_{\rm ff,0}$ 
with $\Sigma$, as the
more massive clouds will convert more of their gas to stars before the
effects of radiative feedback dominate and, additionally, begin forming
stars at a slightly earlier time.  However, as high-$\Sigma$ clouds
also have much shorter freefall times, we find that, generically, cloud
star formation times ($t_{\rm unb} - t_*$ or $t_{\rm 90} - t_*$) 
are between $2$ and $8~$
Myr (see Figure ~\ref{Fig:Lifetimes}c), with more massive clouds being
slightly shorter-lived than their low surface density
counterparts. Since SN feedback can only act more than 3 Myr after
$\sim t_*$, it seems likely that if there were no other source of star
formation suppression, direct radiation pressure would be able to
disperse clouds across a wide range of surface densities before SNe
begin to impact cloud dynamics.  However, this cloud dispersal would
be at the expense of a larger net star formation efficiency than
inferred from observations in many Milky Way clouds.

This apparent conflict with observation may potentially be explained in 
several ways. One possibility is that GMCs are strongly affected by additional 
sources of internal feedback not modeled here 
(such as ionizing radiation\footnote{
High-pressure ionized gas can drive expansion of the surrounding 
neutral gas and can itself directly escape the cloud from 
``blister'' HII regions if the potential well is not too deep.}) 
or external feedback 
(including supernova blast waves from stars formed in other GMCs). 
Another is that GMCs may have effective virial parameters exceeding 2,
e.g., if the outer parts are still condensing even as the inner parts 
begin vigorous star formation.

\section{Summary and Discussion}
\label{Sec:Conclusion}

We have carried out three-dimensional RHD simulations of internal
gravitational collapse, star formation, and destruction of turbulent
models of GMCs.  Our models consider a range of masses, radii, and
initial virial parameters representative of observed Milky Way GMCs,
with initial surface densities in the range 10 - 300 $~M_\odot~{\rm
  pc}^{-2}$.  Each cloud is initialized with power-law turbulence and
simulated in a computational domain twice the initial cloud diameter,
adopting an isothermal equation of state for the gas.  Sink particles
(representing star clusters) formed via gravitational collapse become
sources of radiation, with a constant luminosity-to-mass ratio $\Psi =
2000 {~\rm erg~s^{-1}~g^{-1}}$.  We follow each cloud for four
(initial) freefall times, until all the gas mass is either in stars or
has been expelled from the box.  Cloud destruction is a consequence of
the direct radiation forces applied to the gas and the relatively high opacity
$\kappa = 1000 {~\rm cm^2~g^{-1}}$, appropriate for non-ionizing UV.
We incorporate the effects of radiation, from sink/star particles but not re-emitted IR, 
using \textit{Hyperion}, which provides time-dependent solutions for the radiation energy density and
flux.  Our main goal is to
investigate the effect that radiation forces from distributed stellar
sources have on the SFE and SFR of turbulent, star-forming clouds.

The clouds in our simulations follow similar evolutionary
tracks. Initially, there is a period of structure formation and
fragmentation driven by turbulent compression and self-gravity.  For
clouds with initial $\alpha_{\rm vir,0} = 2$, by
$t\sim 0.6 t_{\rm ff,0}$ this creates a filamentary gas distribution in which $\sim 10 \%$ of
the cloud mass has collapsed to form stars, and a similar fraction has
become unbound by the initial turbulence.  This is followed by a
period of rapid star formation and a transition to cloud expansion
driven by radiative feedback.  Star formation proceeds but at a slower
rate, and it is largely complete by $\sim 2 t_{\rm ff,0}$.

All of our clouds share a number of common features in their gas
density distributions, and show similar SFR scalings and coefficients.
However, the SFE over a cloud's lifetime depends sensitively on a
cloud's mean surface density and initial virial parameter. 
For virialized, turbulent clouds, we show that limits exist on the SFE, 
when it is interpreted in terms of a localized, sequential competition between gravity and
secularly increasing radiation forces in a cloud with a lognormal
distribution of surface densities.


Below, we summarize the similarities and differences among our models,
as well as our key conclusions regarding the roles of radiation feedback in
controlling star formation and GMC evolution.

\begin{itemize}

\item[1.]{\it Surface Density Distribution}

After an initial transient phase (lasting $\sim 0.4
t_{\rm ff,0}$ for our fiducial model), the surface densities in our
model clouds approach lognormal distributions.  The variances
$\sigma_{\ln \Sigma}^2$ in the PDFs of mass and area as a function of
$\ln\Sigma$ are similar, consistent with expectations for a lognormal distribution. Meanwhile, the 
means (PDF peaks) are close to the predicted $\mu_M = \mu_A +
\sigma_{\ln \Sigma}^2$, where $\mu_A$ and $\mu_M$ are, respectively, the 
area- and mass-weighted means of $\ln\Sigma$.  
The mean value of the gas surface density and
PDF peaks slowly decrease over time as gas is accreted onto star
particles and expelled from the cloud by radiation forces, while the
width of the PDF only slightly increases. After $t\sim t_{\rm ff,0}$, by which
time star formation is already well underway, 
$\sigma_{\ln  \Sigma}$ begins to decrease after the radiation field becomes strong
enough to disperse low-density gas (see, e.g., Fig.
\ref{Fig:LogNormalFits}b).  
For all of our models, which have initial Mach number between 10 and 40 
and initial virial parameter between 0.1 and 10,  
the width of the PDF ranges only over
$\sigma_{\ln \Sigma}\sim 1-2$, with $\sigma_{\ln \Sigma}\sim 1-1.5$ 
for the $\Sigma$ series (see Fig.  \ref{Fig:LogNormalAlpha}).

The lognormal distributions of surface density found in our
simulations are generally consistent with observations of a range of
GMCs, which also show power-law tails at high column densities
associated with star-forming regions
\citep[e.g.,][]{Goodman2009,Kainulainen2009,Lombardi2010,Lombardi2015,Schneider2013,Schneider2015}.
Our simulations do not show the emergence of a clear power-law tail in the
PDF as star formation progresses \citep[cf.][]{Klessen2000,
  Vazquez-Semadeni2008, Federrath2008, Kritsuk2011, Collins2012,
  FederrathKlessen2013,Lee2014}, likely because our global cloud
models lack resolution at the highest densities. The measured values
of $\sigma_{\rm ln \Sigma}$ are somewhat larger in our simulations
than in nearby well-studied clouds \citep{Schneider2015}, although
more massive, more turbulent GMCs are likely to have broader PDFs.

The stationary lognormal form of the surface density PDF during the
main star formation epoch has significant implications, as it allows
us to predict the maximum stellar mass that can be formed before 
clouds are dispersed via radiative feedback (see below).

\item[2.]{\it Time Dependence of the Star Formation Rate} 

After an initial transient (ending at $t_{\rm break} \sim 0.8 t_{\rm
  ff,0}$, when $\sim 10\%$ of the gas has collapsed to make stars), the
SFR in our simulations reaches a near-constant value with $M_* \propto
(t-t_{\rm break})^\beta$, for
$\beta \sim 0.8-1.2$ (see Figure
\ref{Fig:FiducialEvolExponent}).  The majority of the stars in the
cloud are therefore formed at near-constant SFR.

Analyses of previous driven-turbulence simulations \citep{Wang2010,
  PadoanNordlund2011, Bate2012, Krumholz2012, FederrathKlessen2012}
have mostly concluded that the SFR is approximately constant, as we
do. Other simulations have also shown an initial slow
phase of stellar growth, which is generally treated as a transient
effect arising from unrealistic cloud initial conditions, where the
velocity structure is not consistent with cloud
self-gravity. However, the idea of a constant SFR has recently come
into question, with \citet{Myers2014,Lee2014} suggesting that the
initial behavior is not simply a transient; instead, self-gravity
alters the global density structure, resulting in an SFE closer to
quadratic than linear in time.

Although our simulations are not perfectly suited to resolving the
discrepancy between these two views (as we start from very artificial
initial conditions), there does appear to be a very distinct break
in the stellar mass evolution at $M_{*} \sim 0.1~M_{\rm cl,0}$,
after which the stellar mass grows linearly.  Future simulations which
start with more realistic initial density and velocity distributions
(extracted from larger-scale galactic-disk models), while also
including physical feedback rather than idealized forcing, should
provide more realistic understanding of histories of star formation in
turbulent clouds.

\item[3.]{\it Star Formation Efficiency per Freefall Time}  

The roughly constant SFR in our simulations (after the calculated
break time) makes it straightforward to calculate the SFE per freefall
time, $\varepsilon_{\rm ff,\bar \rho} \equiv \langle\dot{M_*}\rangle
t_{\rm ff,\bar\rho}/M_{\rm cl, 0} $ (see Equation
\ref{Eq:epsff_lin}). We find
$\varepsilon_{\rm ff,\bar\rho} \sim 0.3 -0.5$ (Figure~\ref{Fig:SFERates}a) for clouds with $\alpha_{\rm vir,0} =
2$, similar to results from other recent simulations of turbulent,
star forming clouds
\citep{PadoanNordlund2011,Padoan2012,FederrathKlessen2012,Myers2014,
  Lee2014}. Our measured $\varepsilon_{\rm ff,\bar \rho}$ values increase
slightly with increased surface density or higher Mach number
but depend more strongly on initial virial parameter. As the initial
$\alpha_{\rm vir,0}$ increases above unity, $\varepsilon_{\rm ff,\bar \rho}$
decreases systematically (Figure~\ref{Fig:SFERates}b).
For the high initial $\alpha_{\rm vir,0}$ models, $\varepsilon_{\rm ff,\bar \rho}$
can be as low as 0.1. These results are
in line with other (driven-turbulence) simulations, in which the Mach
number and magnetization affect the SFR modestly, but the value of
$\varepsilon_{\rm ff}$ depends strongly on $\alpha_{\rm vir}$
\citep[e.g.,][]{Padoan2012}.  
It must be kept in mind, however, that in our models $\alpha_{\rm
  vir}$ is not constant.  This differs from driven-turbulence
simulations, in which strong or weak driving can maintain either a low
or high level for $\alpha_{\rm vir}$, and $\varepsilon_{\rm ff}$ secularly 
decreases with $\alpha_{\rm vir}$. For our low turbulence models,
$\alpha_{\rm vir}$ grows to reach unity well before $t_{50}$,
which explains why $\varepsilon_{\rm ff,\bar \rho}$ is
relatively constant for $\alpha_{\rm vir,0}\lesssim 1$.
For high initial turbulence models, $\alpha_{\rm vir}$ drops but never reaches
$\sim 1$ as clouds disperse from the simulation volume before this
occurs, which explains why $\varepsilon_{\rm ff,\bar \rho}$ decreases for
$\alpha_{\rm vir,0}\gtrsim 2$.

We note that direct radiation feedback does not significantly alter
$\varepsilon_{\rm ff, \bar\rho}$ (Figure~\ref{Fig:SFERates}).  
In comparison to simulations with no
feedback, the SFR is mildly reduced: the suppression is stronger at
low surface density, but is never more than a factor of $\sim
2/3$. This implies that the primary role of the direct radiation force
that we have studied is in truncating star formation by removing gas
from clouds, rather than in altering their internal states and
star-forming properties.

For both our simulations and others, the low values of $\varepsilon_{\rm
  ff} \lesssim 0.1$ inferred from many observations
\citep[e.g.,][]{Krumholz2012} are only achieved for models with large
$\alpha_{\rm vir}$, i.e., unbound rather than bound clouds, or
systems where gas concentrations are dispersed by turbulence faster
than they collapse gravitationally. Traditionally, GMCs have been
believed to be gravitationally bound structures, i.e., with
$\alpha_{\rm vir} \sim 1$ \citep[e.g.,][]{Solomon1987, Fukui2008,
  Bolatto2008, Wong2011}. However, gas masses (and therefore virial
parameters) of clouds are in fact uncertain, because they either rely
on adopting a constant $X_{CO}$ for $^{12}$CO, or using another tracer
such as $^{13}$CO that may not be in LTE in some locations and may be
optically thick in others \citep{Bolatto2013}.  Indeed,
\citet{Roman-Duval2010} find a wide range of $\alpha_{\rm vir}$ from
the Galactic Ring Survey sample. In principle, it would be possible to
reconcile theory with observations if molecular gas cycles through
both high-$\alpha_{\rm vir}$ and $\alpha_{\rm vir}\sim 1$ states, spending most of its
time in the former as ``diffuse'' gas (see below) and creating stars
rapidly only during the latter. To resolve this issue, it will be
crucial to obtain empirical measures of the mass fractions of
molecular gas at different values of $\alpha_{\rm vir}$.

\item[4.]{\it Cloud Lifetimes}

The consequence of their relatively large $\varepsilon_{\rm ff}$ is that our
model clouds evolve quickly, converting some fraction of their gas
mass to stars and dispersing the rest on very short timescales 
(see Fig. \ref{Fig:Lifetimes}).  We
find cloud lifetimes $\sim 1.5 - 2 ~t_{\rm ff,0}$, with the duration of the star
formation epoch $\sim 0.7 - 1.5 ~t_{\rm ff,0}$.  For our range of
parameters, cloud lifetimes are between $2$ and $8~$Myr.

Observationally, GMC cloud lifetimes are estimated to be $\sim
20-40~$Myr \citep{Leisawitz1989, Kawamura2009, Miura2012, Gratier2012,
  Meidt2015}, considerably longer than lifetimes of our model
clouds. These estimates typically involve dividing clouds into three
distinct populations: Type I with no stars, Type II with HII regions,
and Type III with star clusters and HII regions, and then adding
together their individual lifetimes. Because our simulated clouds have
artificial initial conditions, including containing all their gas
initially rather than accreting it over time, they do not properly
model the first two phases that are seen in observed GMCs.  The
duration of the main star formation/cloud dispersal epoch in
our simulations (see Fig.  \ref{Fig:Lifetimes}c) is only a factor 
$\sim 2$ below observed duration estimates of the Type III phase 
$\sim7$~Myr.  

\item[5.]{\it Lifetime Star Formation Efficiency}

In our numerical simulations, we define the net SFE 
over a cloud lifetime $\varepsilon_{\rm final}$ as the fraction of a
cloud's initial mass that ends up in star particles.  We measure
values in the range $\varepsilon_{\rm final} \sim 0.1 - 0.6$, increasing secularly
with the cloud's initial surface density according to 
$\varepsilon_{\rm final} = 0.37~{\rm log}\Sigma -0.26$ 
over a broad range between $10$ and $300~{\rm M_{\odot}
~pc^{-2}}$ (Figure \ref{Fig:Efficiencies}a).

Our simulations also show a decrease in the efficiency with
  increasing virial parameter, $\varepsilon_{\rm final} = -0.45~{\rm
    log}\alpha_{\rm vir,0} + 0.51$ (Figure \ref{Fig:Efficiencies}b).
  However, the role that turbulence plays in setting the star
  formation efficiency is more difficult to interpret than that of the surface
  density, as there are a number of competing effects to disentangle,
  some of which are dependent on our initial conditions. Turbulence
  provides support for the cloud, thereby preventing collapse and
  slowing down star formation, but at the same time, turbulence also 
  broadens the density and
  surface density distributions, making it more difficult for radiation to
  drive gas out of the cloud; hence the final efficiency is
  higher (see item 6 below). 

Naively, our simulations might be taken to suggest that the first effect is
dominant, since the star formation efficiency decreases with
increasing virial parameter. However, it must be stressed that the
lower turbulence clouds start from an artificial state and collapse to
a new state that is higher in both surface density and virial
parameter. This means that the increased star formation efficiency
at low $\alpha_{\rm vir, 0}$ can
be understood entirely in terms of an enhancement in surface density.
Overall, the reduction of star formation by turbulence is relatively
modest, in that 
$\varepsilon_{\rm ff} \sim 0.3-0.5$ for clouds once they have reached
a natural ``virialized'' state.  As discussed below, however,
the limitation of radiation effects by turbulence-induced compression
can increase $\varepsilon_{\rm final}$ by a large factor (more than an
order of magnitude) compared to the case in which the density is uniform.

Except at large values of initial $\alpha_{\rm vir}$ and small $\Sigma$, the
values of $\varepsilon_{\rm final}$ we find are larger than those observed in
Milky Way GMCs (see Section \ref{Sec:Introduction}).  This suggests
that other feedback effects that we have not included may be important
in trucating star formation in individual GMCs.  Yet, other recent
investigations of the effects of photoionization on GMC evolution
\citep{Walch2012,Dale2012,Dale2013a} have also found that star
formation efficiencies are quite high. Potentially, the combined effects of non-ionizing and
ionizing radiation are not simply additive, such that the overall impact on
limiting star formation in a cloud is much greater.  However, it is
also possible that star formation feedback from 
supernovae---either within clouds or originating at nearby
locations---is more important than the combined early feedback 
in unbinding the majority of the mass in a
GMC.  Alternatively, if GMCs have large virial parameters (see below)
or are magnetically subcritical, they could have significantly reduced
SFRs.  It is important to explore all of these
alternatives in future simulations.

If most molecular gas is in GMCs, the SFEs and lifetimes of clouds
combine to determine the overall molecular depletion time in a galaxy 
(or galactic region).  Averaging
over the star-forming epoch of model clouds leads to an effective 
depletion time $t_{\rm dep} = \Delta t/\varepsilon_{\rm final}=4$ to $20$~Myr for the
$\Sigma$-series  (Fig. \ref{Fig:Lifetimes}d), 
where $\Delta t \equiv t_{\rm unb} -t_*$ or $t_{90} -t_*$.  
For the $\alpha$-series, the corresponding range is 
2 to 50~Myr.  These values are very
small compared to the Gyr extragalactic depletion times measured for
CO-emitting gas (see Section \ref{Sec:Introduction}).  
The ratio $t_{\rm dep}/t_{\rm ff,0}$ is 2.0 to 2.9 for
the $\Sigma$-series, and 0.5 to 12 for the $\alpha$ series.  
While inclusion
of additional early feedback effects and magnetic fields would likely
reduce the SFE and increase 
$t_{\rm dep}$ over a cloud lifetime, it is also possible that much of
the CO-emitting gas is in fact not in strongly-bound systems.
These conditions would be more similar to high-$\alpha_{\rm vir}$ models than near-virial
cases. In addition, if
observed CO-emitting gas is at lower density, with longer $t_{\rm ff,0}$
than the range we have considered, it would also tend to increase the
depletion time.  

Maintaining high enough
$\alpha_{\rm vir}$ and/or low enough mean density to match the depletion times
observed in extragalactic systems likely requires much more strongly
driven turbulence than radiation feedback alone can supply.  It is possible
that in molecule-dominated regions of galaxies, similar to
atomic-dominated regions, most of the gas is effectively diffuse, and 
turbulence is primarily driven by late-stage expanding SNRs. In numerical
simulations of diffuse-dominated galactic disk
regions with dynamics governed by
momentum input from SNRs, the values $\varepsilon_{\rm ff}\sim 0.006$ are
indeed found to be quite small, independent of the large-scale
mean gas surface density in the disk \citep{ Kim2013}.

\item[6.]{\it Analytic Limits on the Star Formation Efficiency}

For a given total stellar luminosity as set by the instantaneous
  value of $\varepsilon$, only gas in structures of sufficiently low
  surface density will have the outward radiation force exceed the
  inward gravitational force.  This defines a (time-varying)
  ``Eddington'' surface density $\Sigma_E$ (which depends on
  $\varepsilon$ according to Equation~\ref{Eq:FinalSigma}) below which
  gas can be expelled.  Our simulations (and other work) show that the
  PDF of surface densities in a cloud follows a lognormal distribution
  set by the mean surface density of a cloud and its internal
  turbulence.  For a cluster-forming cloud with a given PDF of
  circumcluster surface densities, there is a maximum fraction of the
  original cloud material that can become super-Eddington.  At low
  SFE, the luminosity is low and only a small fraction of the mass (in
  structures with very low $\Sigma$) can be driven out of the cloud; if
  the SFE is high, all of the remaining gas would be super-Eddington, but
  there would be very little material available.  This suggests that
  clouds may evolve by sequential expulsion of portions of gas at
  increasingly high surface density until the maximum mass in outflowing
  material is reached.  We argue that the maximum final
  SFE
  $\varepsilon_{\rm max}$ for $\alpha_{\rm vir} \sim 1$ clouds would then depend
  only on the
  initial cloud mean surface density and the variance of the lognormal
  (which does not vary much over our
  models). Equation~(\ref{Eq:LognormalEff}a) provides a prediction
  for $\varepsilon_{\rm final}$ based on this formulation,
  which agrees quite well with our numerical results.

\citet{ThompsonKrumholz2014} also developed an analytic model for the
net SFE in a cloud that accounts for the lognormal distribution of
surface densities relative to a critical value that depends on the
total stellar luminosity; it differs from our model in that it
explicitly depends on timescales for star formation and mass ejection
as parameters. While, for their default parameter values, their  
predicted SFE is much lower than we find numerically, it is closer
when a larger value for $\varepsilon_{\rm ff}$ is adopted.  Both
our analytic model and our numerical simulations show that allowing for a
nonuniform (lognormal) surface density distribution leads to much
greater final SFE than in simpler ``Eddington''-type models with a
single instantaneous mean surface density, as in 
Equation~(\ref{Eq:FinalEfficiency})
\citep[see also][]{Murray2010,
  Fall2010, DekelKrumholz2013,Kim2016}.

Finally, we note that in sufficiently dense clouds, at least the upper
end of the lognormal surface density distribution will be optically
thick to IR.  It will be interesting to extend into this regime and
compare numerical and analytic models that allow for both direct and
reprocessed radiation forces.
We also note that Equation~(\ref{Eq:LognormalEff}a) predicts a decrease in 
$\varepsilon_\mathrm{max}$
at lower $\sigma_{\ln \Sigma}$ (see Figure \ref{Fig:EffVarSigma}).  
Potentially, inclusion of magnetic fields could reduce $\sigma_{\ln \Sigma}$
and therefore
the SFE; as real GMCs have significant magnetization, this represents an 
important question to address with future simulations.

\end{itemize}

\acknowledgements
 
We are grateful to the referee for a careful reading of the manuscript and
very thorough report, which helped us to improve the presentation.  
This work was supported by Grant No. AST-1312006 from the National
Science Foundation.  Part of this project was conducted during a visit
to the KITP at U.C. Santa Barbara, which is supported by the National
Science Foundation under Grant No. NSF PHY-1125915. MAS is supported by 
the Max-Planck/Princeton Center for Plasma Physics under Grant No.
NSF PHY-1144374. Simulations were performed 
on the computational resources supported by the PICSciE TIGRESS High
Performance Computing Center at Princeton University.

\appendix

\section{Tests of the Numerical Code}
\label{Sec:Tests}

In this section, we present tests that examine the extent to which
numerical approximations in our code and limited numerical
resolution might affect our results.  In particular, we are
interested in quantifying the regimes of cloud mass and radius (or,
equivalently, surface density) over which we accurately capture the
physics of radiatively-driven expansion.

Both the \textit{Athena} MHD code and the \textit{Hyperion} RHD
extension have been tested extensively in the past. However, the
majority of the tests of the \textit{Hyperion} module were performed
with radiation in the diffusion limit. This would be satisfied in
very optically thick clouds, in which the effects of reprocessed IR
continuum radiation are dominant \citep{SkinnerOstriker2015}.
Here, we are interested in the case where gas is optically thick to UV photons from the source and
optically thin to re-emitted IR radiation. In this context, the
interaction of gas and radiation is very different, as there is strong
local absorption rather than absorption and re-emission over large
volumes. We therefore implement several tests to see how well we
model this behavior in successively more realistic scenarios.

\subsection{Radiative Momentum-Driven Expanding Shell}
\label{SubSec:Expanding Shell}

To test our code behavior in highly idealized conditions,
we consider the expansion of a spherical shell of gas due to the
absorption of radiation momentum from a central source, similar to that
described in \cite{OstrikerShetty2011, SkinnerOstriker2013}.  In this
problem, we imagine an idealized spherical GMC of mass $M_{\rm cl,0}$
that forms stars of total mass $M_*$ with efficiency $\varepsilon=
M_*/M_{\rm cl,0}$. The remaining gas of mass $M_{\rm sh} \equiv
(1-\varepsilon)M_{\rm cl,0}$ is ejected as an expanding, spherical
shell of radius $r$ due to the radiation force from the stellar
component. The stars are modeled here as a centrally-located cluster.

We initialize a shell at an initial radius $r_0$ with zero velocity at
time $t=0$.  In reality, the shell would have some initial velocity,
but this just provides an additive constant. Assuming
the ejected shell is thin and of uniform surface density $\Sigma(r)
= M_{\rm sh} / (4 \pi r^2)$, the optical depth across the shell is
$\tau_{\rm sh}(r) \approx \Sigma(r) \kappa$, where $\kappa$ is the
absorption opacity of the gas to UV photons. If, as discussed earlier,
we are in the limit where we consider only UV radiation, then the flux
at r is given by
\begin{equation}
	F(r) = \frac{L_*}{4\pi r^2}{\rm exp}\left[-\int^r \rho(r')\kappa dr'\right] 
	\equiv \frac{L_*{\rm e}^{-\tau_{\rm sh}(r)}}{4\pi r^2}.
\end{equation}
The total radiation force on the shell is then 
\begin{equation}
	\int F(r) \rho(r) \frac{\kappa}{c}4\pi r^2 dr = \frac{L_*}{c} \int {\rm e}^{-\tau}d\tau
	= \frac{L_*}{c}(1 - {\rm e}^{-\tau_{\rm sh}(r_{\rm max})}).
\end{equation}
Neglecting gravitational and internal pressure forces, the outward acceleration of the 
shell becomes
\begin{equation}
	\ddot{r} = \frac{L_*\left[1 - {\rm exp}\left(-\Sigma(r) \kappa\right) \right]}{M_{\rm sh} c} ,
\end{equation}
which is independent of the shell's thickness. Substituting for the surface density and the 
cluster luminosity, the shell acceleration reduces to
\begin{equation}
	\ddot{r} = \frac{\Psi \varepsilon(1 - {\rm e}^{-\tau_0(r_0 / r)^2})}{c (1 - \varepsilon)} ,
	\label{Eq:rddot}
\end{equation}
where we have introduced the shell optical depth at $r = r_0$, given by
\begin{eqnarray}
	\tau_0 &\equiv& M_{\rm sh} \kappa / (4 \pi r_0^2) \nonumber \\
	&=& 1.67 \left(\frac{r_0}{10~{\rm pc}}\right)^{-2}
	\left(\frac{\kappa}{1000~{\rm cm^{2}~g^{-1}}}\right)
	\left(\frac{M_{\rm sh}}{10^4~M_{\odot}}\right).
	\label{Eq:taudepth}
\end{eqnarray}
We may simplify further by rewriting Equation~(\ref{Eq:rddot}) in terms of the dimensionless variables 
$\tilde{r} \equiv r/r_0$ and $\tilde{t} \equiv t/t_0$, so that 
\begin{equation}
	\frac{d^2\tilde{r}}{d\tilde{t}^2} = \frac{\varepsilon(1 - {\rm e}^{-\tau_0 / \tilde{r}^2})}{(1 - \varepsilon)},
	\label{Eq:rddotdimless}
\end{equation}
where
\begin{eqnarray}
	t_0 &\equiv& \sqrt{\frac{r_0 c}{\Psi}} \nonumber \\
	&=& 0.68~{\rm Myr} \left(\frac{r_0}{10~{\rm pc}}\right)^{1/2}
	\left(\frac{\Psi}{2000~{\rm erg~s^{-1}~g^{-1}}}\right)^{-1/2}.
	\label{Eq:DiffFullShell}
\end{eqnarray}
The presence of the exponential term in inverse radius precludes a general analytic solution to this 
problem. However, in the optically thick ($\tau_{\rm sh} \gg 1$) limit, or close to 
the initial shell radius with $\tau_{\rm sh}(r) \approx \tau_0$, all explicit dependence on radius drops 
out of Equation~(\ref{Eq:rddotdimless}). For the $\tau \approx \tau_0$ case it may then be solved 
trivially to give a quadratic expansion in time
\begin{equation}
	\tilde{r} = \frac{\varepsilon(1 - {\rm e}^{-\tau_0})}{2(1 - \varepsilon)} \tilde{t}^2 + 1.0; 
	\label{Eq:QuadShell}
\end{equation}
in the optically thick case, we instead have $1 - {\rm e}^{-\tau_0} \rightarrow 1$. We note that if the 
shell is optically thick, then to first order the solution depends only on the inital shell radius, star 
formation efficiency $\varepsilon$, and luminosity per unit mass $\Psi$ (see Equation~\ref{Eq:rddot}). 
Otherwise, dependence on the mass and opacity only enters through $\tau_0$ 
(see Equation~\ref{Eq:taudepth}) and is only significant for relatively low optical depths.

\subsubsection{Convergence Tests}
\label{SubSubSec:Convergence}

\begin{figure*}
  \centering
  \epsscale{1}
  \includegraphics{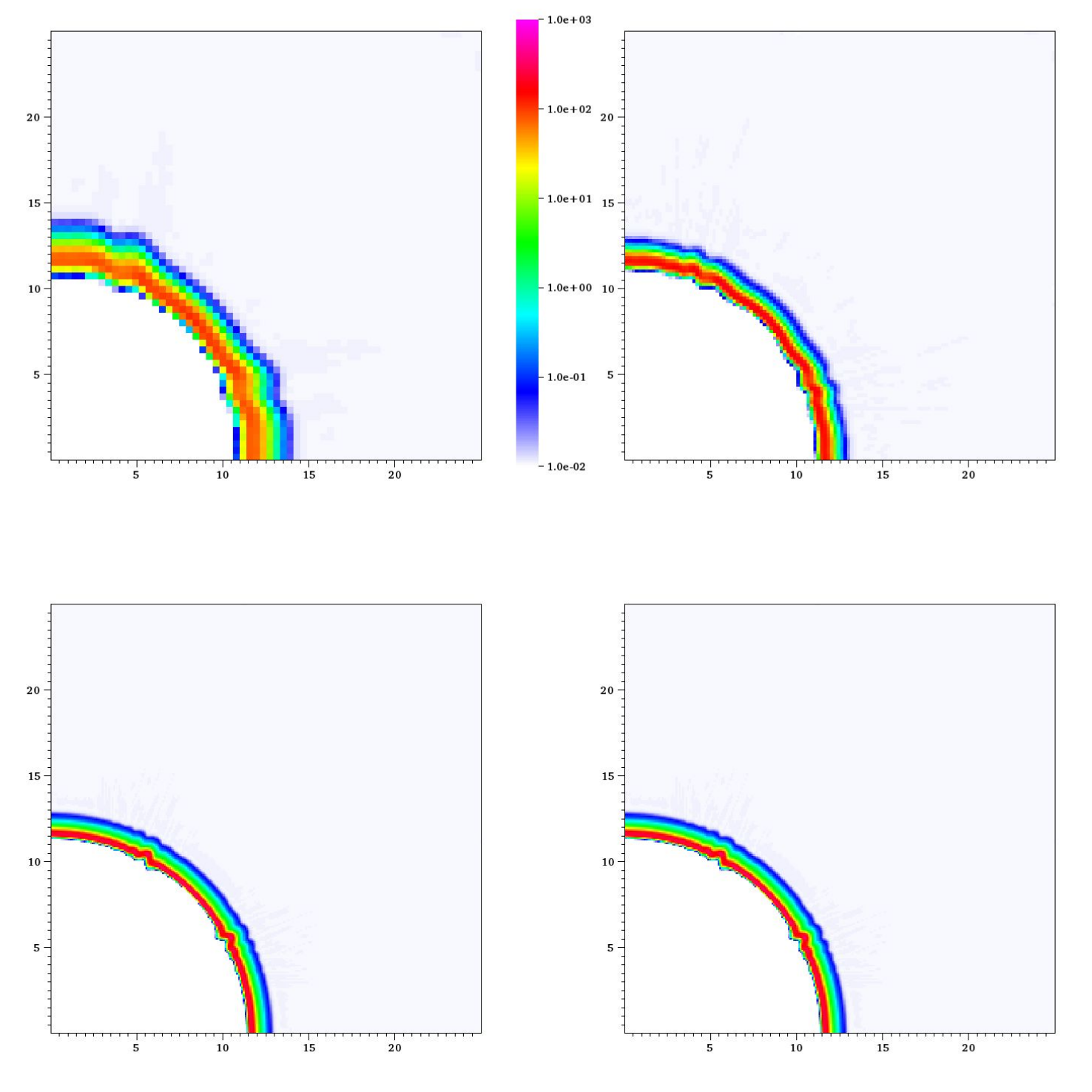}
  \caption{Snapshots of the density at $t = 1.2t_0$, for the fiducial
    spherical shell problem, with cloud mass $M_{\rm cl, 0} = 10^4~M_{\odot}$. 
    We show (reading from left to right and
    top to bottom) simulations with $N = 64, 128, 256$ and $512$
    respectively. In each case, the snapshots show 2D slices through
    the x-y plane of the shells. The color scale for
    the gas density $n_H$ (top) is in units of ${\rm cm^{-3}}$.}
  \label{Fig:RSS_VarNSlice}
\end{figure*}

\begin{figure}
  \centering
  \epsscale{1}
  \includegraphics{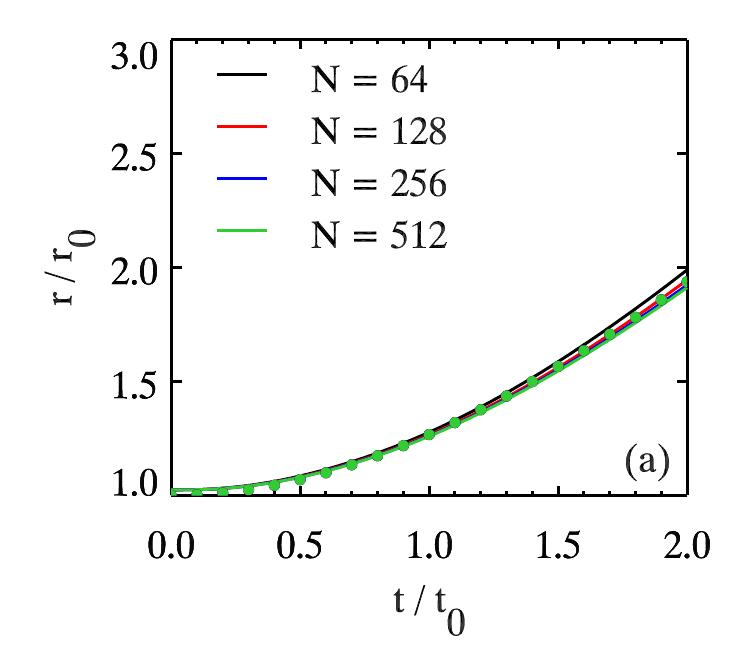} \\
  \includegraphics{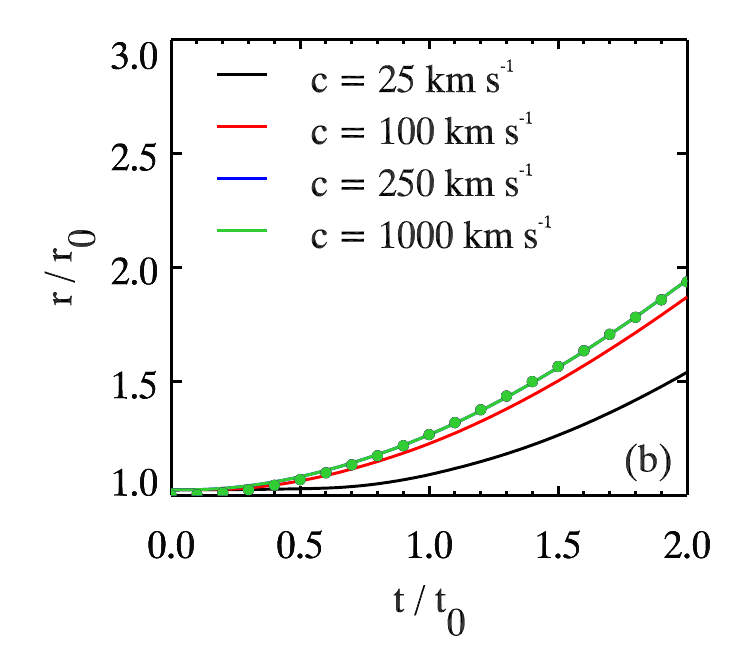} 
  \caption{Mass-weighted shell radius as a function of time for (a)
    varying resolution $N$ and (b) reduced speed of light $\hat{c}$, 
    in the spherical shell problem. In each case, we use
    $M_{\rm cl, 0} = 10^4~M_{\odot}$, $r_0
    = 10$~pc, $\Psi = 2000~{\rm erg~s^{-1}~g^{-1}}$, $\kappa =
    1000~{\rm cm^{2}~g^{-1}}$, and $\varepsilon = 0.5$. The key shows
    (a) the resolution in cells and (b) $\hat{c}$ in ${\rm km
    s^{-1}}$. The dotted curve in each panel shows the analytic solution 
    for comparison.}  
\label{Fig:RSS_VarN}
\end{figure}

\begin{figure}
  \centering
  \epsscale{1}
  \includegraphics{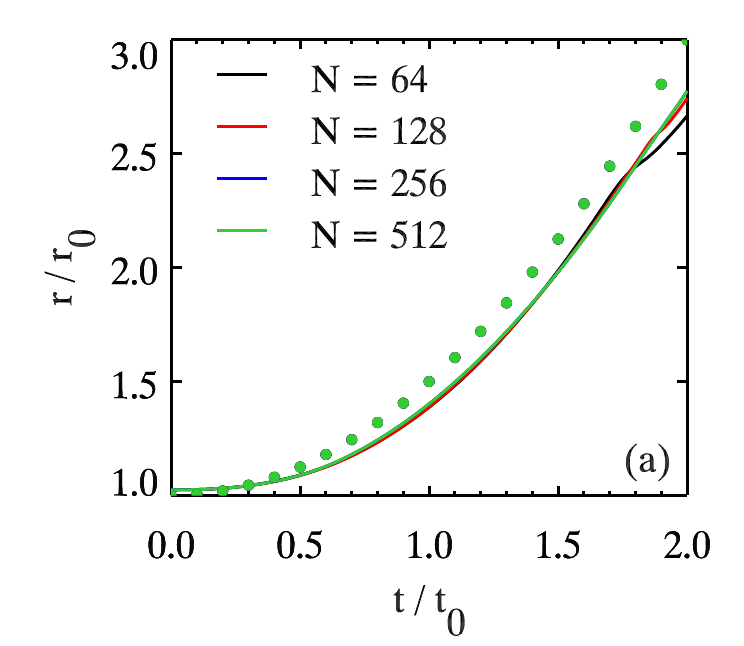} \\
  \includegraphics{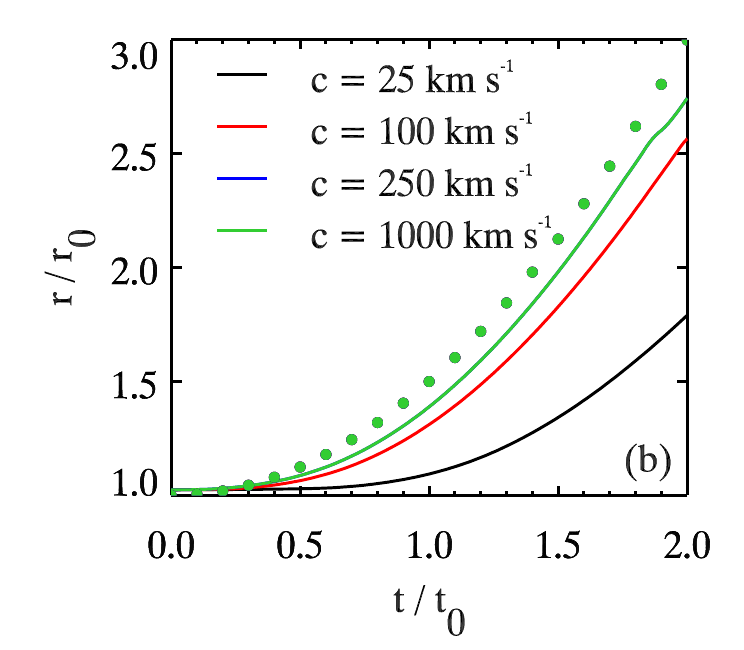} 
  \caption{Same as Figure~\ref{Fig:RSS_VarN}, except for a 
  cloud of mass $M_{\rm cl, 0} = 3 \times 10^5~M_{\odot}$.}  
\label{Fig:RSS_VarN_HighM}
\end{figure}

In this test, we wish to explore the sensitivity of our code to changes in
the key physical parameters $\kappa$ and  $M_{\rm cl,0}$, 
as well as numerical parameters $N$, and
$\hat{c}$. We consider a central
luminous cluster defined by a sink particle at the origin, with
$r_*=1$~pc for the radiation source function given in
Equation~(\ref{Eq:jprofile}). This central source illuminates a thin
shell with Gaussian density profile given by
\begin{equation}
	\rho_{\rm sh}(r) = \frac{M_{\rm sh}}{4\pi r_0^2\,\sqrt{2\pi \sigma_{\rm sh}^2}} \exp \left( -\frac{(r-r_0)^2}{2\sigma_{\rm sh}^2} \right),  
	\label{Eq:rhoprofilegauss}
\end{equation}
where $H_0 = 2\sqrt{2\ln 2}\sigma_{\rm sh}$ is the FWHM of the shell.

For this test, we initialize the radiation field using 
the result of a prior simulation in which we turn on the
cluster and evolve the radiation field without
evolving the gas hydrodynamics. The final radiation field after $\sim
10$ radiation crossing times can then be used to initialize the flux
and energy density for the case where we consider shell expansion.

To prevent the gas time steps from becoming prohibitively small, we
enforce a density floor of $\rho_{\rm min} \equiv 10^{-8}\rho_{\rm
  sh}(r = r_0)$ initially, as well as after each gas integration step.
We employ an $N^3$ grid on the domain $(x,y,z) \in ([0,3\,r_0],
[0,3\,r_0], [0,3\,r_0])$ and enforce outflow boundary conditions along
all faces of the box that do not touch the origin of the cloud. We
note that due to spherical symmetry, we only run the test in a single
octant of the sphere. We run each simulation for a time $t \approx
2t_0$ so that the planar shell reaches an outermost radius of $r
\approx 3r_0$.

For the RSLA, it is necessary to choose a value of $\hat{c}$
such that $v_{\rm max} \ll \hat{c}$ at all times. For each of our
simulations, the shell reaches a maximum velocity of $v_{\rm max} =
\varepsilon v_0 \tilde{t}_{\rm max} / (1 - \varepsilon)$, where
\begin{eqnarray}
	v_0 &\equiv& \sqrt{\frac{\Psi r_0}{c}} \nonumber \\
	&=& 14.4~{\rm km~s^{-1}} \left(\frac{r_0}{10~{\rm pc}}\right)^{1/2}
	\left(\frac{\Psi}{2000~{\rm erg~s^{-1}~g^{-1}}}\right)^{1/2}.
\end{eqnarray}
As we use $\varepsilon = 0.5$ and $\tilde{t}_{\rm max} = 2$, a choice of 
$\hat{c} = 250~{\rm km~s^{-1}}$ should be satisfactory for the RSLA in 
most situations. Of course, increased values of $\Psi$ or the star formation efficiency $\varepsilon$ 
will provide greater accelerations and so require higher values of $\hat{c}$. The robustness of our results 
to variations in $\hat{c}$ is therefore also verified below.

We begin by adopting a set of fiducial parameters roughly 
characteristic of Milky Way clouds. In addition to $\Psi = 2000~{\rm erg~s^{-1}~g^{-1}}$ and 
$\kappa = 1000~{\rm cm^{2}~g^{-1}}$, we choose an initial shell radius of $r_0 = 10$~pc, 
a relatively thin initial shell width of $H_0 = 1.0$~pc, and a cloud mass of 
$M_{\rm cl,0} = 10^4~M_{\odot}$. This mass provides a marginally optically thick shell 
$\tau_0 \approx 2$ (see Equation~\ref{Eq:taudepth}). Finally, we adopt a star formation efficiency of $\varepsilon = 0.5$; although this is high for Milky Way GMCs, it is similar to the upper range of what 
we find in our full cloud simulations. 

For this test we adopt a sound speed of $c_s = 0.5~{\rm km~s^{-1}}$, 
which lies somewhere between the turbulent velocities in our full cloud
simulations and the true sound speed (a factor of 2 lower). This helps ensure that when 
resolution is adequate, the shell does not develop thin-shell instability as
it is accelerated outward, which would compromise our ability to
compare to the analytic spherical solution.

We are initially interested in the numerical parameters $N$ and
$\hat{c}$ required to match the analytic thin shell solution for a
typical GMC. In Figure~\ref{Fig:RSS_VarNSlice} we show the results for
several fiducial simulations run at different numerical resolutions. 
Pictured are snapshots of density 
in slices through the x-y plane at $t = 1.2t_0$ when the mean radius
is $r \sim 1.7r_0$ for resolutions $N = 64, 128, 256$, and $512$.
We note that the shell width remains roughly constant or only
slightly expands to a final shell width of $H \sim 2$~pc. However, in
the lowest resolution simulation, $N = 64$, the spherical shell is
disturbed by grid scale noise. By the time the shell has expanded to
close to twice its initial radius, it is no longer spherical, but
instead, has large scale perturbations with angle caused by the
initial difficulty of resolving a spherical shell on a square grid.

Figure~\ref{Fig:RSS_VarN}a shows the evolution of mass-weighted shell
radius with time as compared to the analytic solution at varying
resolution for a cloud of mass $10^4~M_{\odot}$. Since the optical
depth has a radial dependence that appears through the surface
density, we may only find the analytic solution by numerically
integrating Equation~(\ref{Eq:rddotdimless}). 
Evidently, the numerical solution follows the analytic prediction fairly well 
even at low resolution, although for higher curvature structures, 
the accuracy would be reduced at each given resolution. 
The main conclusion from this test is that at the typical resolution of
our simulations, we are able to satisfactorily capture the predicted
expansion driven by radiation forces.

The primary other numerical parameter that may affect our results is
the reduced speed of light $\hat{c}$. In \textit{Hyperion} simulations
measuring the effect of reprocessed radiation, results can be quite
sensitive to this parameter, since the RSLA static diffusion criterion
requires that the effective radiation diffusion speed remain large
compared to dynamical speeds, i.e., $\hat{c} /\tau_{\rm max}\gg v_{\rm
  max}$, where $\tau_{\rm max}$ is the maximum optical depth across
all cells. However, for direct radiation the RSLA criterion is simply
$\hat{c} \gg v_{\rm max}$, where the highest surface density clouds we
consider typically have $v_{\rm max} \sim v_{\rm esc} \sim 15~{\rm km
~s^{-1}}$.
The shell expansion test with the fiducial cloud parameters described above has 
$v_{\rm max} \sim \sqrt{2} v_0 = 20~{\rm km~s^{-1}}$. Figure~\ref{Fig:RSS_VarN}b, showing results 
for varying $\hat{c}$, demonstrates that there is only a small error with respect to the analytic 
solution for $\hat{c} \sim 100~{\rm km~s^{-1}}$, and we recover it exactly for 
$\hat{c} = 250~{\rm km~s^{-1}}$. As discussed earlier, we have conservatively adopted 
the latter value for all our cloud simulations.

Figure~\ref{Fig:RSS_VarN_HighM}a shows the same evolution of shell radius as 
Figure~\ref{Fig:RSS_VarN}a, but for a more massive cloud with 
$M_{\rm cl,0} = 3 \times 10^5~M_{\odot}$, i.e., thirty times larger than in 
Figure~\ref{Fig:RSS_VarN}a. 
The mass of the central cluster is again half this at 
$M= 1.5 \times 10^5~M_{\odot}$. 
Evidently, the numerical tests no longer agree closely with the 
analytic solutions for higher mass shells, 
for numerical resolution $N$ in the range shown. 
In all cases, the numerical solution systematically underestimates the shell
radius. While increasing the numerical resolution may marginally
improve the agreement between analytic and numerical solutions, the
error is still close to $\sim 10 \%$ once the shell has reached the
edge of the simulation volume.  Meanwhile, if we consider the same
test for varying $\hat{c}$ (Figure~\ref{Fig:RSS_VarN_HighM}b), we see
that increasing the reduced speed of light beyond our adopted value of
$\hat{c} = 250~{\rm km~s^{-1}}$ does not remedy this discrepancy
either.  As further tests show (see below), this discrepancy can 
in fact be traced to inadequate resolution of the flux.

\subsubsection{Flux Resolution}
\label{SubSubSec:TauCell}

\begin{figure}
  \centering
  \epsscale{1}
  \includegraphics{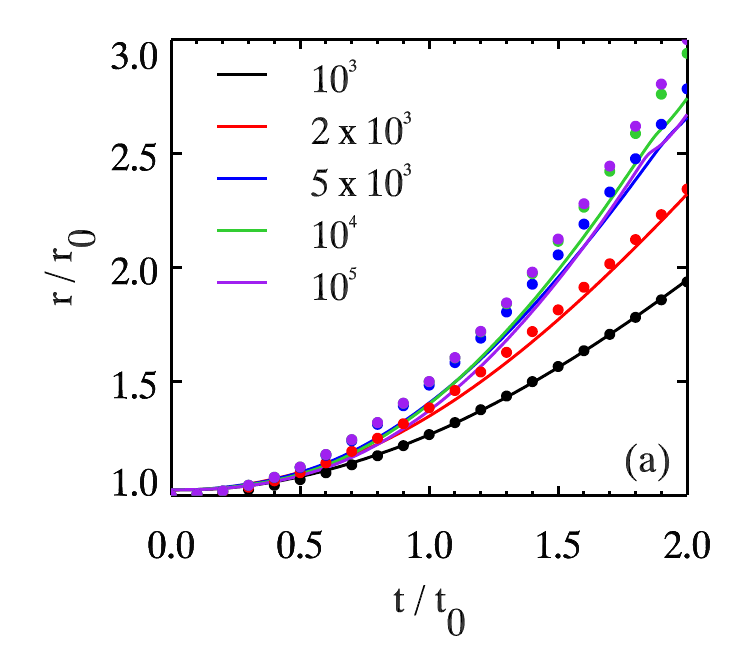} \\
  \includegraphics{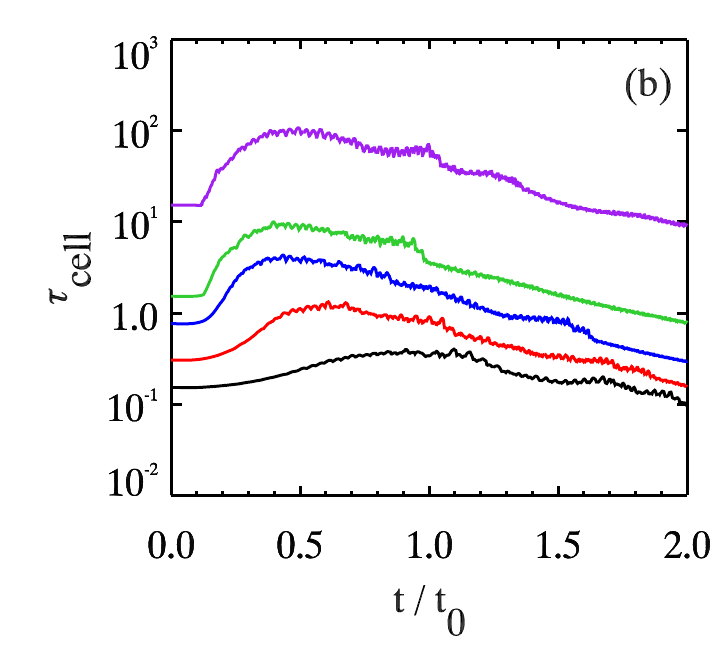}
  \caption{(a) Mass-weighted shell radius, and (b) maximum cell optical
    depth, for the spherical shell test at with varying opacity,
    $\kappa$.  The key shows the opacity
    in ${\rm g~cm^{-2}}$. Dotted curves in the upper panel show the 
analytic expansion solution. 
}  
  \label{Fig:RSS_VarKappa}
\end{figure}

\begin{figure}
  \centering
  \epsscale{1}
  \includegraphics{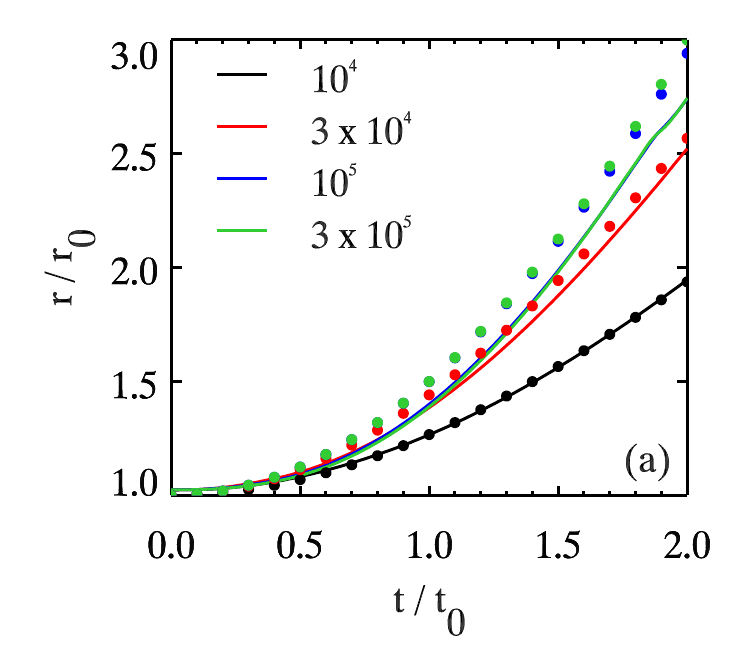} \\
  \includegraphics{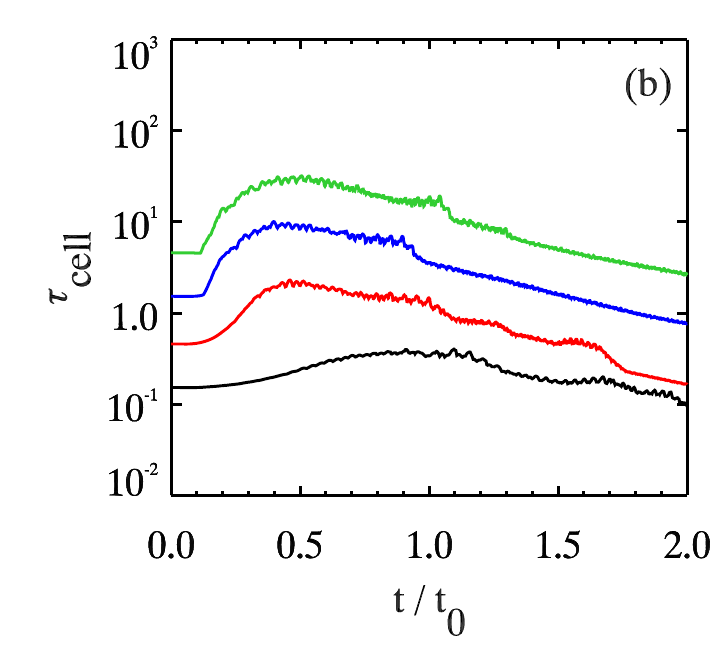}
  \caption{Same as Figure~\ref{Fig:RSS_VarKappa}, but for varying 
   cloud mass (in units $M_\odot$, as shown in the key).}
  \label{Fig:RSS_VarMass}
\end{figure}

The optical depth across individual cells in our grid is $\tau_{\rm cell} = \rho\kappa\Delta x$. 
If this optical depth exceeds unity by a considerable amount, then the flux across cells is not well 
resolved spatially and the impulse provided to the gas is not captured accurately.  For the shell problem, 
with $\rho_{\rm sh}(r) \approx \Sigma / H$ and ${\Delta x} = 4r_0 / N$, 
\begin{equation}
	\tau_{\rm cell} \rightarrow \frac{4 \Sigma \kappa r_0}{H N_x};
\end{equation}
this depends on the opacity, the shell surface density (and hence mass and radius) and the 
numerical resolution.

We may systematically test the dependence of the solution accuracy on
$\tau_{\rm cell}$ by using the shell problem with varying $\kappa$ and
$\Sigma$ (through $M_{\rm cl, 0}$).  Figure~\ref{Fig:RSS_VarKappa}a
shows the result from varying opacity over two orders of magnitude
between $10^3$ and $10^5~{\rm g~cm^{-2}}$.  Above $\kappa \sim 10^4
{\rm g~cm^{-2}}$, the shell is completely optically thick and there
is no variation in the analytic solution with opacity. The numerical
results match the analytic solutions well for opacities $\kappa =
2000~ {\rm g~cm^{-2}}$ and below, but then they deteriorate as the opacity
increases to $\kappa = 5000 ~{\rm g~cm^{-2}}$. Interestingly,
beyond this opacity, even up to $\kappa = 10^5 ~{\rm g~cm^{-2}}$
there is not a significant further increase in the error relative to
the analytic solution, and there is only a $10 \%$ deviation in radius
out to $2$ times the initial shell radius (where a real cloud could
have become gravitationally unbound).

The underlying reason for these deviations can be seen in the maximum
optical depth across individual cells as shown in
Figure~\ref{Fig:RSS_VarKappa}b.  In all cases, this increases by up to a
factor $\sim 10$ from its starting value as the shell is
compressed during expansion, and then decreases as the shell expands
outwards and the mean shell density decreases. More importantly, the
value of the maximum optical depth plays a key role in determining
solution accuracy. We see that for $\kappa = 2000$ and $\kappa =
5000~{\rm g~cm^{-2}}$, the optical depths peak at $\tau_{\rm cell} =
1.3$ and $\tau_{\rm cell} = 4.3$, respectively, and somewhere between
these, there is a transition point at
$\tau_{\rm cell} \sim 2 - 3$
beyond which the flux is not well resolved.  For $\kappa = 5000~{\rm g~
cm^{-2}}$, $\tau_{\rm cell} \gtrsim 2$ for around $0.5 t_0$, and the
numerical simulation differs from the analytic solution by around $5
\%$. For larger $\kappa = 10^4~{\rm g~cm^{-2}}$, $\tau_{\rm cell}
\gtrsim 2$ for the majority of the shell evolution, which leads to the
$10 \%$ errors discussed earlier. Beyond this point, increasing the
opacity does not have a strong effect on the solution accuracy, since
$\tau_{\rm cell} \gtrsim 2$ always.

Figure~\ref{Fig:RSS_VarMass} shows results of similar tests, in which
we vary the surface density (through $M_{\rm cl, 0}$).  Over a range
of two orders of magnitude above the fiducial surface density, we see
the same trends with cell optical depth. For $M_{\rm cl, 0} \lesssim 3
\times 10^4~M_{\odot}$, for which $\tau_{\rm cell} \le 2$ at all
times, we match the analytic solution reasonably well. However, at
larger masses and correspondingly larger $\tau_{\rm cell}$, we again
underestimate the shell expansion velocity at all times.

We conclude that, provided $\tau_{\rm cell}$ remains below $\sim 2$, we
can obtain an accurate solution for this problem. Since $\tau_{\rm
  cell}$ is inversely proportional to $N$, we can in principle capture
the behavior in increasingly high-density clouds by increasing the
numerical resolution, although in practice this becomes numerically
prohibitive for very dense systems. In any case, regardless of how
high $\tau_{\rm cell}$ becomes, we never underpredict the
shell radius by more than $10 \%$ even by the time gas in one of our
turbulent clouds will have become unbound.

Finally, we note that the maximum values of $\tau_{\rm cell}$ depend
strongly on gas compression and hence, on the detailed
problem-specific evolution of turbulent clouds. Therefore, while the
spherical tests show that a resolution of $N = 256$ is generally
sufficient for this problem, we also need to directly test convergence
in our full turbulent models.

\bibliographystyle{apj}
\bibliography{refs}

\end{document}